\documentclass[11pt]{article}

\usepackage{jheppub}
\usepackage[T1]{fontenc}

\usepackage{hyperref} 
\hypersetup{
     colorlinks=true,
     citecolor=red, 
     linkcolor=blue,
     urlcolor=blue,
     filecolor=cyan      
     linktoc=section,
               }

\numberwithin{equation}{section} 
\usepackage{cancel, braket, enumitem, float}

\usepackage[section]{placeins} 

\subheader{\begin{flushright}
\end{flushright}}

\title{ Spread of Entanglement in Non-Relativistic Theories}

\author{\centering{Sagar F. Lokhande}}
\affiliation{\centering{Institute for Theoretical Physics, University of Amsterdam,} \\
\centering{ Science Park 904, 1098 XH Amsterdam, The Netherlands.}\\
\vspace{2mm}
\centering{\textcolor{blue}{Email: sagar.f.lokhande@gmail(dot)com}}\\
\vspace{1cm}
}

\abstract{We use a simple holographic toy model to study global quantum quenches in strongly-coupled, hyperscaling-violating-Lifshitz quantum field theories using entanglement entropy as a probe. Generalizing our conformal field theory results, we show that the holographic entanglement entropy of small subsystems can be written as a simple linear response relation. We use this relation to derive a time-dependent first law of entanglement entropy. In general, this law has a time-dependent term resembling relative entropy which we propose as a good order parameter to characterize out-of-equilibrium-states in the post-quench evolution. We use these tools to study a broad class of quantum quenches in detail: instantaneous, power law and periodic.
\vspace{1cm} 
}


\begin{document}

\maketitle

\flushbottom

\section{Introduction}
\label{intro}

\subsection{Quantum Quenches and Entanglement Entropy}
\label{QQ_intro}

\noindent Studying the evolution of quantum field theories (QFTs) after generic time-dependent perturbations is an important problem. If $H_0$ denotes the time-independent Hamiltonian of a QFT defined on a manifold $\mathbb{R} \times \mathbb{R}^{d-1}$, one is often interested in a time-dependent perturbation
\begin{equation}
H(t) = H_0 + \delta H(t)  \, .
\end{equation} 
where $\delta H(t)$ could correspond to making a coupling in the Hamiltonian time-dependent \cite{Calabrese:2007rg} such as
\begin{equation}\label{eq:ham_quench}
H(t) = H_0 + \int d^{d-1} x \, J(t,x^i) \, O(t,x^i)  ,
\end{equation}
with $O(t,x^i)$ denoting a generic operator in the theory and $J(t,x^i)$ its corresponding source. A perturbation of the above kind, followed by unitary quantum evolution, is called a \textit{Quantum Quench}. If the theory was in a pure state $\ket{\Psi_0}$ before the quench, unitarity of the post-quench evolution implies that it will remain in a pure state. However, experimental studies in ultracold atoms \cite{greiner2002quantum} and theoretical arguments \cite{2008Natur.452..854R, 1977JPhA...10.2083B, PhysRevE.50.888} suggest that the end state of the evolution looks thermal to a very good approximation when studied with local, coarse-grained probes. This process is often termed as \textit{thermalization}. If a quantum quench acts uniformly at all space points such that $J(t,x^i) \equiv J(t)$, it is known as a \textit{Global Quench}. In this paper, we will only discuss quenches of this kind. \\

\noindent Their study was initiated in \cite{Calabrese:2007rg} using \textit{boundary conformal field theory} techniques of \cite{Diehl:1996kd}. For the case of free field theories, \cite{Das:2014hqa} made significant progress using symmetries of the theory. Let us begin by considering a global quench to a more general conformal field theory (CFT). In this case, one-point functions are known to thermalize instantaneously \cite{Chesler:2008hg, Bhattacharyya:2009uu, Caceres:2014pda}, thus forcing one to look for better probes of thermalization in CFTs. Natural choices are non-local observables like two-point correlation functions, Wilson lines and entanglement entropy. Although all of these have been studied recently \cite{Balasubramanian:2011ur}, entanglement entropy is the most attractive of these, and for good reasons. As shown in \cite{Balasubramanian:2011ur}, entanglement entropy equilibriates slower compared to other quantities for a finite-duration global quench, thus determining the physical rate of thermalization of the theory. It has further been used as an order parameter for phases of quantum matter \cite{Klebanov:2007ws}. The entanglement shared between regions codify all possible correlations between them, making it very useful. For example mutual information, a combination of entanglement entropies of two regions, gives an upper bound on all possible connected two-point functions between operators in the two regions \cite{Wolf:2007tdq}. Lastly, entanglement entropy grows in time after a quench (as we will discuss in great detail later), thus making the quench a way to generate entanglement. This is a goal of great interest in quantum information theory \cite{schachenmayer2013entanglement}.\\

\noindent Entanglement entropy of a subregion $A$ (also called subsystem interchangeably) in the CFT is defined as the von Nuemann entropy of $A$
\begin{equation}
S_A \equiv - \text{tr} \, (\rho_A \, \log \rho_A ) \, ,
\end{equation}
where $\rho_A = \text{tr}_{A^c} \, \rho$ is the reduced density matrix on the region $A$. In a QFT, the entanglement entropy is generally calculated using the replica trick \cite{Calabrese:2004eu} but it is UV-divergent for any subregion $A$ and any state $\rho$, due to the short-distance divergences. To obtain a finite answer, one often considers the \textit{difference} between the entanglement entropy of $A$ in two nearby states. \cite{Calabrese:2005in} studied the time-evolution of entanglement entropy of a general subregion in (1+1) dimensional CFT and showed that the entanglement entropy increases linearly in time and saturates after a specific time. However, technical difficulties do not allow one to generalize these calculations to higher dimensions easily\footnote{Recently, some progress has been made on this account in the dilute gas limit in \cite{Casini:2015zua}}.  \\

\noindent In such cases, the AdS/CFT correspondence \cite{Maldacena:1997re, Gubser:1998bc, Witten:1998qj} has been very useful. It maps a strongly-coupled \textit{holographic} CFT to a weakly-coupled theory of gravity in AdS space. Thus quantum quenches in such CFTs correspond to classical time-evolution the semi-classical theory of gravity. A general asymptotically AdS spacetime in the Feffermann-Graham gauge looks like
\begin{equation}
ds^2=\frac{L^2}{z^2} \, \bigg(dz^2 + g_{\mu \nu}(z,x^\mu) \, dx^\mu \, dx^\nu  \bigg) \, ,
\end{equation}
where $x^\mu$ parameterizes the CFT which lives on the boundary of this spacetime $z \to 0$. Operators in the CFT correspond to matter fields in this geometry. Then, if the CFT Hamiltonian is perturbed such as in equation \eqref{eq:ham_quench}, the insertion of the operators $O(t,x^i)$ in the CFT act as sources for the matter fields in AdS and change their boundary conditions. The full matter plus gravity system evolves in time and this evolution can be used to understand the post-quench evolution in the CFT. AdS spacetime can be thought of as a box and thus acts as a potential well for the matter fields. These then fall into the center of AdS and after sufficient time, generically form a black hole \cite{Bizon:2011gg, Banks:1998dd, Danielsson:1999fa} \footnote{However, see \cite{Balasubramanian:2014cja} for an interesting discussion of a class of initial conditions that do not result in black hole formation. Whether a generic perturbative initial condition leads to a collapse or not is still an open problem. See \cite{Dimitrakopoulos:2014ada} and references therein for a discussion about this.}. Thermalization in the CFT is thus described by black hole formation in the dual semi-classical theory of gravity in AdS, as has been noted by several authors \cite{Witten:1998zw, Horowitz:1999jd, Danielsson:1999fa}. Consequently, collapsing solutions to semi-classical gravity in AdS are one of the most widely used tool to study quenches in holographic CFTs \cite{Ebrahim:2010ra, Balasubramanian:2011ur, Galante:2012pv, Caceres:2012em, Balasubramanian:2013oga}. Apart from weak-field analytic study in \cite{Bhattacharyya:2009uu}, they have also been studied using probe D-branes \cite{Das:2010yw, Basu:2013soa, Ugajin:2013xxa} and numerical methods \cite{Chesler:2008hg, Buchel:2014gta}. \\

\noindent In 2006, Ryu and Takayanagi \cite{Ryu:2006bv} proposed that the entanglement entropy of a subregion $A$ in the CFT is given by the minimal area of a bulk co-dimension 2 surface homologous to $A$ at the AdS boundary
\begin{equation}
 S_A =\underset{\gamma_A}{\text{min}} \, \frac{\text{Area}(\gamma_A)}{4 \, G_N} \, ,
\end{equation}
with the \textit{homology condition} $\partial \gamma_A \approx \partial A$. Due to the simplicity of this formula, \textit{holographic entanglement entropy} has received a lot of attention in the past decade (see \cite{Rangamani:2016dms} for a recent review). However, this formula works only for time-independent states. To calculate entanglement entropy of a subregion in a time-dependent state, one needs to use the covariant generalization of this formula, proposed in  \cite{Hubeny:2007xt}. This general proposal is called HRT proposal and it states that
\begin{equation}
S_A =\underset{\gamma_A}{\text{ext}} \, \frac{\text{Area}(\gamma_A)}{4 \, G_N} \, ,
\end{equation}
where $\gamma_A$ now denotes an extremal codimension-2 surface in the AdS bulk i.e. one with vanishing trace of the extrinsic curvature.\\ 

\noindent Using the HRT proposal, \cite{AbajoArrastia:2010yt} initiated the study of time-dependent holographic entanglement entropy, in the context of quenching a holographic (1+1) D CFT. They considered an abrupt (instantaneous) global quench to the CFT, such as
\begin{equation}
H(t) = H_0 + \Theta(t) \,  \int d^{d-1}x \, O(t,x^i) ,
\end{equation}
where $\Theta(t)$ is the Heaviside Theta function. As the operators $O(t,x)$ are uniformly inserted in the CFT, the dual bulk fields get sourced at the boundary of the AdS$_3$. If these operators have conformal dimensions much smaller than the central charge, the bulk fields are light and then the quench is described by a 3D geometry with a thin shell of infalling matter that forms a black hole. This geometry is well-studied and known as the \textit{Vaidya Geometry} \cite{Vaidya:1951zza}
\begin{equation}
ds^2 =  \frac{L^2}{z^{2}} \, \bigg(  \frac{dz^2}{f(t,z)} - f(t,z) \, dt^2+d x^2 \bigg) ,
\end{equation}
where the function $f(t,z)$ determines the horizon(s) of the black hole and is called the \textit{Blackening Function}. The end state of the quench will be a stationary state that looks thermal locally, with a temperature fixed in terms of the blackening function of the black hole. Like the field theory calculation of \cite{Calabrese:2005in} in the (1+1) D CFT, \cite{AbajoArrastia:2010yt} found a linear growth of entanglement entropy with time. \cite{Hartman:2013qma} generalized this holographic calculation to the case where the initial state is not vacuum but a thermal state or a typical pure state. These works for lower-dimensional field theories used the properties of (2+1) D AdS gravity, which is special in various ways. \cite{Albash:2010mv} instead considered the evolution of holographic entanglement entropy in (2+1) D CFT, exploring evolving geometries in AdS$_4$. They also observed a linear behavior for the growth of entanglement entropy; but noticed some novel phenomena, such as a discontinuity in the time derivative of entanglement entropy near the time when it is about to saturate. Motivated by these examples, \cite{Liu:2013iza, Liu:2013qca} considered the case of general dimensional CFTs. For subsystems whose characteristic size $R$ is much greater than the temperature scale $1/T$, they found that the time-evolution of holographic entanglement entropy has a linear regime
\begin{equation}
\delta S_A(t) = v_E \, s_{\text{eq}} \, \mathcal{A}_{\Sigma} \,  t  \, ,
\end{equation}
where $s_{\text{eq}}$ is the entropy density of the subsystem in the final state, $\mathcal{A}_{\Sigma}$ is the area of the \textit{entangling surface} in the CFT and $v_E$ is a subsystem-independent constant called the \textit{Entanglement Velocity}. Furthermore, the rate of growth in this linear regime was found to be bounded by 1 (in units $c=1$). As envisaged by \cite{Calabrese:2005in} using a quasi-particle description for the propagation of this entanglement, \cite{Casini:2015zua} showed that this bound is a consequence of the causality of the $d$ (spacetime) dimensional CFT. \\

\noindent However, these works focused on the limit of large subsystem sizes. \cite{Kundu:2016cgh} proposed a method to explore the entanglement entropies when the subsystem sizes are small compared to the final temperature. They observed a perturbative expansion for the area of the extremal surface and used that to study the post-quench evolution after an instantaneous quench. \cite{OBannon:2016exv} used a similar method to study the very interesting case of a global quench that is linear in time. But it was only in \cite{Lokhande:2017jik} that a comprehensive study of growth of entanglement entropy in the small subsystem regime was undertaken. \cite{Lokhande:2017jik} studied a wide class of global quenches, including instantaneous, power law $t^p$ with arbitrary $p$ and periodic quenches. This was possible due to a novel interpretation that the growth of entanglement entropy after a quench can be understood as a linear response of the subsystem (in time) to the energy the quench injects. This interpretation allows one to rewrite the entanglement entropy (and other related information-theoretic quantities) as a convolution of two functions - a kernel that depends on the shape and size of the subsystem and a source that only depends on the energetics of the quench:
\begin{equation}
\delta S_A(t) = \int_{-\infty}^{\infty} \, dt' \, \mathfrak{m}(t-t') \, \mathfrak{n}(t') \, .
\end{equation}
This interpretation is very useful. As described in detail in \cite{Lokhande:2017jik}, it implies the existence of a \textit{time-dependent first law of entanglement} for small subsystems. Moreover, such a law can be used to define a time-dependent quantity analogous to relative entropy that would measure the distance between out-of-equilibrium states explored during the post-quench evolution and either the initial or the final equilibrium state.

\subsection{Hyperscaling-Violating-Lifshitz Theories}
\label{hvLif_intro}

\noindent In this paper, we would like to study global quantum quenches when the CFT is in an excited state $\ket{\Psi}$ which partially breaks the full conformal symmetry. These states have a finite energy density and charge density $\rho(x^\mu)$. The conserved current for the charge is dual to a $U(1)$ gauge field $A_M(z, x^\mu)$ in the AdS bulk. Finite charge density then implies that this gauge field has a \textit{non-normalizeable} mode at the boundary \cite{Hartnoll:2009sz}, which plays the role of a chemical potential for the charge. Since there is a \textit{finite} energy and charge density, these generically backreact on the AdS geometry, modifying the interior \cite{Gubser:2009cg, Hartnoll:2009ns, Hartnoll:2010gu}. We will further include fermions $\psi^M$ in the bulk and focus on the action\footnote{ In the simplest case, the backreaction is known to uniquely give the AdS-Reissner-Nordstrom metric \cite{Son:2006em}. But this is known to be unstable at low temperatures \cite{Hartnoll:2010gu}. }
\begin{equation}
\label{eq:top-down-S-lif}
\mathcal{S}= \int d^{d+1} x \, \sqrt{-g} \, \bigg(R- 2 \Lambda - \frac{1}{4 \, e^2} \, F_{MN} \, F^{MN} + \mathcal{L}_{\text{f}} \bigg) \, .
\end{equation}
where $\mathcal{L}_{\text{f}}$ denotes an ideal fluid Lagrangian at zero temperature for the fermions. Using this, \cite{Hartnoll:2009ns} showed that the solutions to the equations of motion are
\begin{equation}
\label{eq:Lif_solns_1}
ds^2 = \frac{L^2}{y^2} \, \bigg(d y^2 - \frac{dt^2}{y^{2 \,( z-1)}} + dx_i^2 \bigg) , \qquad  A = \frac{\sigma \, e}{y^{z-1}} dt \, ,
\end{equation}
where the exponent $z$ is fixed in terms of mass of the fermion. AdS radius $L$ appears in the metric because the finite charge density $\rho(x^\mu)$ is smeared in an appropriate finite spatial region in the CFT. If we then zoom-in near this region, that would correspond to zooming-in on the interior part of the geometry. This part is not asymptotically AdS but instead has only part of AdS isometries. If we take this metric as an effective description of some field theory defined at its asymptotic boundary, this field theory will not be fully conformal. In fact, in the background given by the metric \eqref{eq:Lif_solns_1}, time and space scale in an anisotropic way
\begin{equation}
t \to \alpha^z \, t , \qquad x^i \to \alpha \, x^i , \qquad y \to \alpha \, y \, .
\end{equation}
The constant $z$ is a \textit{Dynamical Critical Exponent} in this theory. It determines, for example, how the mass gap scales with respect to the coherence length $\xi$ near the critical UV CFT
\begin{equation}
 \text{Gap} \sim  \frac{1}{\xi^z} \, .
\end{equation}
Field theories with such properties are called \textit{Lifshitz Field Theories}. They have the usual time and space translation generators (the Hamiltonian $H$ and spatial momenta $P_i$), the spatial rotation generators (angular momenta $M_{ij}$) but only an \textit{anisotropic dilatation} $D$ with the following commutation relations
\begin{equation}
 [D, P_i] = i P_i , \qquad [D, H] = i \, z \, H .
\end{equation}
As a consequence, they are also known as \textit{non-relativistic} field theories. Such theories are known to describe general quantum critical points in condensed matter \cite{Ardonne:2003wa} and they can be constructed from $SU(N)$ Yang-Mills theories \cite{2005PhRvL..94n7205F, Horava:2008jf}. Thus, in this effective limit, we would be describing \textit{non-relativistic holography}. \\

\noindent The idea of investigating holography for theories without conformal symmetry is not new. See \cite{Son:2008ye, Balasubramanian:2008dm, Goldberger:2008vg, Adams:2008wt, Herzog:2008wg, Maldacena:2008wh} for older work studying the holographic duals of \textit{Schr\"{o}dinger field theories}. The asymptotically Lifshitz geometry \eqref{eq:Lif_solns_1} was first introduced in \cite{Kachru:2008yh} to study \textit{Lifshitz field theories}. They observed that this metric is nonsingular and all its curvature invariants are finite. However, it has a peculiar behavior near $y \to \infty$ (the interior of the geometry) and in fact it is geodesically incomplete \cite{Kachru:2008yh}. Thus to have a well-defined holography, one needs to find this metric as a solution of string theory. We thus need to embed the action \eqref{eq:top-down-S-lif} in some string model. There has been a lot of work in this direction, \cite{Adams:2008wt, Herzog:2008wg, Maldacena:2008wh, Hartnoll:2008rs, Adams:2008zk, Azeyanagi:2009pr, Hartnoll:2009ns, Narayan:2012hk, Dey:2012fi}, to name a few. Some works have advocated taking the asymptotically Lifshitz metric as a solution of general relativity with some matter \cite{Taylor:2008tg}. Energy conditions on this matter then decide what class of geometries are expected to have a Lifshitz field theory as a holographic dual. In particular, \cite{Hoyos:2010at} showed that the condition $z>1$ needs to hold for the Null Energy Condition (NEC) to be satisfied in the bulk\footnote{ Although these works suggest that there may exist a Lifshitz holography, recently there has been some debate about what exactly the geometry dual to a UV Lifshitz critical point looks like. See \cite{Christensen:2013rfa}. }. \\

\noindent In this paper, we will study a class of excited states in the CFTs that is further qualified as follows. The IR geometry describing these states has an extra critical exponent, called the \textit{Hyperscaling-Violating Parameter} $\theta$. As before, we will zoom-in on the IR and take the asymptotics to define an effective, non-relativistic field theory. Following investigations in the holography of charged dilatonic black holes, it was realized \cite{Charmousis:2010zz, Dong:2012se} that such geometries are good effective holographic descriptions for condensed-matter systems. The metric looks like
\begin{equation}
\label{eq:hsv_eff_holo_metric}
 ds^2 =\frac{L^2}{y^{2 (d-1-\theta)/(d-1)}} \, \bigg( dy^2 - \frac{dt^2}{y^{2(z-1)}} + dx_i^2  \bigg) \, ,
\end{equation}
where we recall that $i=1,2,\cdots,(d-1)$. In addition to Lifshitz symmetries, this metric transforms as
\begin{equation}
 ds \to \alpha^{\frac{\theta}{(d-1)}} \, ds  ,
\end{equation}
under scaling. From now on, we will call this the \textit{Hyperscaling-Violating-Lifshitz} metric and denote it by the acronym \textbf{hvLif}. Hyperscaling laws are well-known in condensed-matter physics. Traditionally, they are defined as those laws where the critical exponents depend on the dimension. In our case, the presence of $\theta$ roughly means that the asymptotic field theory effectively lives in $(d-\theta-1)$ spatial dimensions instead of $(d-1)$ spatial dimensions. This may be a concern for dimensional analysis, but the issue is resolved when one posits the existence of a length scale $y_F$ that does not decouple in the IR.\\

\noindent Hyperscaling violating metrics can be obtained from the action \cite{Charmousis:2010zz} 
\begin{equation}
 \mathcal{S}= \int d^{d+1} x \, \sqrt{-g} \, \bigg(R- \frac{e^{\alpha \phi}}{4 e^2} \, F_{M N} \, F^{MN} - \frac{1}{2} \, (\partial \phi)^2 - V_0 \, \cosh(\eta  \phi)  \bigg)  \, ,
\end{equation}
where the parameters $\alpha$ and $\eta$ determine the exponents $z$ and $\theta$. The potential $\cosh (\eta  \phi)$ gives an asymptotically AdS solution in the limit $\phi \to 0$ and an asymptotically \textbf{hvLif} solution in the limit $\phi \to \infty$. For a range of exponents $\alpha$ and $\eta$, the metric can be shown to arise in a UV-complete theory like string theory \cite{Dong:2012se, Narayan:2012hk, Perlmutter:2012he, Dey:2013oba}. With these pieces of evidence, it is natural to study further the holography of hyperscaling-violating solutions in AdS spacetime. The Null Energy Condition (NEC) in the bulk imposes some constraints on the class of non-relativistic field theories dual to the hyperscaling-violating backgrounds, namely that the critical exponents must satisfy \cite{Dong:2012se, Chemissany:2014xsa}
\begin{align}
\begin{split}
\label{eq:NEC_hsv}
(d-1-\theta) \bigg((d-1)(z-1)-\theta \bigg) &\ge 0 ,\\
(z-1)(d-1-\theta+z) &\ge 0 .
\end{split}
\end{align}
We will study global quantum quenches in these field theories. These quenches correspond to classical evolution of the hyperscaling-violating background, possibly with some matter fields. In analogy with AdS/CFT correspondence, changing coupling constants in the field theory will correspond to turning on the non-normalizeable modes of some bulk matter fields. Owing to the natural gravitational potential well in the AdS spacetime, these matter fields will then collapse towards the center of AdS to eventually form a black hole. Formation of the black hole will correspond to thermalization in the dual field theory, which we assume will happen generically thanks to the strong coupling and chaotic dynamics. The simplest model to study this is to assume that the bulk matter is light and that it falls as a homogenous, thin, spherical shell. Its backreaction on the bulk geometry is then given by the Vaidya-like metric \footnote{Vaidya-like solutions in asymptotically Lifshitz backgrounds were first studied by \cite{Keranen:2011xs}}
\begin{equation}
 \label{eq:HSV-Vaidya}
 ds^2 = \frac{L^2}{y^{2(d-1-\theta)/(d-1)}} \, \bigg(\frac{dy^2}{f(t,y)} - \frac{f(t,y) \, dt^2}{y^{2(z-1)}} + dx_i^2 \bigg) ,
\end{equation}
with $f(t,y)$ being the blackening function. We display it explicitly in equation \eqref{eq:black_func_HSV}. We will study the time-dependent holographic entanglement entropy in such a background and infer characteristics of the quenches in the asymptotic non-relativistic field theory. The entangling region in the boundary can have any geometry. For the case of strip geometries, this problem was also studied in \cite{Alishahiha:2014cwa, Fonda:2014ula} but in a regime complementary to what we will study. They studied the evolution of entanglement entropy when $\ell  T \gg 1$, where $T$ denotes the temperature of the final state after thermalization\footnote{Thermalization of mutual information between two widely separated regions in a hyperscaling-violating field theory was studied in \cite{Fischler:2012uv, Tanhayi:2015cax}.} . We will instead study the complementary regime $\ell T \ll 1$. Unlike \cite{Alishahiha:2014cwa, Fonda:2014ula}, we will also consider more general quantum quenches: instantaneous, power-law with any power and periodic in time. \\

\subsection{Reader's Map}

\noindent  The plan of the paper is as follows. In Section \ref{area_exp}, we discuss the perturbative expansion of the HRT area functional in the presence of a small parameter. In \ref{holo_renorm}, we make some general comments on holographic renormalization for asymptotically \textbf{hvLif} backgrounds. In Section \ref{Vaidya_model}, we discuss our Vaidya model of global quantum quenches in detail. In \ref{spread_EE}, we use the perturbative expansion of the area functional in the Vaidya model to calculate a simple integral expression for time-dependent holographic entanglement entropy (equation \eqref{eq:EE_general}) which holds for any global quantum quench. In Section \ref{linear_response}, we interpret this equation as a linear response relation and use it to derive a time-dependent generalization of the first law of entanglement entropy for small subsystems in subsection \ref{first_law}. Such a law allows one to study a time-dependent analogue of relative entropy, which we discuss in \ref{relative_entropy}. In Section \ref{special_cases}, we start studying specific quantum quenches using the general machinery we have developed so far. Subsection \ref{instantaneous} discusses in detail the instantaneous global quantum quench to the \textbf{hvLif} field theory. We once again study small subsystems and display explicitly various time-dependent quantities related to entanglement entropy. In Subsection \ref{power-law}, we study a finite-duration global quench that is a power law in time, with an arbitrary power. We discuss the method to obtain results in this general case and for convenience discuss the case of integer powers in detail. In Subsection \ref{linear-pump}, we elaborate on the case of a global quench linear in time. This is a special case of the power law quench but this is interesting in itself due to earlier work \cite{OBannon:2016exv, Lokhande:2017jik}. We then come to Subsection \ref{Floquet} where we study a Floquet quench, a global quench that is periodic in time. This is a case of particular importance in condensed-matter community and we disucss thermalization of entanglement entropy for small subsystems for a Floquet quench. Finally, in Section \ref{summary}, we conclude with a summary of our results and some directions for future work. There is an Appendix \ref{stress_tensor_HSV} where we compute the stress tensor for \textbf{hvLif} backgrounds.

\section{Perturbation Theory for Small Subsystems}
\label{area_exp}

From now on, we will work in an asymptotically \textbf{hvLif} spacetime and use that to study the quenches in the asymptotic non-relativistic field theories. Our results will be an approximation to quenching the specific class of excited states in CFTs that we have discussed in the Introduction \ref{intro}.

\subsection{Holographic Renormalization}
\label{holo_renorm}

In this subsection, we will discuss holographic renormalization for asymptotically \textbf{hvLif} spacetimes. This is a very important problem, albeit outside the purview of this paper. Thus, we will be brief. \\

\noindent For a review of holographic renormalization in the case of CFTs, see \cite{Skenderis:2002wp}. Discussion of holographic renormalization for asymptotically Lifshitz backgrounds was initiated in \cite{Kachru:2008yh, Taylor:2008tg}. It was studied in more detail in \cite{Ross:2009ar, Ross:2011gu, Baggio:2011cp, Mann:2011hg, Keeler:2012mb, Andrade:2012xy, Andrade:2013wsa}, to name a few. In this subsection, we will follow the excellent article \cite{Chemissany:2014xsa}. They study holographic renormalization in \textbf{hvLif} backgrounds using radial Hamilton-Jacobi method. In this method, one starts with a general action for the gravity-matter system
\begin{equation}
 \mathcal{S}= \int d^{d+1} x \, \sqrt{-g} \, \bigg(R- \frac{F_1(\phi)}{4 e^2} \, F_{M N} \, F^{MN} - \frac{1}{2} \, (\partial \phi)^2 -F_2(\phi) - F_3(\phi) \, A^2 \bigg)  \, ,
\end{equation}
along with the usual Gibbons-Hawking boundary term. Then one takes an ADM-like ansatz for the metric and finds the Hamiltonian $\mathcal{H}$ as well Hamilton's Principal Function $\mathcal{W}$. These are related by the Hamilton-Jacobi equation
\begin{equation}
-\frac{\partial \mathcal{W}}{\partial y} = \mathcal{H}  \, .
\end{equation}
Solving this equation and using the definition of $\mathcal{W}$, one obtains the normalizeable and non-normalizeable modes for the fields. After an appropriate canonical transformation that diagonalizes the symplectic form on the space of these solutions, the normalizebale modes of the fields are identified as sources for some operators in the asymptotic field theory and the non-normalizeable modes are identified with the expectation values of these operators. Moreover, the asymptotic behavior of the solution to the Hamilton-Jacobi equation, including the finite terms, is useful in obtaining an analogue of Feffermann-Graham expansion for the fields on the \textbf{hvLif} backgrounds. The Feffermann-Graham expansion for the metric looks like
\begin{equation}
   ds^2 = \frac{L^2}{y^{2 d_\theta/(d-1)}} \, \bigg( dy^2 + g_{\mu \nu}(y, x^\mu; z, \theta) \,  dx^\mu dx^\nu \bigg) \, ,
\end{equation}
where, for economy of notation, we have defined
\begin{equation}
d_\theta \equiv d-1-\theta ,
\end{equation}
and $\mu=0,1, \cdots, d-1$. The metric $g_{\mu \nu}(0,x^\mu;z,\theta)$ defines the metric on the asymptotic boundary. We demand that this be given by
\begin{equation}
g_{\mu \nu}(0,x^\mu;z,\theta) \, dx^\mu  dx^\nu = \lim_{y \to 0} \bigg( -\frac{dt^2}{y^{2(z-1)}} + dx_i^2 \bigg) .
\end{equation}
We are interested in studying quantum quenches in the asymptotic field theory by calculating entanglement entropy of a subregion holographically. If we consider a subregion $A$ in the field theory with a characteristic size $\ell$, let $y_*$ denote the turning point in the bulk upto which the extremal homologous surface homologous goes. \cite{Dong:2012se} have argued that the UV of the asymptotic field theory gets mapped to the IR of the gravity theory on the \textbf{hvLif} background. Thus if $\ell$ is small (compared to the temperature scale), $y_*$ will be small (compared to the radius of the spacetime $L$). The bulk surface will then probe only the near-boundary region of the geometry. \\

\noindent When we do a global quantum quench in the field theory, this will in general change the geometry by a finite amount but because the subregion $A$ is small, the reduced density matrices on $A$ before and after the quench will differ by a small amount. In the bulk this implies that the area of the HRT surface dual to $A$ will get corrections from leading perturbations to the asymptotic \textbf{hvLif} metric, such as
\begin{equation}
  g_{\mu \nu}(y, x^\mu; z, \theta) =  g_{\mu \nu}(0, x^\mu; z, \theta) + \delta g_{\mu \nu} .
\end{equation}
The leading term in the perturbation with the perturbation is a simple function of the expectation of the stress tensor
\begin{equation}
 \delta g_{\mu \nu} = y^{d_\theta+z} \,  \langle T_{\mu \nu}\rangle ,
\end{equation}
with the power of $y$ fixed by symmetry. Higher powers of $T_{\mu \nu}$ are subleading for small $y$. If the quantum quench involved changing the coupling constant of some operator $O(t,x^i)$ in the field theory, the metric perturbation $\delta g_{\mu \nu}$ may also get corrections from the expectation value of $O(t,x^i)$ as well as from the source. We will assume that the (aniosotropic) scaling dimension $\Delta$ of this operator is such that its contribution to the metric perturbation appears at subleading orders compared to that of $\langle T_{\mu \nu} \rangle$. \\

\noindent But the area of the HRT surface will be divergent, reflecting the fact that entanglement entropy of $A$ in any state is UV-divergent. This divergence, however, is easy to remove: we subtract the entanglement entropy of $A$ in vacuum from all our answers. There will in general be contributions to entanglement entropy from the source $J(t,x^i)$ directly. Understanding them will need a source-dependent renormalization, so we will subtract this as well. Thus we will calculate
\begin{equation}
 \delta S_A(t) \equiv S_A(t) - S_A(0) - S_J (0) ,
\end{equation}
where $S_J$ is the contribution to the holographic entanglement entropy from the source.

\subsection{Perturbative Expansion of the Area Functional}
\label{perturb_expansion}

In the previous subsection, we have made precise the class of universal corrections we will aim to capture by calculating entanglement entropy holographically. In this subsection, we briefly review the results of \cite{Kundu:2016cgh} where they set up a perturbative calculation of holographic entanglement entropy of a subsystem in the presence of a small parameter (see also \cite{Fischler:2012ca}). They considered the case of a homogeneous and instantaneous quench to the CFT ground state, while \cite{Lokhande:2017jik} generalized this to any homogenous quench. We will discuss such a calculation for the area of a bulk surface in asymptotically \textbf{hvLif} spacetime. We start with the HRT formula \cite{Ryu:2006bv, Hubeny:2007xt}
\begin{equation}
 S_A =\underset{\gamma_A}{\text{ext}} \, \frac{\text{Area}(\gamma_A)}{4 \, G_N}  ,
\end{equation}
where $G_N$ is $(d+1)$ the Newton's constant in $(d+1)$ dimensional bulk and $\gamma_A$ is a $(d-1)$ dimensional bulk surface that is homologous to boundary subregion $A$. We will assume that the characteristic size $\ell$ of $A$ is small compared to any other scale in the field theory, following \cite{Kundu:2016cgh, Lokhande:2017jik}. To be precise, the small parameter in our case will be
\begin{equation}
\lambda_A = \ell^z \, T  ,
\end{equation}
with $T$ denoting the temperature of the final state. We will discuss how to calculate it in Section \ref{spread_EE}. Let $\phi_A(\xi)$ denote all the embedding fields for the surface $\gamma_A$ and $\mathcal{L}[\phi_A(\xi); \lambda_A]$ denote the Lagrange functional for the area of the surface $\gamma_A$
\begin{equation}
 \mathcal{A} \equiv \text{Area}(\gamma_A) = \int d\xi \,  \mathcal{L}[\phi_A(\xi); \lambda_A] .
\end{equation}
Assuming the dimensionless parameter $\lambda_A \ll 1$, we can expand the embedding functions and the Lagrange function as follows
\begin{align}
\begin{split}
 \mathcal{L}[\phi_A(\xi); \lambda_A] &= \mathcal{L}^{(0)}[\phi_A(\xi); \lambda_A] + \lambda_A \, \mathcal{L}^{(1)}[\phi_A(\xi); \lambda_A] + \mathcal{O}(\lambda_A^2) ,\\
\phi(\xi) &= \phi^{(0)}(\xi) + \lambda_A \, \phi^{(1)}(\xi) + \mathcal{O}(\lambda_A^2) ,
\end{split}
 \end{align}
where the embedding functions that extremize the area at any given order in $\lambda_A$ can in principle be obtained by solving Euler-Lagrange equations order by order in $\lambda_A$. However, as first noted in \cite{Kundu:2016cgh}, the calculation of the area to order $\mathcal{O}(\lambda_A)$ becomes particularly simple. One gets
\begin{align}
 \begin{split}
\mathcal{A}_{\text{on-shell}}[\phi_A(\xi)] &= \int \, d\xi \,    \mathcal{L}^{(0)}[\phi_A(\xi); \lambda_A] + \lambda_A \, \int d\xi \, \mathcal{L}^{(1)}[\phi_A(\xi); \lambda_A] \\
&\quad + \lambda_A \int d\xi \, \phi_A^{(1)} \, \cancelto{0}{\bigg[\frac{d}{d\xi} \, \frac{\partial \mathcal{L}_A^{(1)}}{\partial \phi'_A(\xi)} - \frac{\partial \mathcal{L}^{(0)}}{\partial \phi_A(\xi)} \bigg]} + \cdots ,
 \end{split}
\end{align}
where we have used the equations of motion at zeroth order in $\lambda_A$. Hence the first order correction to holographic entanglement entropy is given by
\begin{equation}
\mathcal{A}^{(1)}[\phi_A(\xi)] =  \lambda_A \, \int d\xi \, \mathcal{L}^{(1)}[\phi_A(\xi); \lambda_A] .
\end{equation}
This is the term we will calculate, and as we will see, will be universal.

\section{The Vaidya Model for Global Quenches}
\label{Vaidya_model}

Let us start with the \textbf{hvLif} metric is \eqref{eq:hsv_eff_holo_metric}
\begin{equation}
 ds^2 = \frac{1}{y^{2 d_\theta/(d-1)}} \, \bigg(dy^2 -\frac{dt^2}{y^{2(z-1)}} + dx_i^2 \bigg) ,
\end{equation}
where $y>0$ is the bulk radial direction in the Schwarzschild frame, with the boundary at $y=0$ and we have set the radius of the spacetime to 1. We will assume that the critical dynamical exponents $z$ and $\theta$, apart the constraints \eqref{eq:NEC_hsv}, also satisfy
\begin{equation}
z \ge 1 \, , \qquad  \qquad \theta \ge 0 \,  .
\end{equation}
As discussed in the Introduction \ref{intro}, the metric \eqref{eq:HSV-Vaidya} (reproduced below) describes a simple model of a thin, homogenous shell of infalling matter to form a black hole with asymptotically \textbf{hvLif} geometry
\begin{equation}
  ds^2 = \frac{1}{y^{2 d_\theta/(d-1)}} \, \bigg(\frac{dy^2}{f(t,y)} -\frac{f(t,y) \, dt^2}{y^{2(z-1)}} + dx_i^2 \bigg)
\end{equation}
We will call this the \textbf{hvLif}-Vaidya geometry and use this to model our quenches holographically. In what follows, it will be useful to work in tortoise coordinates, also called Eddington-Finkelstein coordinates, defined by
\begin{equation}
du \equiv dy \, , \qquad \qquad d v \equiv d t - \frac{d y}{y^{(1-z)} \, f(t,y)}   \, .
\end{equation}
In these cordinates, the metric takes the form
\begin{equation}\label{HSV_metric}
 ds^2 = \frac{1}{u^{2 d_\theta/(d-1)}} \,  \bigg(-\frac{2 du \, dv}{u^{2(z-1)}} -\frac{f(u,v) \, dv^2}{u^{2(z-1)}} + dx_i^2 \bigg) .
\end{equation}
The class of global quenches we study can be parameterized in terms of the blackening function
\begin{equation}
\label{eq:black_func_HSV}
 f(u,v) = 1 - g(v) \, \bigg(\frac{u}{u_H}\bigg)^{(z+d_\theta)}  ,
\end{equation}
where the time-dependent function $g(v)$ corresponds to the quench in the boundary. We can always choose it to be bounded such that $0 \le g(v) \le 1$ and the leading term $1$ indicates the time-independent vacuum. The constant $u_H$ is the horizon radius and it encodes the equilibrium properties of the final state, a long time after the time-dependent perturbation. In case of pure AdS backgrounds, it precisely coincides with the horizon radius of the black hole formed as a result of perturbing the boundary CFT. Depending on the functional form of $g(v)$, we will distinguish between two kinds of global quenches:
\begin{itemize}[leftmargin=5mm, itemindent=0mm]
 \item \textbf{Quenches of finite duration}\\
We will denote the duration of the quench by $t_q$, with reference to the Schwarzschild coordinate natural to a lab. For quenches with finite duration, $g(v)$ interpolates smoothly between two values at $t=0$ and $t=t_q$. Denoting $g(v(t))$ by $g(t)$ for a moment
\begin{equation}
g(t=0) \to 0 \,  , \qquad \qquad g(t=t_q) \to 1 \,  ,
\end{equation}
where we have normalized the value of $g$ when the quench ends. From a geometrical viewpoint, at $t=0$ we have the AdS-HSV geometry, we turn on the quench and as a result at $t=t_q$ end up in a static black hole solution with AdS-HSV asymptotics \cite{Gouteraux:2011ce, Huijse:2011ef, Dong:2012se, Alishahiha:2012qu, Pedraza:2018eey}.

\item \textbf{Quenches of infinite duration}\\
In this case, we keep on perturbing the system indefinitely in time and as a result keep on inserting energy. One can still formally expand in a small parameter $\lambda_A$, but the expansion invariably becomes bad at sufficiently late times. We will not discuss very late time dynamics. 
\end{itemize}
In this paper, we will mainly study quenches of finite duration.

\subsection{Spread of Entanglement Entropy}
\label{spread_EE}

In \cite{Lokhande:2017jik}, we considered boundary subregions with spherical and strip geometry to study the time evolution of entanglement. But in this paper, we will only study subregions with a strip geometry. This is because, in a non-relativistic setting, it is very difficult to solve for the embedding functions with spherical geometry that extemize the area.  The strip geometry will be defined on a time-slice so as to have $(d-1)$ coordinates $(x, x^i_p)$ such that 
\begin{equation}
 -\frac{\ell}{2} \le x \le \frac{\ell}{2} \,  , \qquad \qquad 0 \le x^i_p \le \ell_p \,  .
\end{equation}
We will assume that $\ell$ is the smallest energy scale in the theory but, as we will see, will not need to assume anything about $\ell_p$. The strip has translation invariance along the $x^i_p$ directions since $\ell \ll \ell_p$. This translation invariance can be used to constrain the homologous surface in the bulk, which as a result is completely specified by one function $x \equiv x(u)$. We follow the usual procedure to calculate the area of this homologous surface in the bulk. It can be shown to be
\begin{equation}
 \mathcal{A}[x(u), v(u)] = \int_0^{u_*} du \, \mathcal{L}[x(u), v(u)] ,
\end{equation}
where $u^*$ denotes the tip of the point in the bulk where the surface curves i.e. the depth upto which the surface falls inside. The Lagrange function is given by
\begin{equation}
 \mathcal{L}[x(u), v(u)] = \frac{2 \, \ell_p^{(d-2)}}{u^{d_\theta}} \, \sqrt{  x'^2 - \frac{2 v'}{u^{(z-1)}} - \frac{v'^2 \, f(u,v)}{u^{2(z-1)}} } .
\end{equation}
Before we proceed to calculate this, we remark the exact expansion parameter. As we argue in Appendix \ref{stress_tensor_HSV}, the blackening function can be used to define an ``effective temperature''
\begin{equation}
 T = \frac{1}{4 \pi} \bigg| \frac{df(u,v)}{du} \bigg|_{u=u_H} .
\end{equation}
Explicitly evaluating this, we see that the parameter $ \ell^z \, T$ is dimensionless and we define it to be $\lambda_A$. It can also be written differently using the bulk distance $u_*$ as 
\begin{equation}
\lambda_A \equiv \frac{(d_\theta+z)}{4 \pi} \, \bigg( \frac{u_*}{u_H} \bigg)^{z}  .
\end{equation}
The Lagrange function $\mathcal{L}[x(u), v(u)]$ now can be written as a function of this parameter, and can be expanded around $\lambda_A=0$. The zeroth order Lagrange function and its first order correction are then given by
\begin{equation}
 \mathcal{L}^{(0)} [ x(u), v(u)] = \frac{2 \ell_p^{(d-2)}}{u^{d_\theta}} \quad \sqrt{x'^2 - \frac{2 v'}{u^{(z-1)}} - \frac{v'^2}{u^{2(z-1)}}} ,
\end{equation}
\begin{equation}
 \mathcal{L}^{(1)} [x(u), v(u)] = \frac{\ell_p^{(d-2)} \, v'^2 \, u^{(2-d_\theta-2z)} \, g(v) }{ \sqrt{x'^2 - \frac{2 v'}{u^{(z-1)}} - \frac{v'^2}{u^{2(z-1)}}}} .
\end{equation}
The Euler-Lagrange equations from the zeroth order function can be solved to obtain the solution for the extremal surface in the time-independent case. This solution\footnote{I thank Juan F. Pedraza and Gerben W. J. Oling for help in deriving this equation.} is
\begin{equation}
 x(u) = \frac{\sqrt{\pi } \, \Gamma \big[\frac{d_\theta+1}{2d_\theta}\big] \,  u_* }{2 d_\theta \,  \Gamma \big[\frac{2d_\theta +1}{2 d_\theta} \big]}  -   \frac{u^{(d_\theta+1)}}{(d_\theta+1) \, u_*^{d_\theta}} \, _2F_1 \bigg[\frac{1}{2}, \frac{(d_\theta+1)}{2 d_\theta}, \frac{(3 d_\theta +1)}{2 d_\theta}; \bigg[\frac{u}{u_*}\bigg]^{2 d_\theta}\bigg] \, ,
\end{equation}
where $_2F_1(a, b, c; x)$ is the hypergeometric function and
\begin{equation}
 v(u) = t - \frac{u^z}{z} .
\end{equation}
Evaluating the first correction to the area at these solutions, the time-dependent change in the entanglement entropy becomes
\begin{equation}
\label{eq:EE_general}
\delta S_A(t) = \frac{ \ell_p^{d-2}}{4 \, G_N \, u_H^{d_\theta+z}} \, \int_0^{u_*} d u \, u^z \, \sqrt{1 - \left[\frac{u}{u_*}\right]^{2d_\theta}} \, g \Big(t - \frac{u^z}{z} \Big) ,
\end{equation}
where the turning point $u_*$ can be calculated from $x(u)$ to be
\begin{equation}
\label{eq:rel_tstar_ell}
 u_* = \frac{\ell \, d_\theta \, \Gamma \left[\frac{2 d_\theta+1}{2 d_\theta}\right]}{\sqrt{\pi } \, \Gamma \left[\frac{d_\theta +1}{2 d_\theta}\right]} .
\end{equation}

\section{Entanglement as a Linear Response}
\label{linear_response}

The equation \eqref{eq:EE_general} for time-evolution of entanglement entropy has a very interesting structure. In particular, if we define a time-like coordinate
\begin{align}
 t' &\equiv \frac{u^z}{z} \, ,
\end{align}
the time-dependent change in entanglement entropy can be written as the convolution equation
\begin{equation}
\label{eq:EE_lin_resp}
 \delta S_A(t) = \int_{-\infty}^{\infty} \, dt' \, \mathfrak{m}(t-t') \, \mathfrak{n}(t') \equiv \mathfrak{m}(t) * \mathfrak{n}(t) \, ,
\end{equation} 
for some appropriate functions $\mathfrak{m}(t)$ and $\mathfrak{n}(t)$. In the theory of linear response, one of these functions, say $\mathfrak{m}(t)$, is called the input or \textit{Source} function and the other one is called a \textit{Response}.  However, the role of $\mathfrak{m}$ and $\mathfrak{n}$ are interchangeable due to the properties of the convolution operation.\\

\noindent As we argue in equation \eqref{eq:HSV_energy_density} in Appendix \ref{stress_tensor_HSV}, the energy density in AdS-HSV Vaidya spacetime is
\begin{equation}
\epsilon(t) =  \frac{d_\theta \, g(t) }{16 \, \pi \, G_N \, u_H^{d_\theta+z}}  \, .
\end{equation}
Thus, without loss of generality, we identify the source with the energy density
\begin{equation}
\label{eq:source_func}
 \mathfrak{m}(t) \equiv \epsilon(t) = \frac{d_\theta \, g(t) }{16 \, \pi \, G_N \, u_H^{d_\theta+z}}  \, .
\end{equation}
This is a natural choice for the source because this function depends only on the parameters in the quench. The response function $\mathfrak{n}_A(t)$ will then contain all the information about the geometry of the entangling subregion:
\begin{equation}
\label{eq:response_func}
 \mathfrak{n}(t) =\frac{2 \, \pi \, \mathcal{A}_{\Sigma} \,  (z \, t)^{\frac{1}{z}} }{d_\theta} \, \, \bigg[1-\bigg( \frac{t}{t_*} \bigg)^{\frac{2 \, d_\theta}{z}} \bigg]^{\frac{1}{2}} \, \bigg[ \Theta(t) - \Theta(t - t_*) \bigg] \, ,
\end{equation}
where $\mathcal{A}_\Sigma = 2 \ell_p^{d-2}$ is the area of the \textit{entangling surface}, the boundary of the subregion $A$ along the perpendicular directions and $t_*$ is defined in terms of $u_*$. Notice that we have unbounded limits in equation \eqref{eq:EE_lin_resp}, but the actual integral for entanglement entropy \eqref{eq:EE_general} has bounded integration domain. To make this change, we have made the response function $\mathfrak{n}(t)$ an explicit function of these limits. The upper limit $t_*$ also provides a natural reference for the domain of the response function. \\

\noindent For $t<0$, there was no quench and as expected, the response function vanishes as well. Thus, the spread of entanglement is causal in our model. For finite quenches, the source function increases only upto $t=t_q$. In these cases, owing to the properties of the convolution integral, the entanglement growth saturates at a time
\begin{equation}
 \label{eq:sat_time}
t_{\text{sat}} \equiv t_* + t_q .
\end{equation}
As was noted in \cite{Lokhande:2017jik}, writing the growth of entanglement entropy as a convolution has the added benefit that convolution integrals enjoy the following nice properties:
\begin{itemize}[leftmargin=5mm, itemindent=0mm]
 \item \textbf{Linearity}\\
If a source is a linear combination of two independent sources $\mathfrak{m}(t) = c_1 \, \mathfrak{m}_1(t) + c_2 \, \mathfrak{m}_2(t)$, the convolution is the same linear combination of the individual convolutions
\begin{equation}
 \label{eq:linearity}
\delta S_A(t) = c_1 \, \mathfrak{m}_1(t)*\mathfrak{n}(t) + c_2 \, \mathfrak{m}_2(t)*\mathfrak{n}(t) .
\end{equation}

\item \textbf{Time-translation invariance} \\
A convolution is left invariant by translating it
\begin{equation}
 \label{eq:translation_conv}
\delta S_A(t+t_0) = \mathfrak{m}(t+t_0)*\mathfrak{n}(t) = \mathfrak{m}(t)*\mathfrak{n}(t+t_0) \, .
\end{equation}

\item \textbf{Differentiation rule} \\
If $\delta S_A(t) = \mathfrak{m}(t)*\mathfrak{n}(t)$, differentiation follows the simple rule 
\begin{equation}
 \label{eq:diff_rule_conv}
\frac{d \, \delta S_A(t)}{d t} = \frac{d \mathfrak{m}(t)}{d t}*\mathfrak{n}(t) = \mathfrak{m}(t)*\frac{d \mathfrak{n}(t)}{d t} \, .
\end{equation}

\item \textbf{Integration rule}\\
If $\delta S_a(t) = \mathfrak{m}(t)*\mathfrak{n}(t)$, the integral of the convolution is the product of integrals.
\begin{equation}
 \label{eq:int_rule_conv}
\int dt \, \delta S_A(t) = \bigg[ \int dt \, \mathfrak{m}(t)\bigg] \cdot \bigg[ \int dt \, \mathfrak{n}(t)\bigg] \, .
\end{equation}

\end{itemize}

\noindent One can use these properties and study a class of source functions $\mathfrak{m}(t)$ which are relatively simple. If the source function is complicated, but decomposable into a series of such simpler functions, these properties will prove useful in studying evolution of entanglement entropy in such a case.

\subsection{Time-Dependent First Law of Entanglement}
\label{first_law}

There always exists a First Law of Entanglement Entropy for small subregions. For the kind of quenches we are studying, the reduced density matrix for the subregion $A$ after the quench can be written as $ \rho_A = \rho^{(0)}_A + \lambda_A \, \Delta_\lambda \rho_A$ and so can the entanglement entropy:
\begin{equation}
\delta S_A(t) =\delta S^{(0)}_A + \lambda_A \, \Delta_\lambda S_A .
\end{equation}
In fact, the change in the entanglement entropy can be shown to be
\begin{equation}
\label{eq:time-ind-first-law}
 \Delta_\lambda S_A = \text{tr} \bigg( \Delta_\lambda \rho_A \, K^{(0)}_A \bigg)  ,
\end{equation}
where the \textit{Modular Hamiltonian} $K_A^{(0)}$ is defined as $ K^{(0)}_A \equiv -\log ( \rho^{(0)}_A) $. Equation \eqref{eq:time-ind-first-law} is called the First Law of Entanglement Entropy. \\

\noindent In this subsection, we will show that there exists such a first law even in the time-dependent case for the small subregions. In equation \eqref{eq:source_func}, if we assume for a moment that the source is constant in time $g(t)=g_0$, then the entanglement growth \eqref{eq:EE_lin_resp} simplifies to
\begin{equation}
 \delta S_A(t) = \frac{d_\theta \, g_0 }{16 \, \pi \, G_N \, u_H^{d_\theta+z}} \, \int_{-\infty}^{\infty} dt' \mathfrak{n}(t') .  
\end{equation}
Observe that we can explicitly calculate the indefinite integral of the non-trivial part of the response function \eqref{eq:response_func}:
\begin{equation}
\label{eq:indef_int_resp_func}
  \int dt \,  (zt)^{\frac{1}{z}} \, \bigg[1-\bigg( \frac{t}{t_*} \bigg)^{\frac{2 \, d_\theta}{z}} \bigg]^{\frac{1}{2}} =  \frac{ (z \, t)^{1+\frac{1}{z}}}{ (z+1)} \,  \,   {}_2F_1 \bigg[ -\frac{1}{2}, \frac{z+1}{2 d_\theta}, \frac{2 d_\theta+z+1}{2 d_\theta}; \bigg( \frac{t}{t_*} \bigg)^{\frac{2 d_\theta}{z}}  \, \bigg]  \, .
\end{equation}
Using this, we define the function
\begin{equation}
\label{eq:int_resp}
\mathfrak{B}(t) \equiv \frac{2 \, \pi \, \mathcal{A}_{\Sigma} \,  (z \, t)^{1+\frac{1}{z}} }{d_\theta \, (z+1)}  \,   {}_2F_1 \bigg[ -\frac{1}{2}, \frac{z+1}{2 d_\theta}, \frac{2 d_\theta+z+1}{2 d_\theta}; \bigg( \frac{t}{t_*} \bigg)^{\frac{2 d_\theta}{z}}  \, \bigg]  \, ,
\end{equation}
which can be thought of as the indefinite integral of the full response function. Taking the limits $t \to 0$ and $t \to t_*$ of the function $\mathfrak{B}(t)$, we get the entanglement growth to be
\begin{align}
\label{eq:EE_const_quench} 
\delta S_A(t \le t_*) &=  \frac{  \mathcal{A}_{\Sigma} \, g_0 \,  }{8 \, G_N \, u_H^{d_\theta+z}} \, \frac{ \, (z t)^{1+\frac{1}{z}}}{ (z+1)} \,  \,   {}_2F_1 \bigg[ -\frac{1}{2}, \frac{z+1}{2 d_\theta}, \frac{2 d_\theta+z+1}{2 d_\theta}; \bigg( \frac{t}{t_*} \bigg)^{\frac{2 d_\theta}{z}}  \, \bigg]  \, , \\ 
\delta S_A(t \ge t_*) &= \frac{  \mathcal{A}_{\Sigma}  \, g_0 \,}{8 \, G_N \, u_H^{d_\theta+z}} \, \frac{ \sqrt{\pi} \, (z t_*)^{1+\frac{1}{z}} \, \Gamma\big[\frac{2 d_\theta+z+1}{2d_\theta}\big]}{2 \,  (z+1) \, \Gamma \big[ \frac{3 d_\theta + z+1}{2 d_\theta} \big] } \,  .
\end{align}
We can rewrite the latter expression as
\begin{equation}
 \label{eq:first_law_const_energy}
\delta S_A(t \ge t_*) = \frac{\delta E_A}{T_A} \, ,
\end{equation}
where by $\delta E_A$ we denote the total energy inside the entangling surface:
\begin{equation}
 \label{eq:total_energy}
\delta E_A = \frac{d_\theta \, g_0 \, V_A}{16 \, \pi \, G_N \, u_H^{d_\theta+z}}  =   \frac{d_\theta \, g_0 \,   \ell_p^{d-2} \, \ell \, }{16 \, \pi \, G_N \, u_H^{d_\theta+z}} \, .
\end{equation}
Using equation \eqref{eq:rel_tstar_ell} for $\ell$ in terms of $t_*$, we can write the parameter $T_A$ as
\begin{equation}
 \label{eq:const_ent_Temp}
T_A =  \frac{ (z+1) \, \Gamma \big[ \frac{3 d_\theta + z+1}{2 d_\theta} \big] \, \Gamma \big[\frac{d_\theta+1}{2 d_\theta} \big] \, }{ \, 2 \pi \, z \,  \Gamma\big[\frac{2 d_\theta+z+1}{2d_\theta}\big] \, \Gamma \big[ \frac{2 D_\theta+1}{2 d_\theta} \big] \, t_*   } \, .
\end{equation}
It is called the \textit{Entanglement Temperature} and has been extensively studied in literature \cite{Bhattacharya:2012mi, Allahbakhshi:2013rda, Wong:2013gua, Roychowdhury:2016fgf}. Observe that the entanglement temperature only depends on the shape of the entangling surface (through $t_*$) and not on the quench. Thus, it is a characteristic of the subregion. In the relativistic limit ($ z \to 1, \theta \to 0$), it reduces to the well-known expression \cite{Bhattacharya:2012mi, Lokhande:2017jik}
\begin{equation}
 \label{eq:CFT_ent_temp}
T_A^{\text{CFT}} = \frac{(d+1) \, \Gamma \left[\frac{d+1}{2(d-1)}\right] \,  \Gamma \left[\frac{d}{2 (d-1)}\right] }{ \pi \, t_* \,  \Gamma \left[\frac{1}{2 (d-1)}\right] \, \Gamma \left[\frac{d}{d-1}\right] } .
\end{equation}
This equation is a manifestation of the standard First Law of Entanglement Entropy.\\

\noindent We now generalize this law to the case of adiabatic sources. In this case, the function $\mathfrak{m}(t)$ varies slowly and is approximately constant for time intervals of the order of $t_*$. To derive the first law, let us start by partially integrating the entanglement entropy integral \eqref{eq:EE_lin_resp}, while using the source function \eqref{eq:source_func}. We get 
\begin{equation}
 \delta S_A(t) = \bigg[ \epsilon(t-t') \, \mathfrak{B}(t') \bigg]_{t'=0}^{t'=t_*}  - \int_0^{t_*} dt' \, \frac{d \epsilon(t-t')}{d t'} \, \mathfrak{B}(t')  \, ,
\end{equation}
where $\mathfrak{B}(t)$ is defined in equation \eqref{eq:int_resp}. Note that one cannot cancel the integration measure $dt'$ in the numerator and the denominator of the second term. One is supposed to think of the derivative of $\epsilon(t-t')$ as an independent function of $t'$. \\

\noindent Naively, this gives us the full expression for the time-dependent change in entanglement entropy. However, there is still a choice of the integration constant in the definition of the function $\mathfrak{B}(t)$ \eqref{eq:int_resp}. For calculating the entanglement growth after a constant quench (equation \eqref{eq:EE_const_quench}), this choice did not matter because the integration constant got cancelled when one implemented the limits of the definite integral in the end. Here, the presence of the integration constant matters because it gives a non-vanishing term as $\epsilon(t)$ changes across an interval of size $t_*$. So, how do we fix this constant? We would like to reproduce the time-independent first law of entanglement entropy \eqref{eq:first_law_const_energy} when we make the source $\epsilon(t)$ time-independent. This fixes the constant to be $-\frac{V_A}{T_A}$. In particular, this implies the condition $\mathfrak{B}(t_*)=0$, which is not surprising for a integral of a function with compact support. We also have $\mathfrak{B}(0)=-\frac{V_A}{T_A}$. Using these, we get
\begin{equation}
\label{eq:general_first_law}
 \delta S_A(t) =  \frac{\delta E_A(t)}{T_A} -   \int_0^{t_*} dt' \,  \frac{d \epsilon(t-t')}{d t'} \, \mathfrak{B}(t')  \, .
\end{equation}
This is then our generalization of the First Law of Entanglement to adiabatic time-dependent cases. We can identify the precise condition when this first law will hold. The adiabatic approximation is true if the integral in the equation above is not very large compared to the first term. One can thus show that if
\begin{equation}
 \frac{d \, \epsilon(t)}{d t} \ll \frac{\epsilon(t)}{t_*} ,
\end{equation}
then the above first law holds.

\subsection{An Analogue of Relative Entropy}
\label{relative_entropy}

In time-independent cases, relative entropy between two density matrices $\rho$ and $\sigma$ can be shown to be \cite{Blanco:2013joa, Lashkari:2014yva},  
\begin{equation}
S_{\text{rel}}(\rho|\sigma) = \Delta \langle K_\sigma \rangle - \Delta S  \, ,
\end{equation}
where 
\begin{equation}
 \Delta K \equiv - \text{tr} \bigg(\rho \, \log \sigma \bigg) + \text{tr} \bigg(\sigma \, \log \sigma \bigg) \,  , \qquad \Delta S = \text{tr} \bigg(\rho \log \rho \bigg) - \text{tr} \bigg(\sigma \, \log \sigma \bigg) \, .
\end{equation}
We see that equation \eqref{eq:general_first_law} is analogous to this. In particular, we can define a time-dependent analogue of relative entropy between the vacuum of non-relativistic field theories and thermal states produced by action of a global quench
\begin{equation}
\label{eq:rel_entrop_def}
\delta S_{\text{rel}}(t) \equiv \frac{\delta E_A(t)}{T_A} - \delta S_A(t) ,
\end{equation}
where by $\delta$ we mean the term that remains after subtracting the vacuum quantity as well as source-dependent terms. This quantity has some nice properties. When there is no quench, it vanishes owing to the time-independent first law. For an adiabatic quench, it is very small, implying that the time-dependent excited state is not very far from the vacuum. For quenches of finite duration, it is non-zero only for $0 \le t \le t_{\text{sat}}$. Moreover, like the relative entropy, it is positive definite and hence must have an extremum in the interval $0 \le t \le t_{\text{sat}}$. 
All these properties suggest that this quantity can be thought of as an order parameter for out-of-equilibrium states. That is, this quantity tells us how far an out-of-equilibrium state is at time $t$ compared to an equilibrium state with the same energy density $\epsilon(t)$. \\

\noindent We can in fact show that for $0 \le t \le t_{\text{sat}}$, $\delta S_{\text{rel}}(t) \ge 0$.  To see this, first we recall the integral expression for the relative entropy
\begin{equation}
\delta S_{\text{rel}}(t) = \int_0^{t_*} \, dt' \, \frac{d \epsilon(t-t')}{d t'} \, \mathfrak{B}(t') = \int_{-\infty}^{\infty} dt' \, \frac{d \epsilon(t-t')}{d t'} \, \mathfrak{B}(t') \, \big[\Theta(t')-\Theta(t'-t_*) \big] .
\end{equation}
Now, we observe that
\begin{equation}
 \frac{d \epsilon(t-t')}{d t'} = -\frac{d \epsilon(t-t')}{d t} .
\end{equation}
Hence we get
\begin{equation}
 \delta S_{\text{rel}}(t) = - \int_{-\infty}^{\infty} dt' \, \frac{d \epsilon(t-t')}{d t} \, \mathfrak{B}(t') \, \big[\Theta(t')-\Theta(t'-t_*) \big] = - \frac{d \epsilon(t)}{d t}*\tilde{\mathfrak{B}}(t) \, ,
\end{equation}
with $\tilde{\mathfrak{B}}(t) \equiv \mathfrak{B}(t) \, \big[\Theta(t)-\Theta(t-t_*) \big]$. Now referring to the discussion above equation \eqref{eq:general_first_law}, $\mathfrak{B}(t) \le 0$ for $0 \le t \le t_*$ because of the choice of the integration constant. Also, $\frac{d \epsilon(t)}{dt} \ge 0$ due to the Null Energy Condition in the bulk. Hence, $\delta S_{\text{rel}} \ge 0$ for $0 \le t \le t_*$ .  \\

\noindent Apart from providing an order parameter to understand out-of-equilibrium states, the time-dependent relative entropy also helps us to organize the post-quench time-evolution of the field theory. To see how this is done, let us first calculate the time derivative of the relative entropy using the differentiation rule of convolution
\begin{align}
\begin{split}
\label{eq:der_rel_entr}
 \frac{d \, \delta S_{\text{rel}}}{d t} &= -\frac{d \epsilon(t)}{d t}* \frac{d \tilde{\mathfrak{B}}(t)}{d t} \\ 
&=  \frac{d \epsilon(t)}{d t}*\frac{V_A}{T_A} - \frac{d \epsilon(t)}{d t}*\mathfrak{n}(t) \, ,
\end{split}
\end{align}
where the first term is a boundary term from the boundary at $t=t_*$. We can now define the following regimes during the post-quench evolution
\begin{itemize}[leftmargin=5mm, itemindent=0mm]
 \item \textbf{Driven regime} ( $ 0 \le t \le t_q $ ) \\ 
For $ 0 \le t \le t_q$, the system is being driven by the quench as $\frac{d \epsilon(t)}{d t} \ge 0$. Since $\mathfrak{n}(t) \ge 0$, the second term is negative wrt first in equation \eqref{eq:der_rel_entr}. However, both the terms contribute in general and there is a change in relative entropy as a function of time. \\

\item \textbf{Transient regime} ( $ t_q \le t \le t_{\text{sat}} $  )\\
For $t_q \le t \le t_{\text{sat}}$, the quench has stopped acting and hence $\frac{d \epsilon(t)}{dt }=0$. Thus, only the second term in equation \eqref{eq:der_rel_entr} contribute. As a result, we get
\begin{equation}
 \frac{d \, \delta S_{\text{rel}}}{d t} \le 0 .
\end{equation}
The distance in the Hilbert space between the vacuum and the excited state thus keeps decreasing with time in this regime.

\end{itemize}

\section{Special Cases}
\label{special_cases}

In this section, we will study the growth of entanglement for small subsystems for some explicit quenches. As we will see, they cover a wide range of time-dependent perturbations to the field theory.

\subsection{Instantaneous Quench}
\label{instantaneous}

As a first example, we will study the instantaneous global quantum quench defined by $g(t) = \Theta(t)$. This will elucidate the use of the convolution formula \eqref{eq:EE_lin_resp} for entanglement entropy. Further, our discussion fills-in a gap in the literature \cite{Alishahiha:2014cwa, Fonda:2014ula} for entanglement growth after instantaneous quenches, which has focused more on large subsystems. From equation \eqref{eq:EE_general}, the entanglement entropy in this case is
\begin{equation}
\label{eq:theta_quench_EE}
 \delta S_A =  \frac{\ell_p^{(d-2)}}{4 \, G_N \, u_H^{(d_\theta+z)}}   \int_0^{u_*} du \,    u^z \, \sqrt{1 - \left[ \frac{u}{u_*} \right]^{2d_\theta}} \,    \Theta \bigg (t - \frac{u_*^z}{z} \bigg) .
\end{equation}
The Heaviside Theta function can be used to naturally divide the evolution of entanglement entropy in the following three regimes,
\begin{enumerate}[leftmargin=5mm, itemindent=0mm]
\item \textbf{Pre-quench regime} \\
When $t<0$ the integrand vanishes for the entire integration domain. Hence
$$\delta S_A(t<0)=0 . $$

\item \textbf{Post-saturation regime} \\
When $t>t_*=\frac{u_*^z}{z}$ , the integrand is nonzero for the entire integration domain. Furthermore the final result then does not depend on time. Thus the growth of entanglement saturates in this regime and gives us the value $\delta S_{eq}$ of entanglement entropy in the final equilibrium state. We will use this value later to normalize our plots. 
\begin{equation}
\label{eq:deltaS-equilibrium-value}
\delta S_{eq} = \frac
{\sqrt{\pi} \ell_p^{d-2} \, \, u_*^{1+z} \, \Gamma \big[\frac{2d_\theta+z+1}{2 d_\theta} \big]}
{8 \, G_N \, (1+z) \, u_H^{d_\theta+z} \, \Gamma \big[\frac{3 d_\theta+z+1}{2d_\theta}\big]} .
\end{equation}

\item \textbf{Time-dependent regime} \\ 
For $0<t<t_*$, the value of $\delta S(t)$ is actually time dependent. This is thus an out-of-equilibrium regime. To study this regime, we split off the equilibrium entanglement entropy of equation \eqref{eq:deltaS-equilibrium-value} and observe that we can define a dimensionless parameter
 \begin{equation}
\label{eq:x_def}
x \equiv  \bigg( \frac{t}{t_*} \bigg)^{\frac{1}{z}}, \quad\text{with}\quad  t_*=\frac{u_*^z}{z} .
\end{equation}
In terms of this parameter, the time evolution of the entanglement entropy is
\begin{align}
\label{eq:deltaS(t)}
\delta S_A(t) &= \delta S_{eq} \, \mathcal{F}(x) \, ,  \\ 
\begin{split}
\label{eq:Fx-def}
\mathcal{F}(x) &\equiv  \frac{\Gamma \big[\frac{3d_\theta + z+ 1}{2 d_\theta} \big] }{\Gamma \big[\frac{3}{2} \big] \, \Gamma \big[ \frac{z+1}{2 d_\theta}\big]} \, \cdot \beta \bigg[x^{2 d_\theta} , \frac{z+1}{2 d_\theta}, \frac{3}{2} \bigg] \, .
\end{split}
\end{align}
where $\beta(z,a,b)$ is the Incomplete Beta Function. 

\end{enumerate}

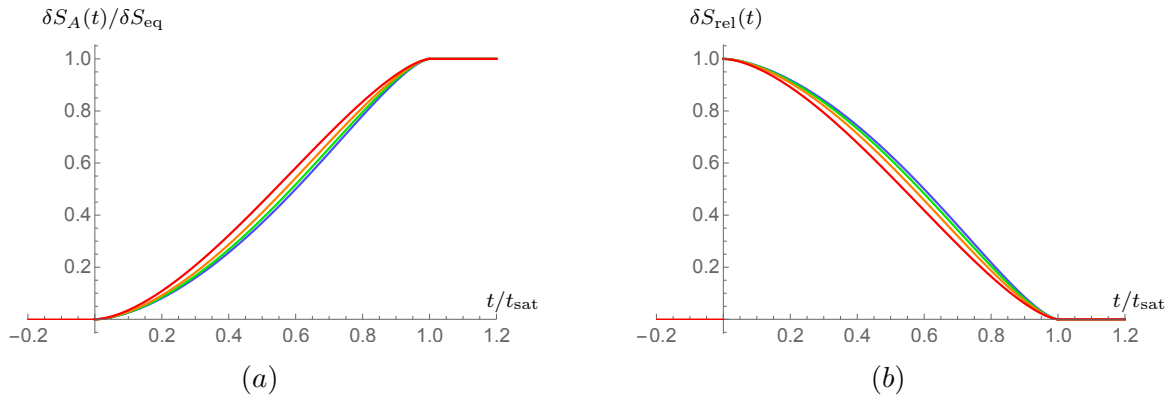
\begin{figure}[H]
$$
\begin{array}{cc}
  \includegraphics[angle=0,width=0.43\textwidth]{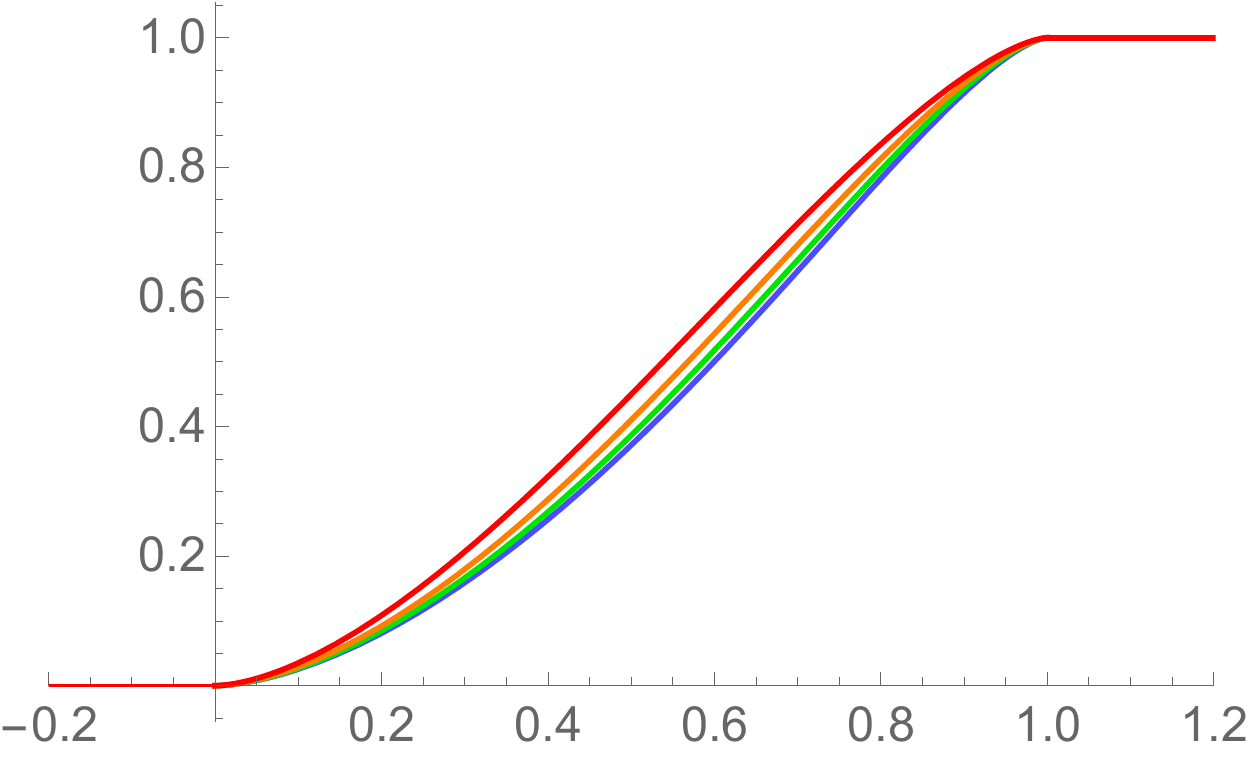} \qquad\qquad & \includegraphics[angle=0,width=0.43\textwidth]{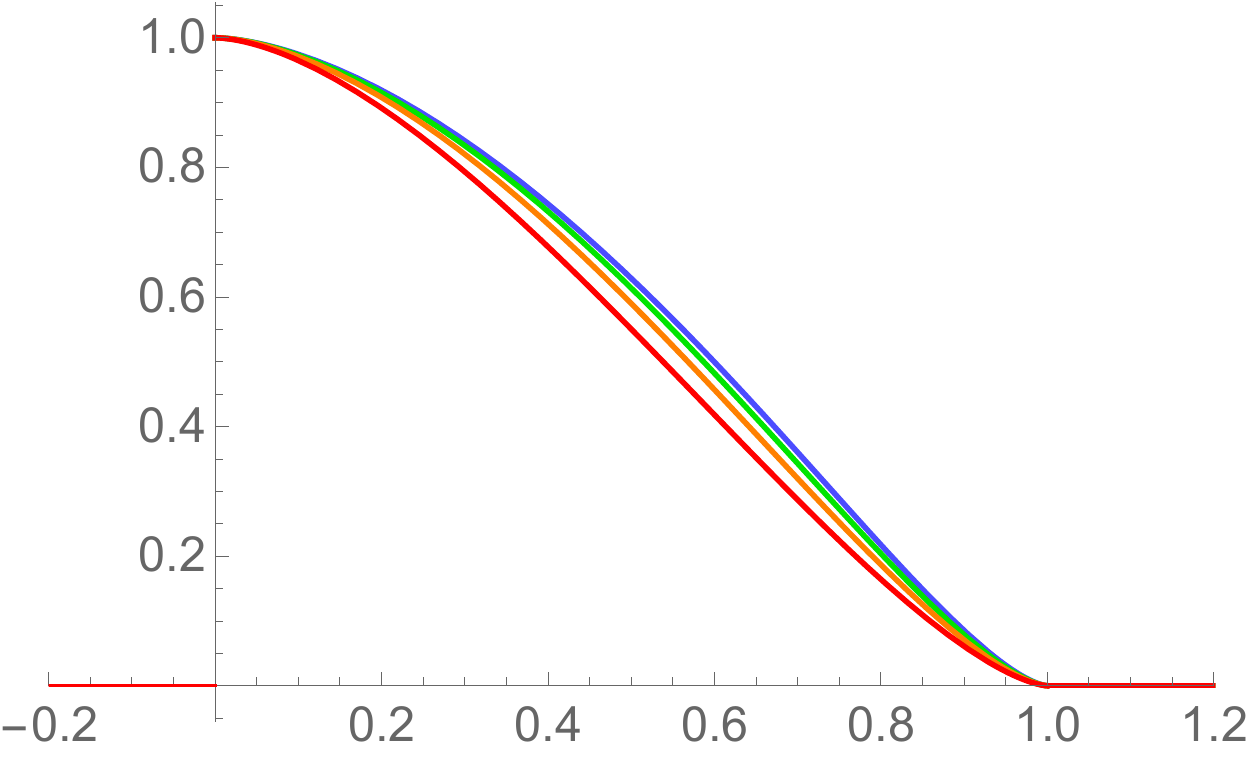}\\
  (a) \qquad\qquad & (b)
\end{array}
$$
\begin{picture}(0,0)
\put(20,154){{\scriptsize $\delta S_A (t)/\delta S_{\text{eq}}$}}
\put(265,154){{\scriptsize $ \delta S_{\text{rel}}(t)$}}
\put(188,47){{\scriptsize $t/t_{\text{sat}}$}}
\put(425,47){{\scriptsize $t/t_{\text{sat}}$}}
\end{picture}
\vspace{-0.7cm}
\caption{Entanglement entropy $(a)$ and relative entropy $(b)$ after an instantaneous quench in $d=\{2,3,4,5\}$ dimensions, depicted in red, orange, green and blue, respectively. We have set $z=1.5$, $\theta=0.1$ and $t_q+t_*=1$ to obtain the plots.}
\label{fig:HSV_instant_EE_RE}
\end{figure}

\noindent Having described the regimes, we now observe that the problem of studying the evolution of entanglement entropy simplifies to the problem of studying the behavior of $\mathcal{F}(x)$ for different time scales. Before we study this in detail, we depict in Figure \ref{fig:HSV_instant_EE_RE} the evolution of entanglement entropy for instantaneous global quenches. 

\subsubsection{Nature of Entanglement Growth}

We now study in detail the time-dependent regime. To organize our study, we observe that the time-dependent regime can be further classified into three different sub-regimes. The existence of these sub-regimes can also be seen from Figure \ref{fig:HSV_instant_EE_RE} , where as a function of time, the entanglement growth has three different functional forms. We now describe each of these sub-regimes in detail.

\subsubsection*{1. Early Time Growth}

The first sub-regime is the one where time is very close to zero, $t \approx 0$. This implies that the dimensionless parameter $x \approx 0$. This motivates us to expand $\mathcal{F}(x)$, given by equation \eqref{eq:Fx-def}, near $x=0$ . We expect that this simplifies the expression for the entanglement growth near $t \approx 0$. The expansion of $\mathcal{F}(x)$ near $x=0$ is
\begin{equation}
\mathcal{F}(x) = \frac{2 d_\theta \, \Gamma \big[\frac{3d_\theta + z+ 1}{2 d_\theta} \big] \, x^{z+1} }{(z+1) \, \Gamma \big[\frac{3}{2} \big] \, \Gamma \big[ \frac{z+1}{2 d_\theta}\big]} \, \bigg(1-\frac{z+1}{2(2 d_\theta+z+1)} x^{2 d_\theta} + \mathcal{O}(x^{4 d_\theta}) \bigg) ,
\end{equation}
where we have assumed $z \ge 1$ and $d_\theta \ge 0$ following \cite{Kachru:2008yh}. Thus the time growth of entanglement entropy $\delta S(t)$ near $t \approx 0$ becomes
\begin{equation}
\label{eq:deltaS_quadratic}
 \delta S_A(t) = \frac{ \ell_p^{d-2} \, (z \, t)^{1+\frac{1}{z}} }{4 \, G_N \, (z+1) \, u_H^{d_\theta+z}}  \, \bigg(1- \frac{z+1}{2 (2 d_\theta+z+1)} \bigg[\frac{t}{t_*} \bigg]^{\frac{2 d_\theta}{z}}  \bigg) .
\end{equation}
If we take the limit $z \to 1$, we recover the Vaidya solution with an asymptotically AdS background. In this limit, the above equation reproduces the well-known results for early time growth of entanglement \cite{Kundu:2016cgh} , in particular the quadratic function of time. It was also argued that this time dependence is  universal irrespective of the size of the subsystem \cite{Liu:2013iza, Liu:2013qca, Kundu:2016cgh}. The argument is that it is the UV part of the CFT which largely determines the early time growth and hence depends only on the symmetries \cite{Kundu:2016cgh}. We expect the same to be true of the early time entanglement growth for non-relativistic field theories under investigation here.

\subsubsection*{2. Quasilinear Growth}

\begin{figure}[!h]
$$
  \includegraphics[angle=0,width=0.43\textwidth]{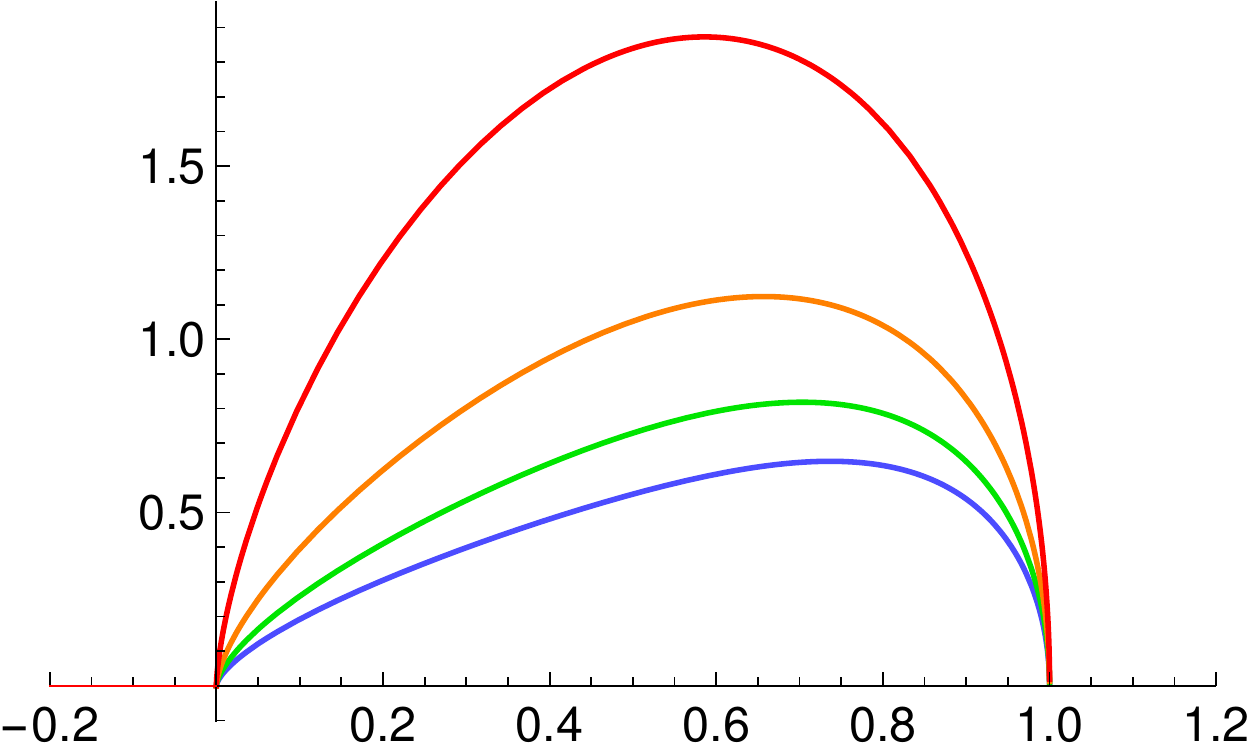} 
$$
\begin{picture}(0,0)
\put(143,143){{\scriptsize $\mathfrak{R}_{\text{HSV}}(t)$}}
\put(320,32){{\scriptsize $t$}}
\end{picture}
\vspace{-0.7cm}
\caption{\small Entanglement velocity as a function of time for different values of $d$. The colors blue, green, red and orange represents $d=\{5,4,3,2\}$ respectively. We have set $\theta=0.1, z=1.5, t_*=1$. }
\label{fig:EEvelVSd}
\end{figure}
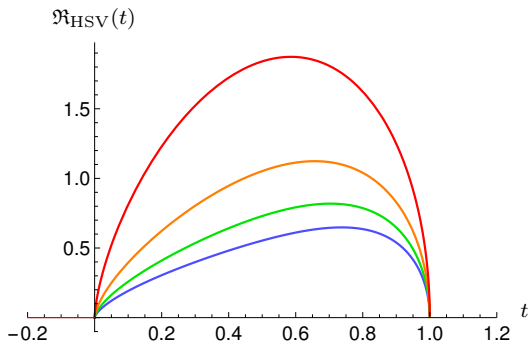

The time dependence of entanglement entropy is not universal for intermediate time scales
\begin{equation}
  0 \ll t \ll \frac{u_*^z}{z} \quad .
\end{equation}
It was shown in \cite{Liu:2013iza, Liu:2013qca, Alishahiha:2014cwa, Fonda:2014ula} that there is a linear regime for large subsystems. But as was argued in \cite{Kundu:2016cgh, Lokhande:2017jik}, there is no such regime for small subsystems. These results were for CFTs, but even in the case of non-relativistic field theories, we observe similar features. In particular, Figure \ref{fig:HSV_instant_EE_RE} suggests that we can model the growth of entanglement as quasilinear. Then to study such a regime, \cite{Liu:2013iza, Liu:2013qca, Kundu:2016cgh, Lokhande:2017jik, Fonda:2014ula} identified a parameter called \textit{Entanglement Velocity}. For CFTs, it was defined as \cite{Liu:2013iza, Liu:2013qca} 
\begin{equation}
 \mathfrak{R}_{\text{CFT}} \equiv \frac{V}{\delta S_{\text{eq}} \, \mathcal{A}_\Sigma} \, \frac{d \,  \delta S_A(t)}{d t} .
\end{equation}
Following \cite{Fonda:2014ula}, we define entanglement velocity for the case of non-relativistic theories to be 
\begin{align}
\begin{split}
 \mathfrak{R}_{\text{HSV}}(t) &= \frac{V}{\mathcal{A}_{\Sigma}} \, \frac{d \, \mathcal{F}(x(t))}{d \, t} \, .\\
  \end{split}
\end{align}
Recall that from equation \eqref{eq:rel_tstar_ell}, we have 
\begin{equation}
\frac{V}{\mathcal{A}_{\Sigma}} = \frac{\Gamma \big[\frac{3}{2} \big] \, \Gamma \big[ \frac{d-\theta}{2 d_\theta} \big] \, u_*}{d_\theta \, \Gamma \big[ \frac{d_\theta+1}{2 d_\theta} \big]}  \, .
\end{equation}
A simple calculation then shows that the entanglement velocity is equal to
\begin{equation}
\label{eq:EE_velocity_HSV}
  \mathfrak{R}_{\text{HSV}}(t) =\frac{2 \, \Gamma \big[ \frac{3 d_\theta + z +1}{2 d_\theta} \big] \, \Gamma \big[ \frac{d-\theta}{2 d_\theta} \big]}{z \, t_* \, \Gamma \big[ \frac{z+1}{2 d_\theta} \big] \, \Gamma \big[ \frac{2 d_\theta+1}{d_\theta} \big]} \, (z \, t)^{\frac{1}{z}} \, \sqrt{1- \big[ \frac{t}{t_*} \big]^{\frac{2 d_\theta}{z}}} .
\end{equation}
As we can see, this is not independent of the subsystem. A similar feature was observed in \cite{Kundu:2016cgh, Lokhande:2017jik}, contrary to the large subsystem limit, where the entanglement velocity is universal \cite{Liu:2013iza, Liu:2013qca}. In Figure \ref{fig:EEvelVSd}, we plot the entanglement velocity as a function of time for different dimensions.\\

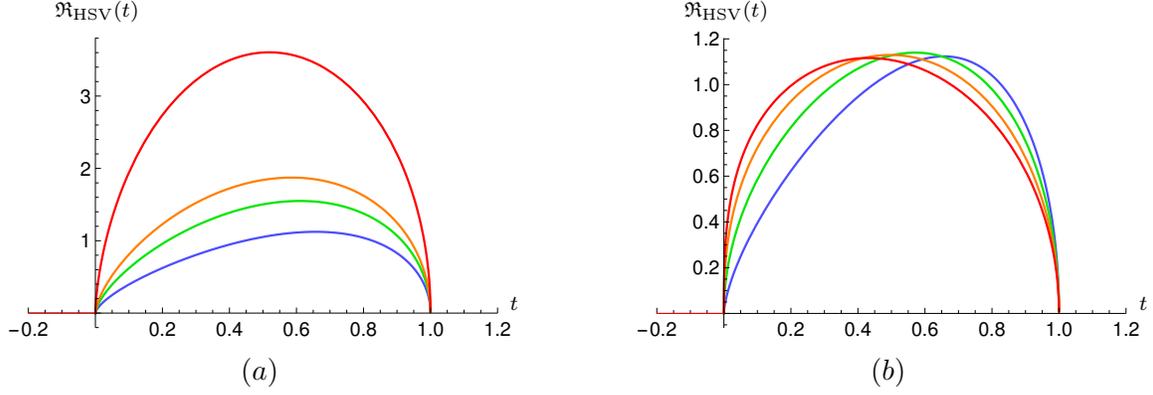
\begin{figure}[!h]
$$
\begin{array}{cc}
  \includegraphics[angle=0,width=0.43\textwidth]{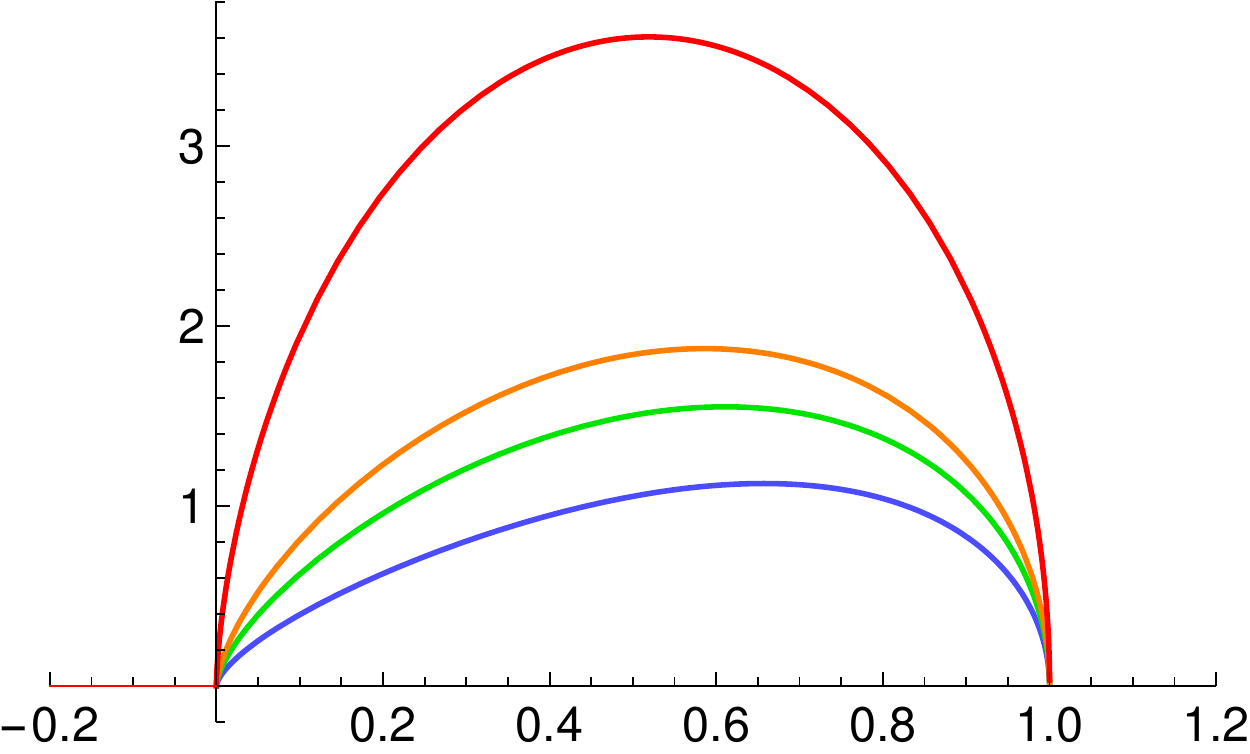} \qquad\qquad & \includegraphics[angle=0,width=0.43\textwidth]{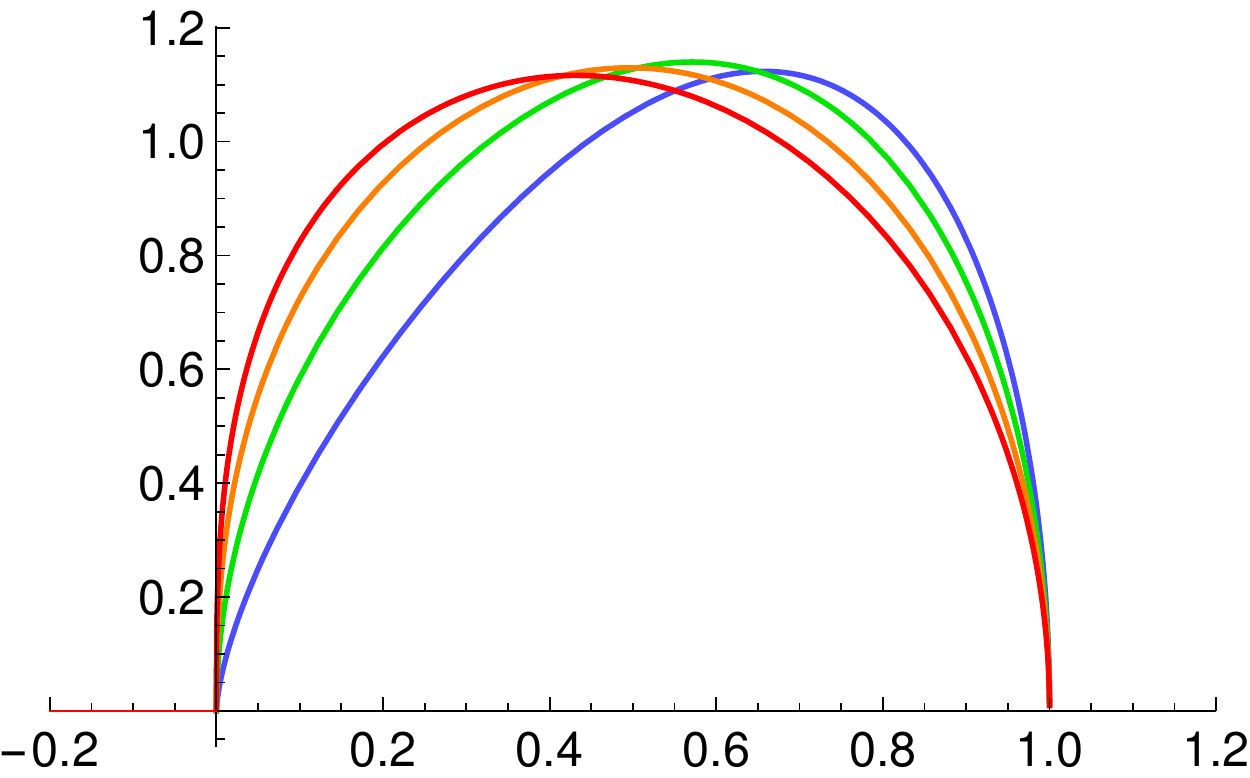}\\
  (a) \qquad\qquad & (b)
\end{array}
$$
\begin{picture}(0,0)
\put(25,154){{\scriptsize $\mathfrak{R}_{\text{HSV}}(t)$}}
\put(263,154){{\scriptsize $\mathfrak{R}_{\text{HSV}}(t)$}}
\put(197,43){{\scriptsize $t$}}
\put(435,43){{\scriptsize $t$}}
\end{picture}
\vspace{-0.7cm}
\caption{\small Time-dependence of entanglement velocity for different values of $\theta$ and $z$. The colors blue, green, red and orange represents $\theta=\{0.1, 0.8, 1.1, 1.7\}$  in part $(a)$ and $z=\{1.5,2,2.5,3\}$ in part $(b)$ respectively. Common parameters are $d=3, t_*=1$. }
\label{fig:EEvelVSthetaZ}
\end{figure}

\noindent To understand this instantaneous velocity better, we study it in some limits. First we set $z=1.5$ but keep $\theta$ arbitrary.  This is the limit of relativistic but non-conformal field theories or the purely hyperscaling-violating theories. The time dependence of the velocity then is shown in part (a) of Figure \ref{fig:EEvelVSthetaZ}. In part (b) of the same figure, we keep $\theta=0.1$ and plot the time-dependence of the velocity for different values of $z$.
For large subsystems, the entanglement velocity was found to be connected to the velocity of quasiparticles produced during the quench \cite{Calabrese:2005in}. It was argued later that the spread of entanglement is causal and that this velocity is bounded by 1 in units of $c$ \cite{Liu:2013iza, Liu:2013qca, Casini:2015zua}. However, the instantaneous velocity does not satisfy this bound \cite{Kundu:2016cgh, Lokhande:2017jik, Ghaffarnejad:2018aui, Ghaffarnejad:2018vry}, although the time-average of the velocity is bounded. This suggests that there is no quasi-particle picture for the production and propagation of entanglement for small subsystems. In the case of non-relativistic theories, the situation becomes more muddled. We will comment on these issues by studying the maximum and average velocity. \\

\begin{figure}[!h]
$$
\begin{array}{cc}
  \includegraphics[angle=0,width=0.43\textwidth]{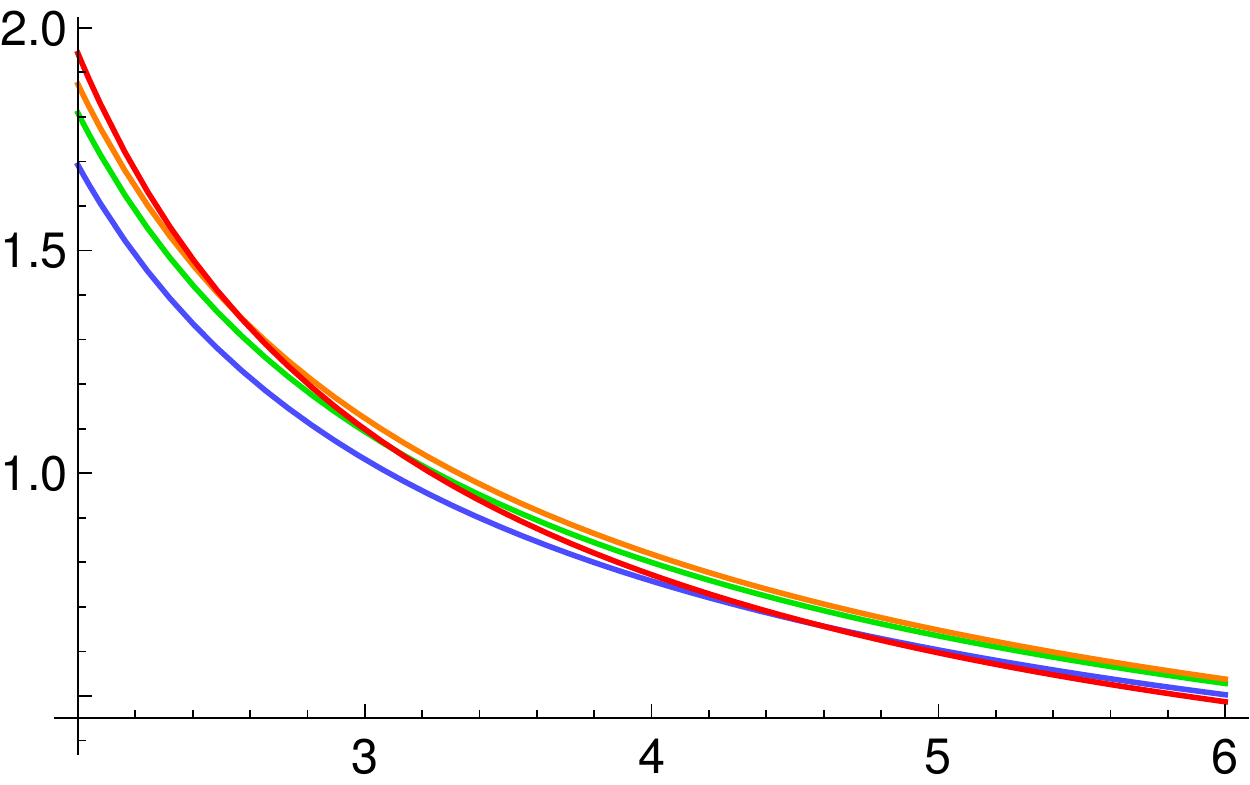} \qquad\qquad & \includegraphics[angle=0,width=0.43\textwidth]{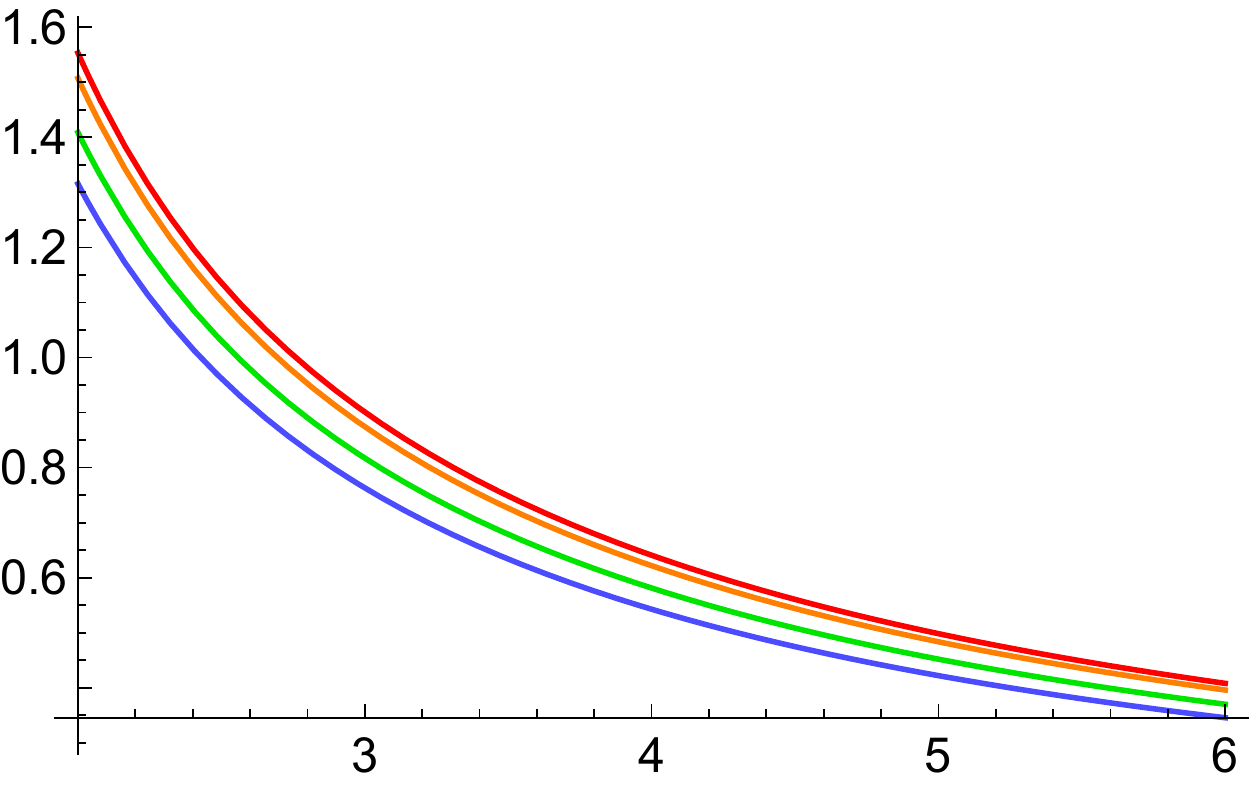}\\
  (a) \qquad\qquad & (b)
\end{array}
$$
\begin{picture}(0,0)
\put(6,155){{\scriptsize $\mathfrak{R}_{\text{HSV}}(t_{\text{max}})$}}
\put(246,155){{\scriptsize $\mathfrak{R}_{\text{avg}}$}}
\put(200,43){{\scriptsize $d$}}
\put(435,43){{\scriptsize $d$}}
\end{picture}
\vspace{-0.7cm}
\caption{\small (a) maximum and (b) average entanglement velocities as a function of $d$. The colors blue, green, orange and red correspond to $z=\{1.3, 1.5, 1.9, 2.7\}$ respectively and we have set $\theta=0.1, t_*=1$.}
\label{fig:VelsVSd}
\end{figure}
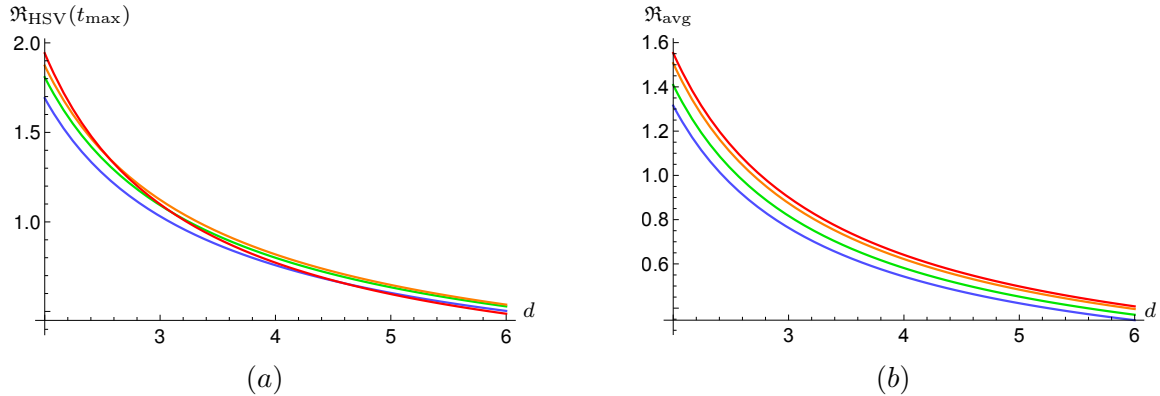

\noindent The maximum entanglement velocity in our case can be found using
\begin{equation}
 	\frac{d \mathfrak{R}_{\text{HSV}(t)}}{dt} =0 \quad\implies\quad
 		t_{\text{max}} = t_* \,  (d-\theta)^{\frac{-1}{2d_\theta}} ,
 \end{equation}
where we have assumed that $z \ge 1$ and $\theta>0$. This has the value
\begin{equation}
\label{eq:F_prime_max}
\mathfrak{R}_{\text{HSV}}(t_\text{max})= \frac{2 \, \Gamma \big[ \frac{3 d_\theta + z +1}{2 d_\theta} \big] \, \Gamma \big[ \frac{d-\theta}{2 d_\theta} \big]}{ \Gamma \big[ \frac{z+1}{2 d_\theta} \big] \, \Gamma \big[ \frac{2 d_\theta+1}{d_\theta} \big]} \, \bigg[\frac{d_\theta}{d-\theta} \bigg]^{\frac{1}{2}} \, (d-\theta)^{\frac{-1}{2 d_\theta}}  \, (z \, t_*)^{\frac{1-z}{z}} ,
\end{equation}
In the AdS limit $\theta\to0$, $z \to 1$, we reproduce the known result \cite{Kundu:2016cgh}
\begin{equation}
\mathfrak{R}_{\text{CFT}}(t_{\text{max}})= \frac{4 \, \Gamma \big[ \frac{3}{2}-\frac{1}{d-1} \big] \, \Gamma \big[ \frac{d}{2(d-1)} \big] \, (d-1)^{\frac{3}{2}} \,   }{\Gamma \big[ \frac{1}{2(d-1)} \big] \, \Gamma \big[ \frac{1}{d-1} \big] \, d^{\frac{d}{2(d-1)}} \,  } ,
\end{equation}
which gives, for example, a maximum velocity of $\frac{3}{2}$ for $d=2$. In part (a) of Figure \ref{fig:VelsVSd} we plot the maximum as a function of dimension.\\

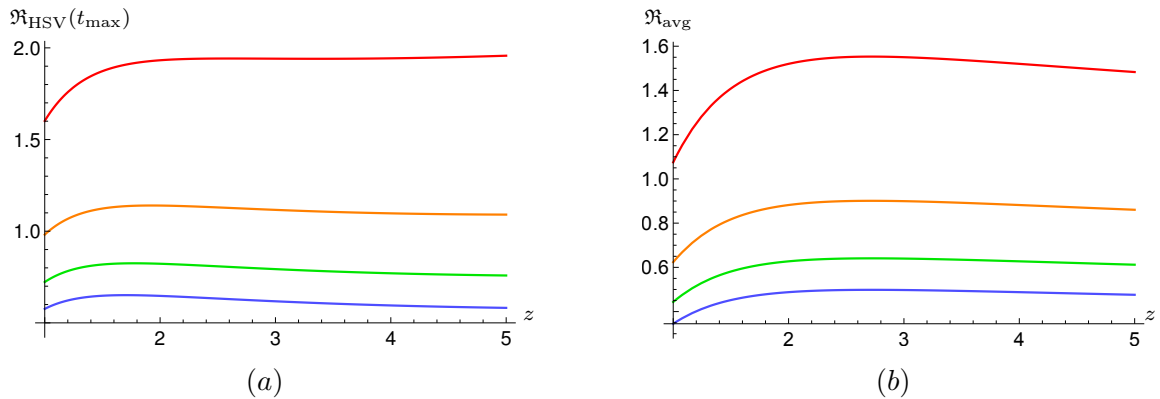
\begin{figure}[!h]
$$
\begin{array}{cc}
  \includegraphics[angle=0,width=0.43\textwidth]{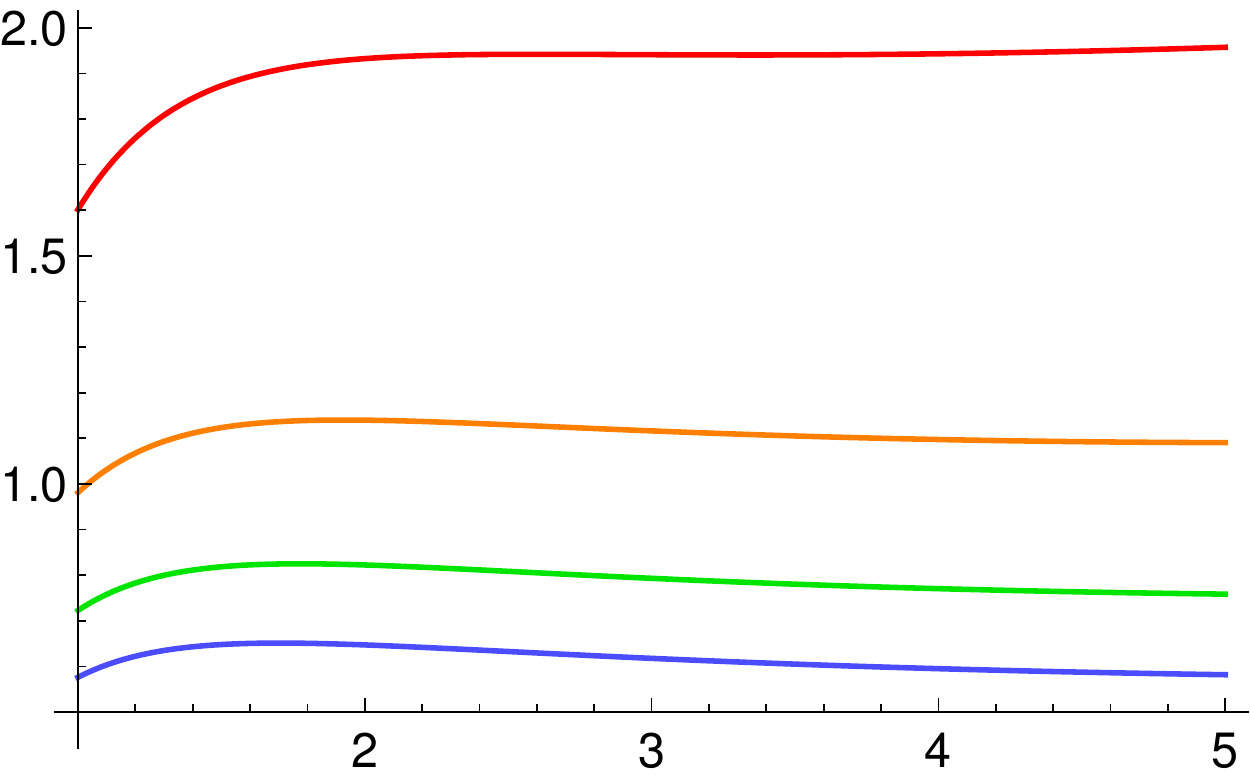} \qquad\qquad & \includegraphics[angle=0,width=0.43\textwidth]{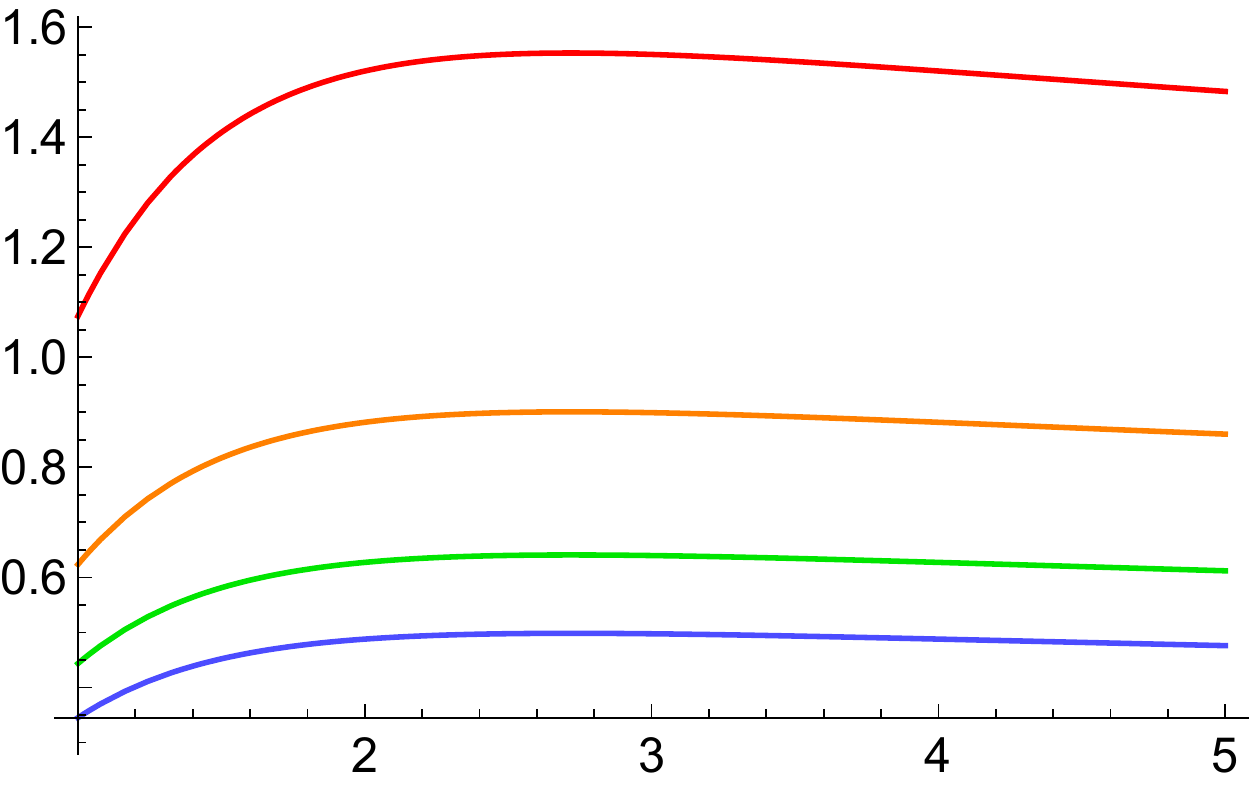}\\
  (a) \qquad\qquad & (b)
\end{array}
$$
\begin{picture}(0,0)
\put(6,155){{\scriptsize $\mathfrak{R}_{\text{HSV}}(t_{\text{max}})$}}
\put(246,155){{\scriptsize $\mathfrak{R}_{\text{avg}}$}}
\put(200,43){{\scriptsize $z$}}
\put(435,43){{\scriptsize $z$}}
\end{picture}
\vspace{-0.7cm}
\caption{\small (a) maximum and (b) average entanglement velocities as a function of $z$. The colors blue, green, orange and red correspond to $d=\{5,4,3,2\}$ respectively and we have set $\theta=0.1, t_*=1$.}
\label{fig:VelsVSz}
\end{figure}

\noindent It is also interesting to consider the maximum entanglement velocity as a function of $z$. This problem was studied numerically in \cite{He:2017wla,MohammadiMozaffar:2017nri} recently. They found that the entanglement entropy for small subsystems is a linear function of $z$. We observe the same linear behavior, above $z=1$, as we show in part (a) of Figure \ref{fig:VelsVSz}.
Now we turn to the average entanglement velocity. It is given by   
\begin{equation}
\mathfrak{R}_{\text{avg}} \equiv \frac{1}{t_*} \, \int_0^{t_*} \, \mathfrak{R}_{\text{HSV}}(t) \, dt  .
\end{equation}
Using equation \eqref{eq:EE_velocity_HSV}, this can be shown to be
\begin{equation}
 \label{eq:avg_ent_vel}
\mathfrak{R}_{\text{avg}} = \frac{\sqrt{\pi} \, \Gamma \big[ \frac{d-\theta}{2 d_\theta} \big] \, (z \, t_*)^{\frac{1}{z}} \, }{ t_* \,  \Gamma \big[ \frac{1}{2 d_\theta} \big] } .
\end{equation}
As expected, it reduces to the known expression \cite{Kundu:2016cgh, Lokhande:2017jik} in the limit of AdS-Vaidya
\begin{equation}
\mathfrak{R}_{\text{CFT}}^{\text{avg}} = \frac{\sqrt{\pi} \, \Gamma \big[ \frac{d}{2(d-1)} \big] \, }{\Gamma \big[ \frac{1}{2(d-1)} \big] } .
\end{equation}
In part (b) of Figure \ref{fig:VelsVSd} we show the average velocity as a function of  the dimension. And we display the average velocity as a function of $z$ in part (b) of Figure \ref{fig:VelsVSz}.  
We observe that both the maximum as well as the average entanglement velocities violate the bound. However, the bound was derived from relativistic considerations. The state that we work with are actually excited states that break conformal invariance. As a result, like a wave traveling in a matter that is refracting with respect to vacuum, the velocities are greater than 1.

\subsubsection*{3. Near-Saturation Regime}
We now study the last sub-regime of the growth of entanglement after the instantaneous quench. This is defined to be when the time $t$ is close to the saturation time, $t \approx t_*$. Recall that $t_q=0$ for instantaneous quenches and hence $t_{\text{sat}}=t_*$. Thus, we would like to expand $\mathcal{F}(x)$ as defined in \eqref{eq:Fx-def} around $x=1$, approaching from below.  The expansion is given by
\begin{align}
\begin{split}
\label{eq:near_sat}
\mathcal{F}(x) = \begin{cases}
1-\frac{2^{\frac{5}{2}} \, d_\theta^{\frac{3}{2}} \, \Gamma \big[\frac{3 d_\theta + z+1}{2 d_theta} \big] \,  }{3 \, \Gamma \big[ \frac{3}{2} \big] \, \Gamma \big[ \frac{z+1}{2 d_\theta} \big] \, }  \,  (1-x)^{\frac{3}{2}}  ,  \quad x^{2 d}<x^{2(1+\theta)} ,  \\
1+\frac{2^{\frac{5}{2}} \, d_\theta^{\frac{3}{2}} \, \Gamma \big[\frac{3 d_\theta + z+1}{2 d_theta} \big] \, }{3 \, \Gamma \big[ \frac{3}{2} \big] \, \Gamma \big[ \frac{z+1}{2 d_\theta} \big] \, }   \,  (1-x)^{\frac{3}{2}} ,  \quad x^{2 d}>x^{2(1+\theta)}  .
\end{cases}
\end{split}
\end{align}
This is consistent with the known results in the AdS-Vaidya limit \cite{Kundu:2016cgh, Lokhande:2017jik}.

\subsection{Power Law Quench}
\label{power-law}

\begin{figure}[H]
\begin{tabular}{ccc}
  Case I: $t_q < t_\ast$   &{}\qquad\qquad&
  Case II: $t_\ast < t_q$ \vspace{1mm}\\
    \includegraphics[width=.42\textwidth]
      {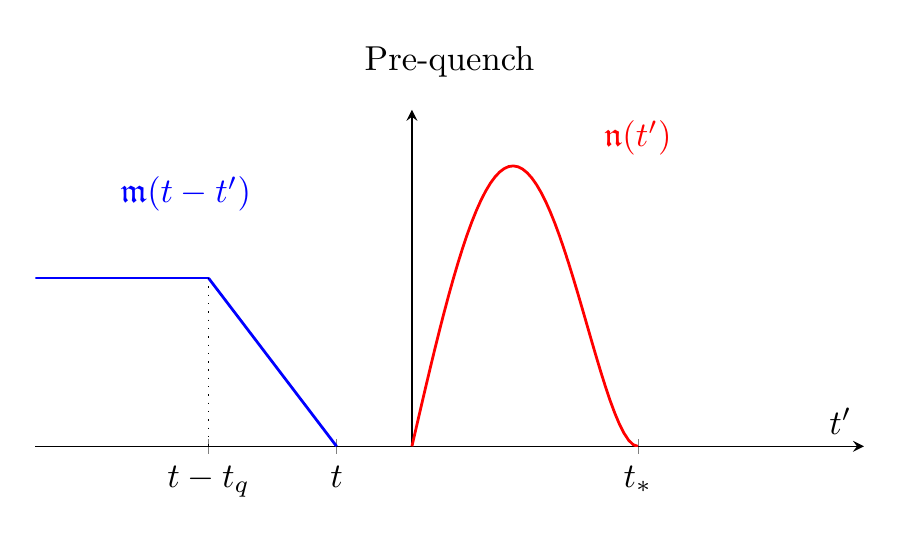} &&
    \includegraphics[width=.42\textwidth]
      {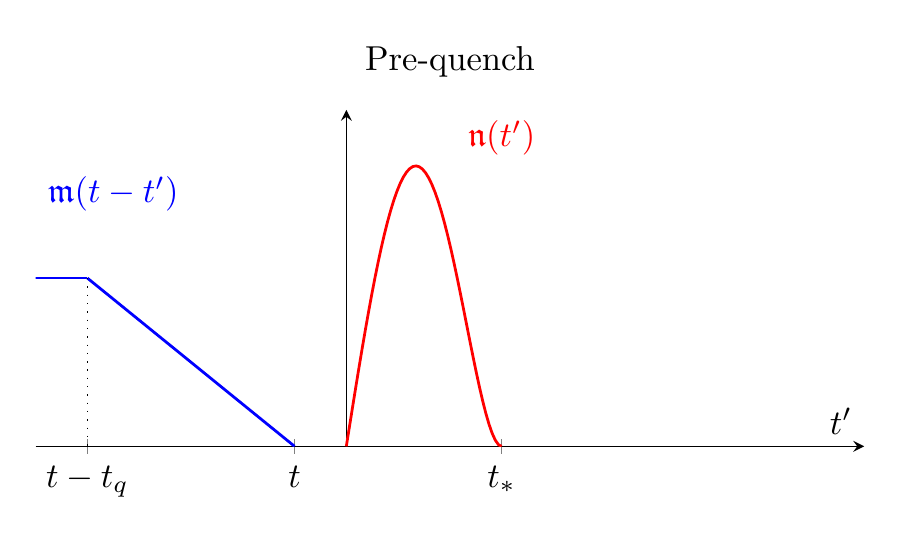} \vspace{-3mm}\\
    \includegraphics[width=.42\textwidth]
      {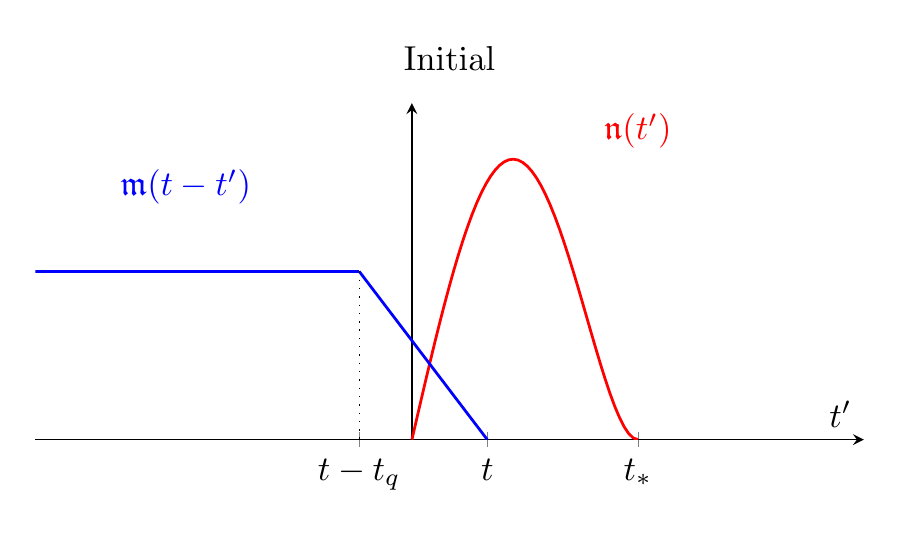} &&
    \includegraphics[width=.42\textwidth]
      {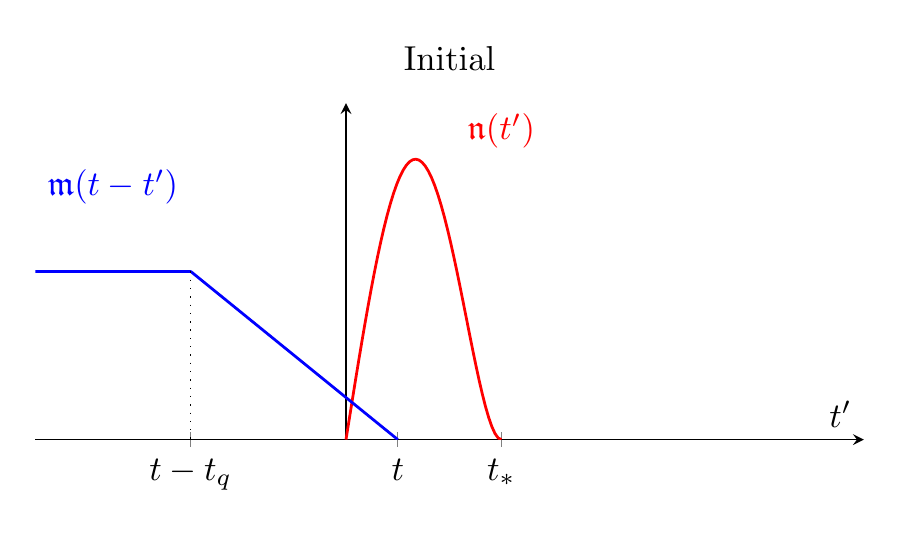} \vspace{-3mm}\\
    \includegraphics[width=.42\textwidth]
      {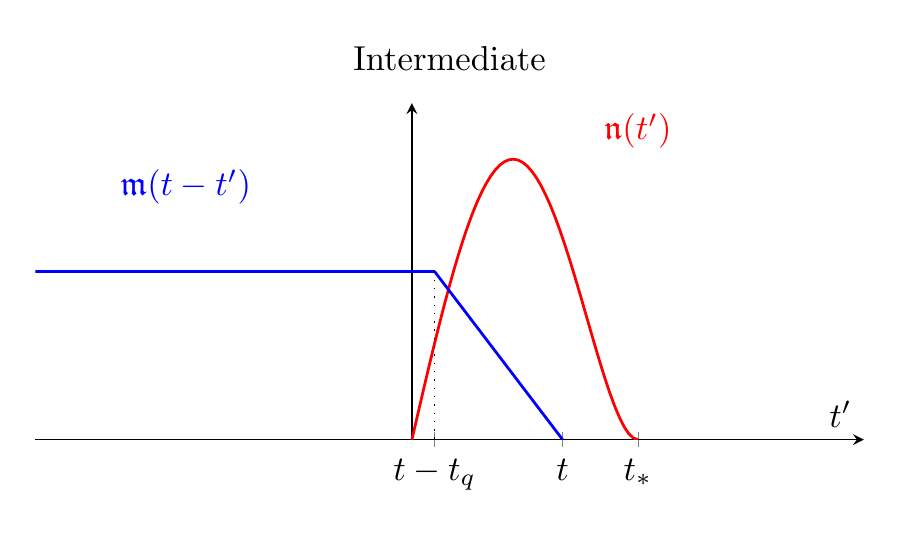} &&
    \includegraphics[width=.42\textwidth]
      {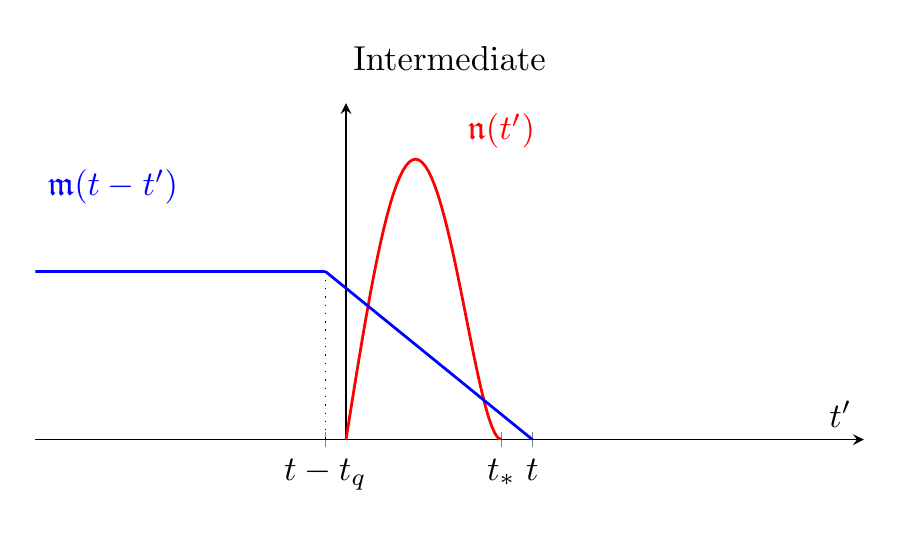}  \vspace{-3mm}\\
    \includegraphics[width=.42\textwidth]
      {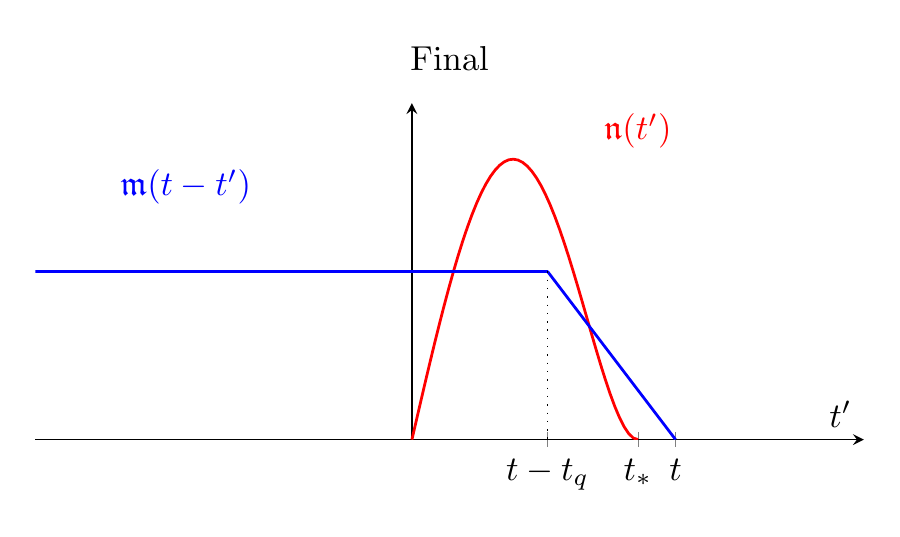} &&
    \includegraphics[width=.42\textwidth]
      {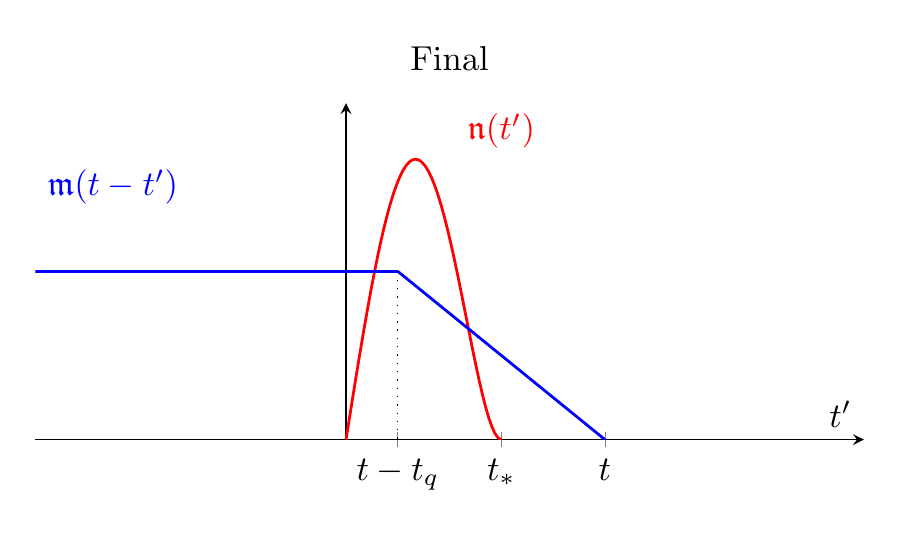}  \vspace{-3mm}\\
    \includegraphics[width=.42\textwidth]
      {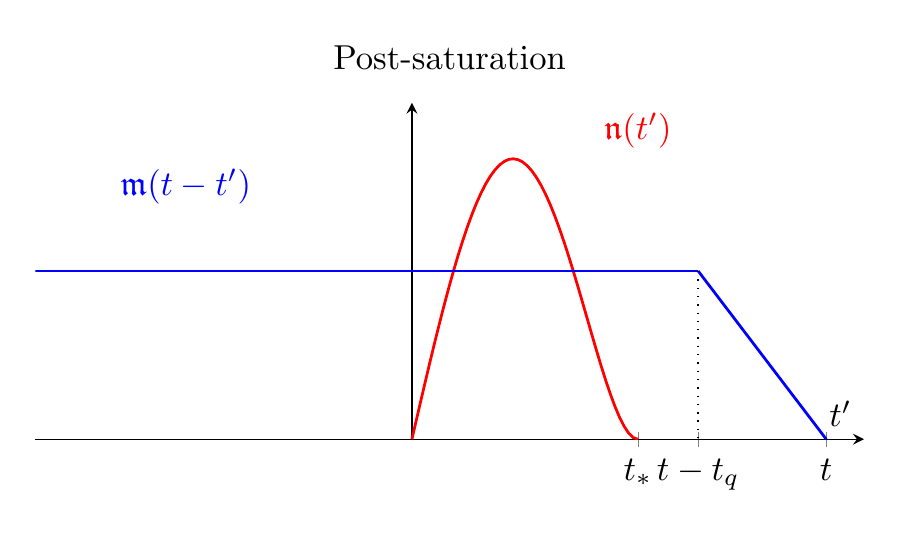} &&
    \includegraphics[width=.42\textwidth]
      {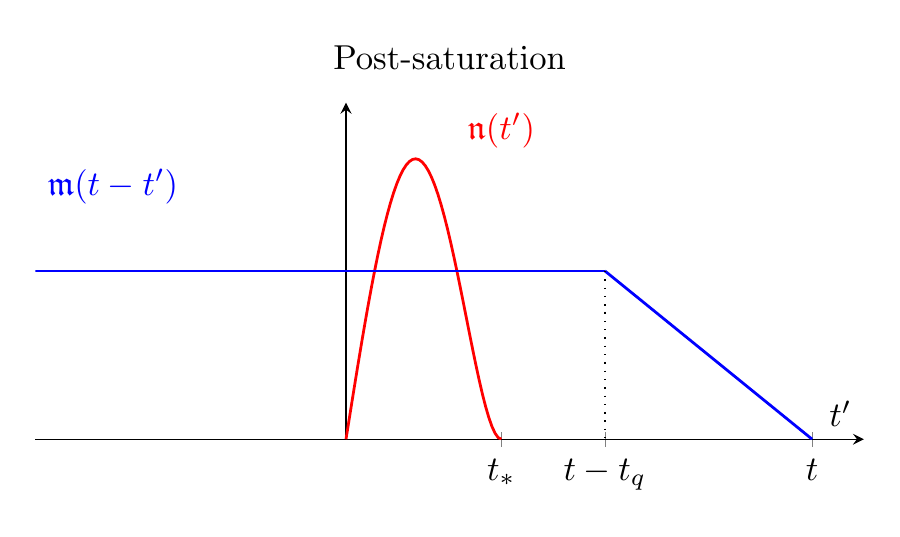}
\end{tabular}
\vspace{-1mm}
\caption{Schematic representation of the convolution integral for a power-law quench with $p=1$.
The right and left columns show the two possible cases I: $t_q < t_\ast$ and II: $t_\ast < t_q$, respectively.
In the pre-quench  and post-saturation regimes the integral is a constant. In the initial, intermediate and final growth regimes the integral is time-dependent and can be performed by splitting it in various intervals, as shown in equations \eqref{eq:HSV_S1_power} and \eqref{eq:HSV_S2_power}.
\label{fig:SchemConv}}
\end{figure}

We now consider global quench that is a power law with respect to time. We will denote the power by $p$ and keep it arbitrary. Once we have obtained the evolution of entanglement entropy in this general case, we could study the evolution of entanglement entropy for any \textit{smooth} quench, by using for example Remez algorithm to rewrite the quench in a basis of polynomials. The perturbation we will consider is
\begin{equation}
g(t) = \sigma \, t^p \, \big[\Theta(t)-\Theta(t-t_q) \big] + \epsilon_0  \, \Theta(t-t_q) \, ,
\end{equation}
where $p \in \mathbb{Z}$ and $\epsilon_0=\sigma \, t_q^p$ is the final energy density when the quench stops at $t=t_q$. This perturbation defines the source function $\mathfrak{m}(t)$ in the convolution equation \eqref{eq:EE_lin_resp}. Using the kernel \eqref{eq:response_func}, we get the entanglement growth to be given by the integral
\begin{align}
 \begin{split}
\delta S_A(t) &= \frac{\sigma \, \mathcal{A}_{\Sigma} \, z^{\frac{1}{z}} \, }{8 \, G_N \, u_H^{d_\theta+z}} \, \int\limits_{-\infty}^{\infty} \, d\tau \, \tau^{\frac{1}{z}} \,  \sqrt{1- \big[\frac{\tau}{t_*}\big]^{\frac{2 d_\theta}{z}} \, } \, \big[\Theta(\tau)-\Theta(\tau-t_*) \big] \, \times  \\
&\qquad \bigg[  (t-\tau)^p \,  \big[\Theta(t-\tau) - \Theta(t-t_q-\tau) \big] + t_q^p \, \Theta(t-t_q-\tau)  \bigg] .
 \end{split}
\end{align}
We naturally encounter two cases: 1) $t_q < t_*$ , and 2) $t_q>t_*$ . In both these cases, the evolution of entanglement entropy is very similar, namely the system is driven by the quench upto some time resulting in entanglement growth; followed by a transient regime which finally leads to saturation to an equilibrium value. In both the scenarios, the saturation time is given by $t_{\text{sat}}=t_q+t_*$ and the evolution can be split and analyzed in various intervals as follows:
\begin{center}
 \begin{tabular}{ |c|c|c|c|c|c|}
 \hline
 Regime: &  Pre-quench & Initial & Intermediate & Final & Post-saturation \\
 \hline
 \hline
 Case I: $t_q<t_*$ & $t<0$ & $0<t<t_q$&$t_q<t<t_*$&$t_*<t<t_{\text{sat}}$ &$t>t_{\text{sat}}$  \\
\hline
 Case II: $ t_* < t_q$& $t<0$  & $0<t<t_*$&$t_*<t<t_q$&$t_q<t<t_{\text{sat}}$ & $t>t_{\text{sat}}$ \\
\hline
\end{tabular}
\end{center}
\vspace{0.4cm}
\noindent In Figure \ref{fig:SchemConv}, we show schematically\footnote{This figure is a modification of a code written by Gerben W. J. Oling. I am thankful to them.} the use of convolution integrals \eqref{eq:EE_lin_resp} to calculate the entanglement entropy in each of these regimes.

\noindent The pre-quench and post-saturation regimes are in equilibrium. In particular
\begin{align}
 \delta S_A(t) &= 0 \qquad \qquad  \text{for} \qquad t<0 \, , \\
 \delta S_A(t) &= \frac{\epsilon_0 \, V_A}{T_A} \qquad \text{for} \qquad t>t_{\text{sat}} \, .
\end{align}

\noindent as expected, with $T_A$ being given by equation \eqref{eq:const_ent_Temp}. Due to equilibrium, the time-dependent relative entropy vanishes in both cases, $\delta S_{\text{rel}}(t)=0$. The initial, intermediate and final regimes are generally time-dependent. \\

\noindent We now derive analytic expressions for entanglement growth. To write them succinctly, we define the following indefinite integral that depends on the subsystem:
\begin{align} \label{eq:p-indef-int1}
 \begin{split}
\mathcal{I}^{(p)}(t,\tau) &\equiv  \frac{ \mathcal{A}_{\Sigma} \,  }{8 \, G_N \, u_H^{d_\theta+z}} \, \int d\tau \, (t-\tau)^p \,   (z \, \tau)^{\frac{1}{z}} \,  \sqrt{1- \big[\frac{\tau}{t_*}\big]^{\frac{2 d_\theta}{z}} \, } \,  ,
 \end{split}
\end{align}
The entanglement growth in different regimes will be given by this integral evaluated at appropriate limits. 

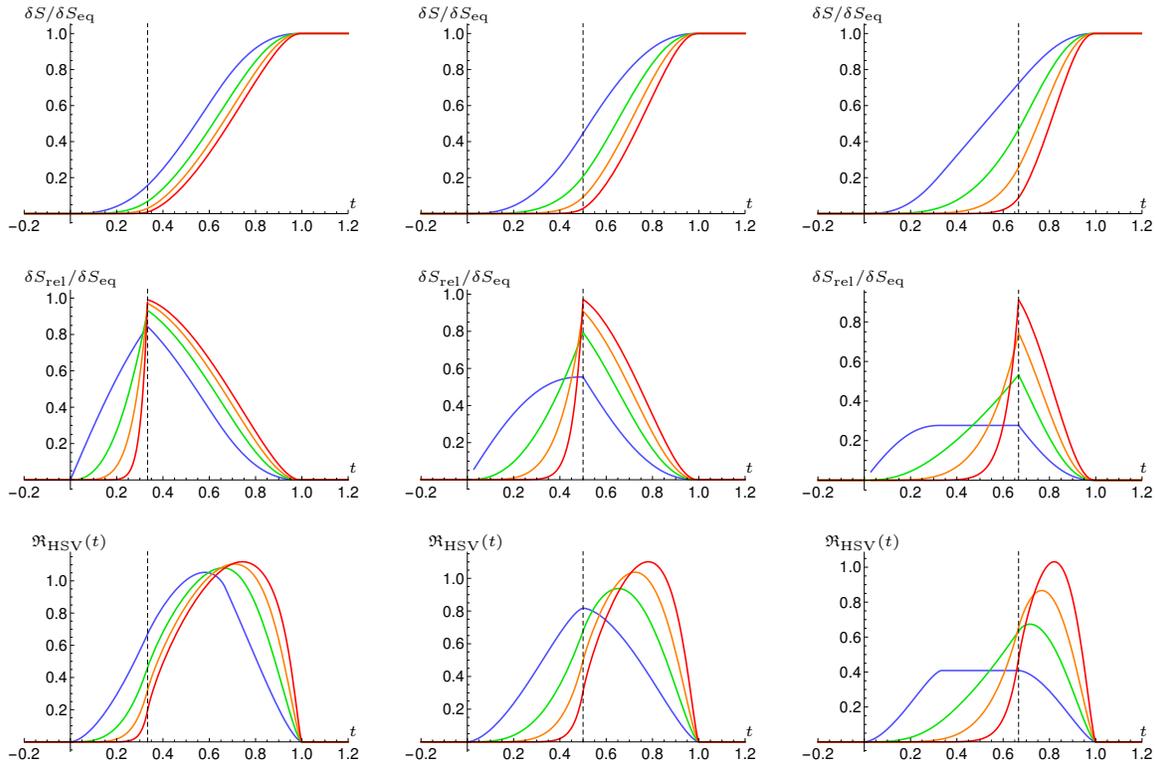
\begin{figure}[!htb]
\begin{tabular}{ccc}
    \includegraphics[width=0.3\textwidth]{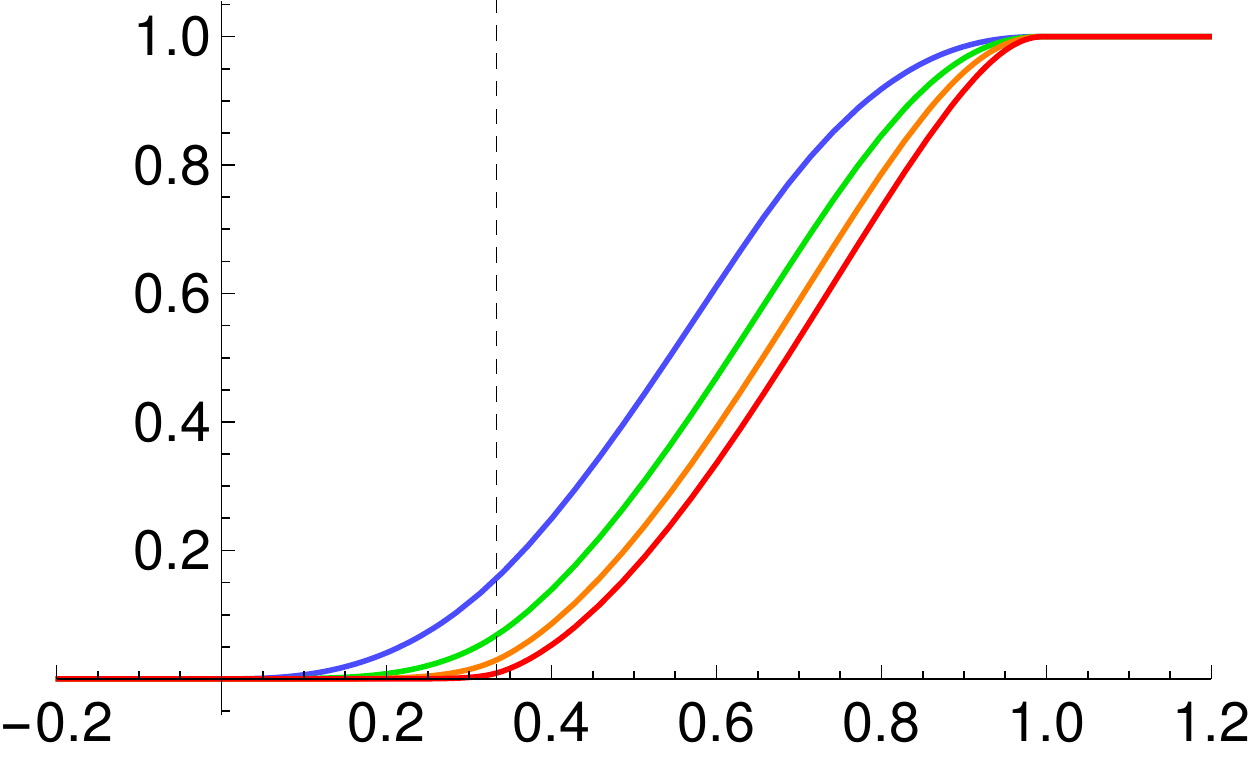} &$\,$ 
    \includegraphics[width=0.3\textwidth]{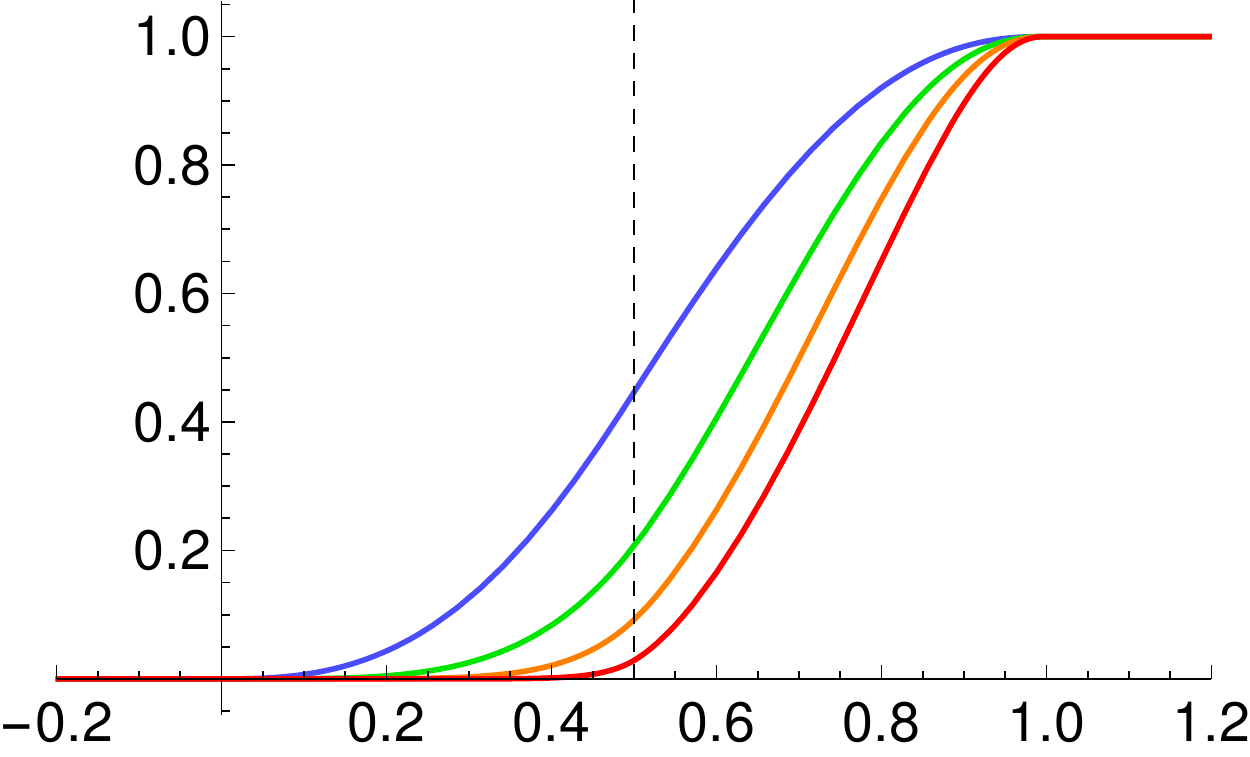} &$\,$ 
    \includegraphics[width=0.3\textwidth]{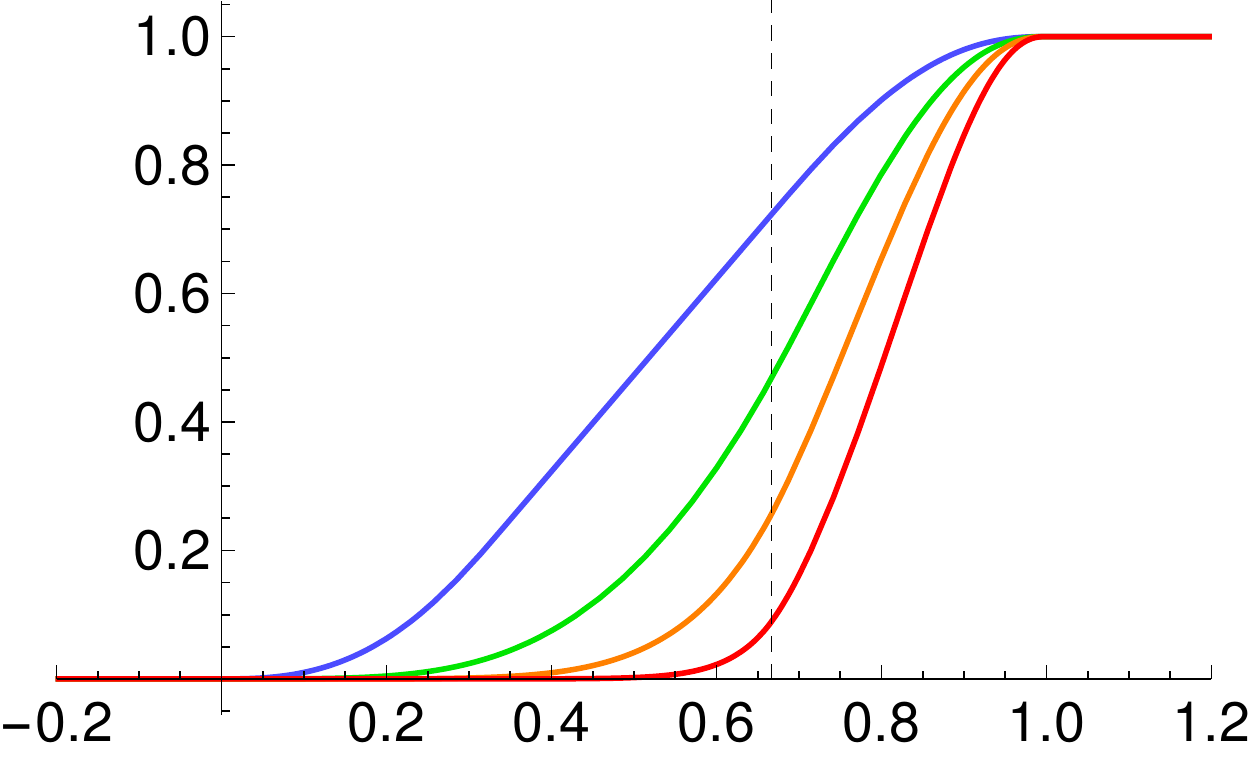} \vspace{5mm} \\
    \includegraphics[width=0.3\textwidth]{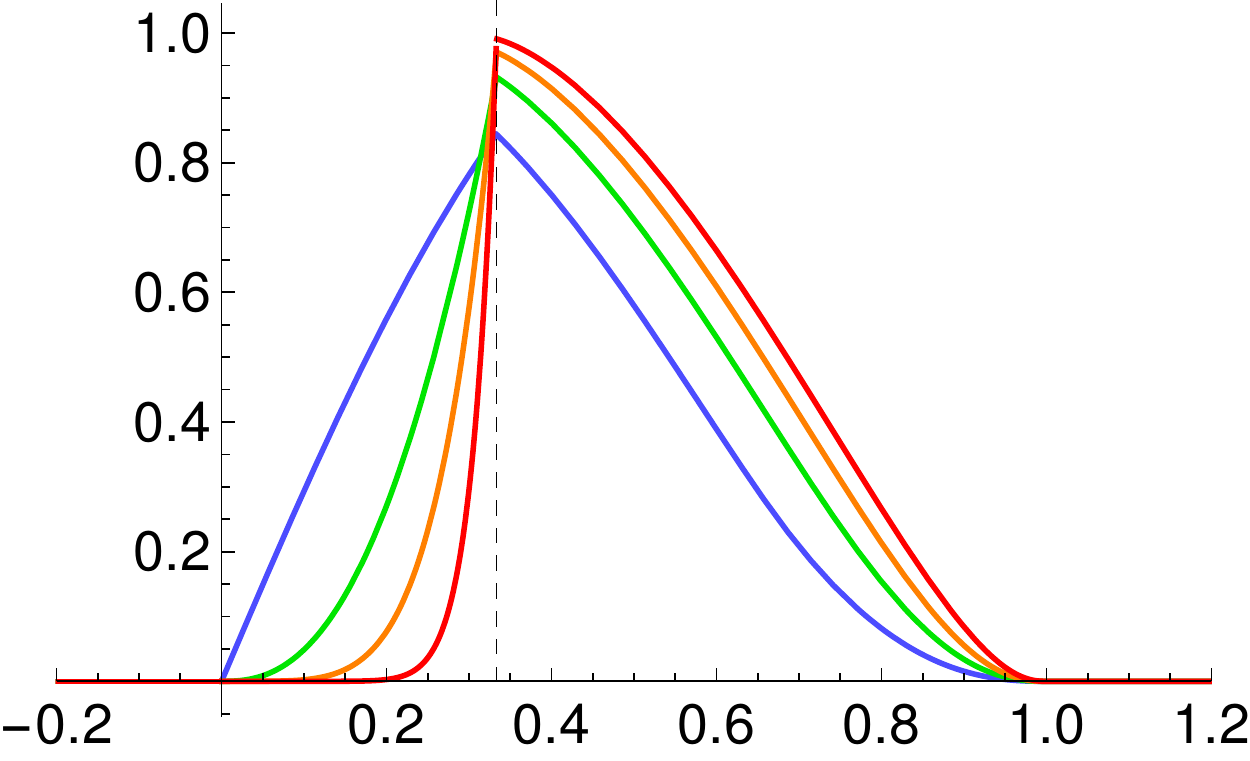} &$\,$ \includegraphics[width=0.3\textwidth]{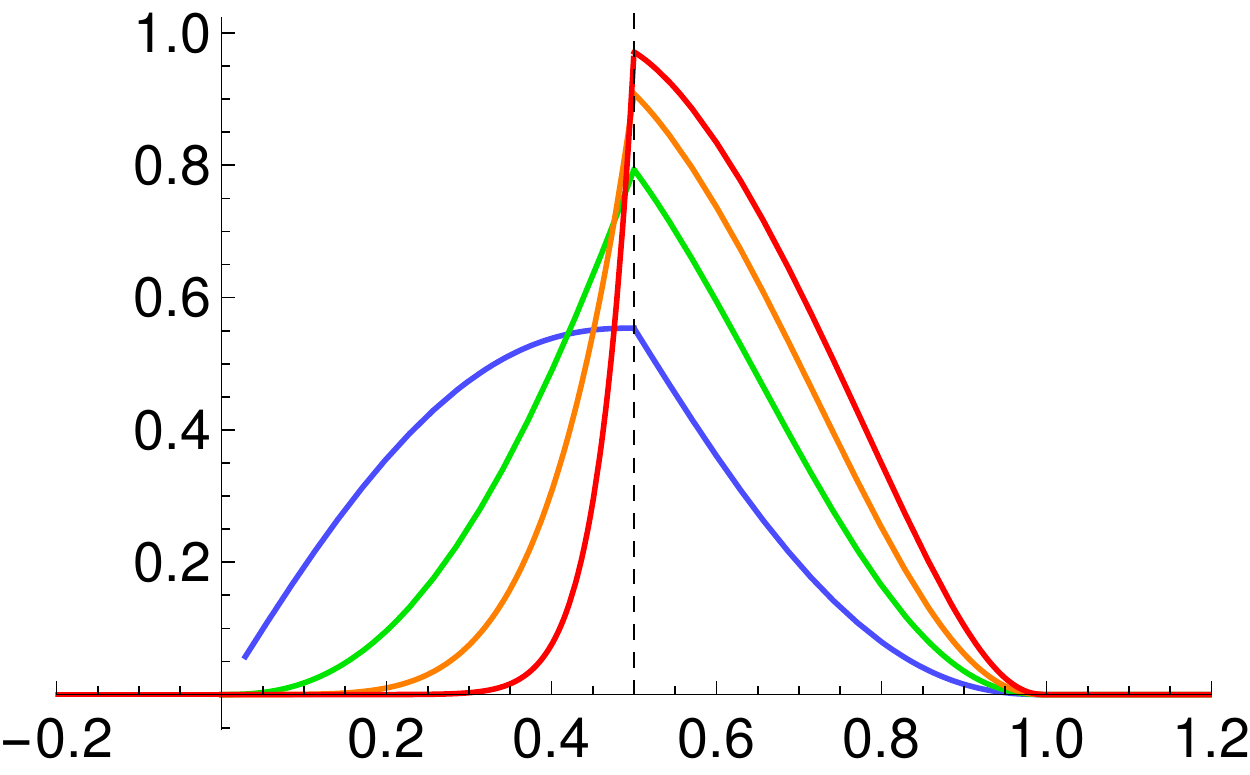} &$\,$ \includegraphics[width=0.3\textwidth]{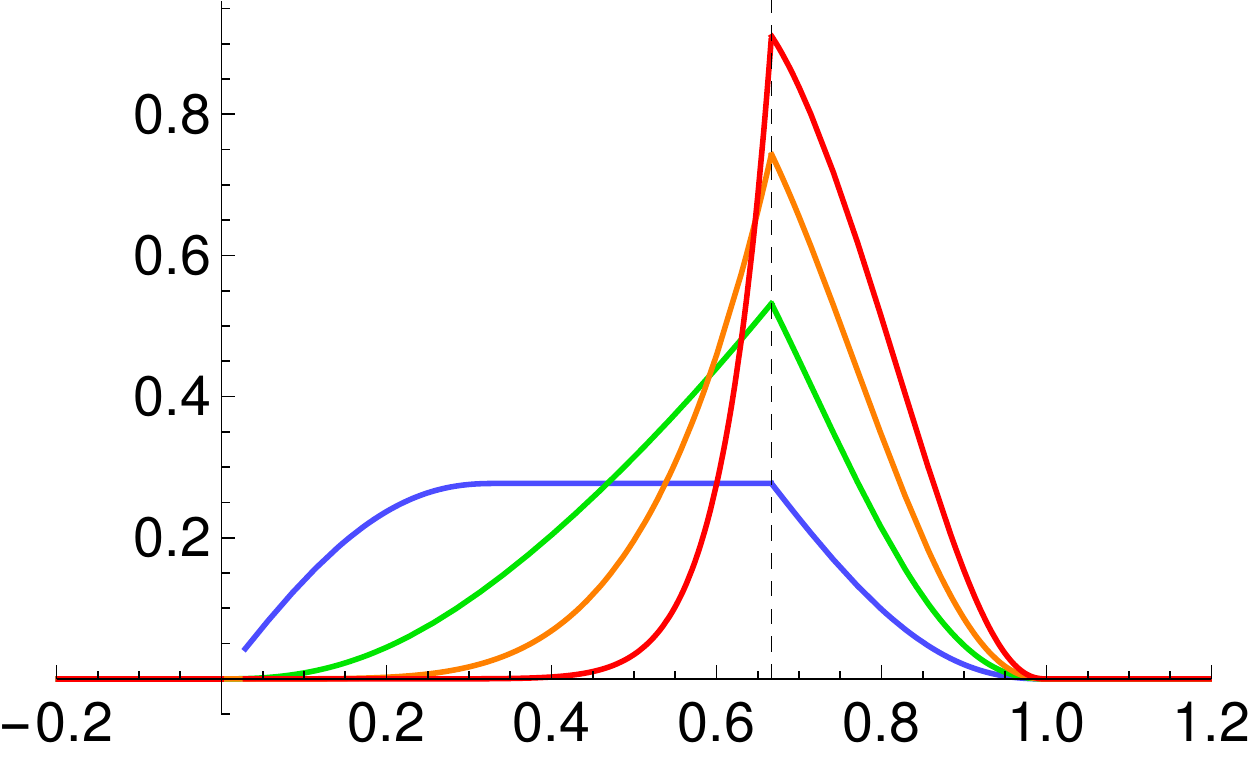} \vspace{5mm}\\
    \includegraphics[width=0.3\textwidth]{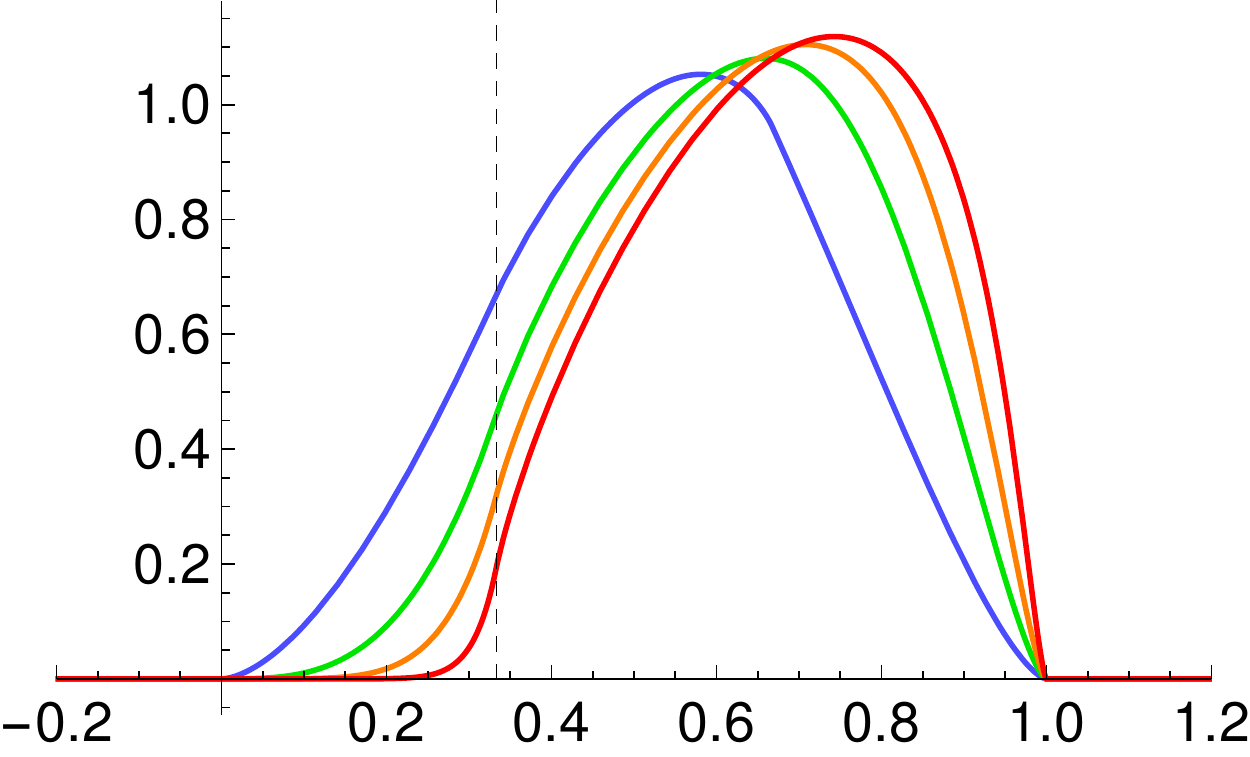} &$\,$ 
    \includegraphics[width=0.3\textwidth]{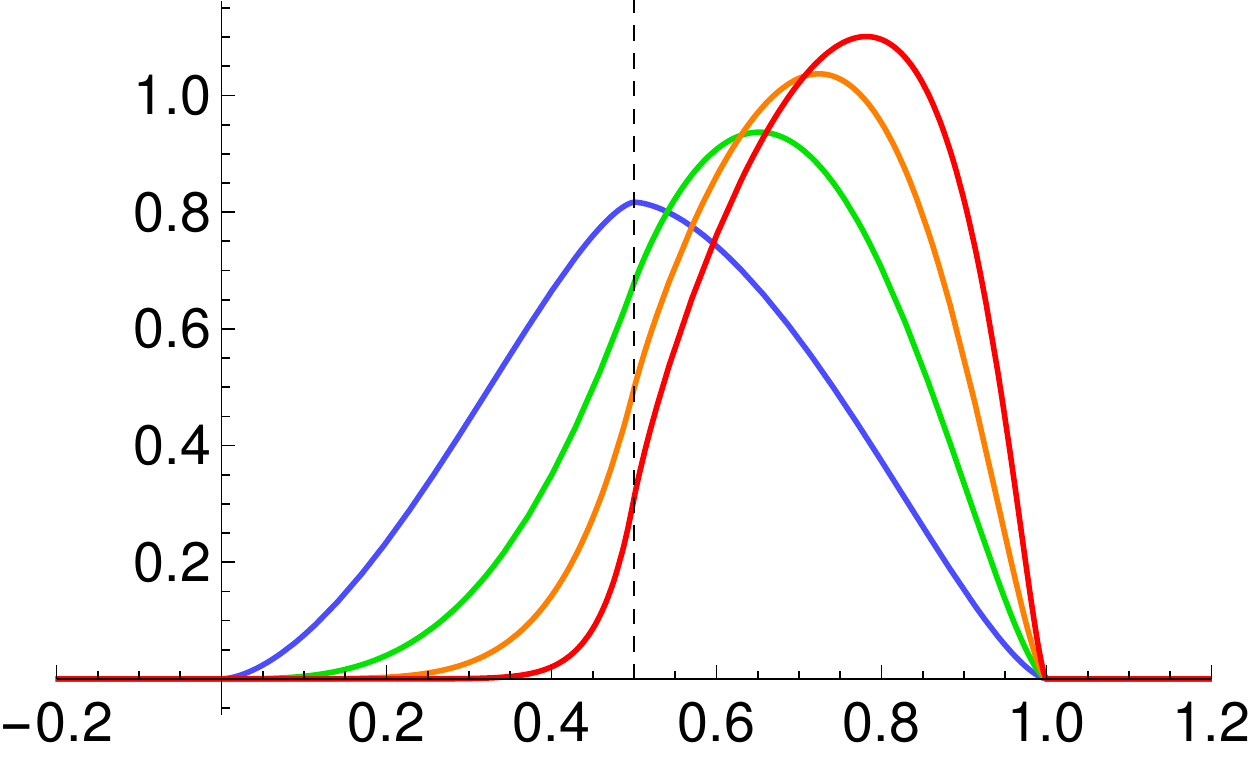} &$\,$ 
    \includegraphics[width=0.3\textwidth]{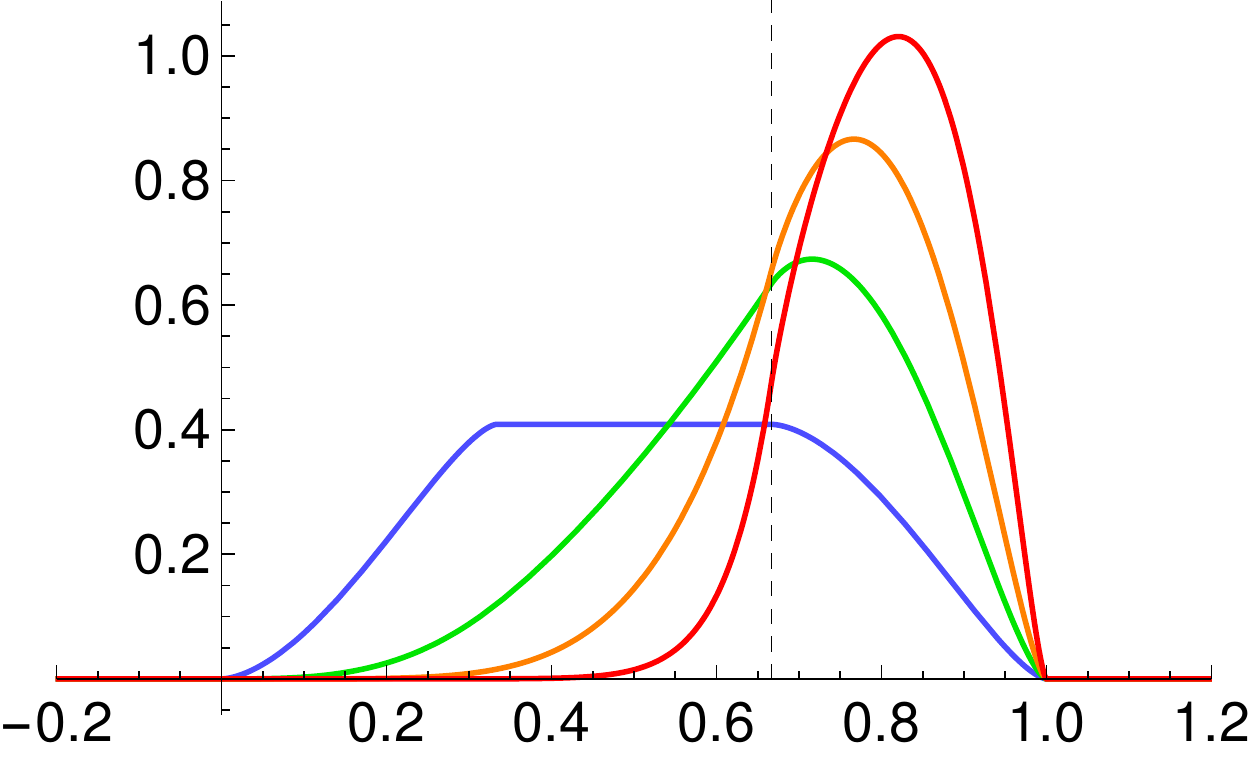} \vspace{5mm}
\end{tabular}
\begin{picture}(0,0)(-460,-145)
\put(-448,157){{\tiny $\delta S/\delta S_{\text{eq}}$}}
\put(-299,157){{\tiny $\delta S/\delta S_{\text{eq}}$}}
\put(-150,157){{\tiny $\delta S/\delta S_{\text{eq}}$}}
\put(-325,85){{\tiny $t$}}
\put(-175,85){{\tiny $t$}}
\put(-26,85){{\tiny $t$}}
\put(-448,57){{\tiny $\delta S_{\text{rel}}/\delta S_{\text{eq}}$}}
\put(-299,57){{\tiny $\delta S_{\text{rel}}/\delta S_{\text{eq}}$}}
\put(-150,57){{\tiny $\delta S_{\text{rel}}/\delta S_{\text{eq}}$}}
\put(-325,-15){{\tiny $t$}}
\put(-175,-15){{\tiny $t$}}
\put(-26,-15){{\tiny $t$}}
\put(-445,-43){{\tiny $\mathfrak{R}_{\text{HSV}}(t)$}}
\put(-295,-43){{\tiny $\mathfrak{R}_{\text{HSV}}(t)$}}
\put(-145,-43){{\tiny $\mathfrak{R}_{\text{HSV}}(t)$}}
\put(-325,-115){{\tiny $t$}}
\put(-175,-115){{\tiny $t$}}
\put(-26,-115){{\tiny $t$}}
\end{picture}
\vspace{-0.6cm}
\caption{\small{Entanglement entropy, relative entropy and entanglement velocity for different powers after a power-law quench. Colors blue, green, orange and red represent $p=\{1, 2.5, 5, 11.5\}$ respectively. Plots from left to right have $t_q/t_*=\{0.5,1,2\}$. In all plots, the dashed vertical line denotes the end of the driven phase $t=t_q$. We have set $d=3, \theta=0.1, z=1.5$.}
\label{fig:QuantitiesvsP}}
\end{figure}

\noindent Now, using the binomial series
\begin{align}
 \begin{split}
(t-\tau)^p &= \sum\limits_{k=0}^\infty \begin{pmatrix} 
                                       p \\
                                       k
                                      \end{pmatrix} \, t^{p-k} \, (-\tau)^k ,\\ 
 \end{split}
\end{align}
where $p$ is a real number, we can do the integral explicitly to obtain
\begin{align}\label{eq:p-indef-int2}
 \begin{split}
  \mathcal{I}^{(p)}(t,\tau) &=  \frac{  \mathcal{A}_{\Sigma} \, }{8 \, G_N \,  \,  u_H^{d_\theta+z}} \, \sum\limits_{k=0}^\infty \begin{pmatrix} 
                                       p \\
                                       k
                                      \end{pmatrix} \, t^{p-k} \,  \,\frac{ (z \, \tau)^{1+\frac{1}{z}} \, (-\tau)^{k}}{(z+1+kz)} \,   \\
&\qquad \times {}_2F_{1}\bigg[-\frac{1}{2}, \frac{z+1+k z}{2 d_\theta},  \frac{2 d_\theta + z+1+k z}{2 d_\theta}; \big[\frac{\tau}{t_*} \big]^{\frac{2 d_\theta}{z}} \bigg]  .
 \end{split}
\end{align}
\noindent This can be evaluated for any given $p$. For $p \in \mathbb{R}$, we get an infinite series but when $p$ is a non-negative integer, the series is finite and gives a closed-form expression. In terms of this integral, the entanglement growth for different regimes can be written as follows
\begin{equation}\label{eq:HSV_S1_power}
  \delta S_{A}^{\text{(I)}}(t)=
 \begin{cases}
  \displaystyle 0\,, & \quad t<0 \ ,\\[0.0ex]
  \displaystyle \sigma \, \mathcal{I}^{(p)}(t,\tau)|_0^t\,, & \quad 0<t<t_q \ ,\\[0.0ex]
  \displaystyle \epsilon_0 \, \mathcal{I}^{(0)}(t,\tau)|_0^{t-t_q}+ \sigma \, \mathcal{I}^{(p)}(t,t')|_{t-t_q}^t\,,   & \quad t_q<t<t_*\ ,\\[0.0ex]
  \displaystyle \epsilon_0 \, \mathcal{I}^{(0)}(t,\tau)|_0^{t-t_q}+\sigma \,  \mathcal{I}^{(p)}(t,t')|_{t-t_q}^{t_*}\,, & \quad t_*<t<t_{\text{sat}}\ ,\\[0.0ex]
  \displaystyle \epsilon_0 \,  \mathcal{I}^{(0)}(t,\tau)|_0^{t_*}\,, & \quad t>t_{\text{sat}}\ ,
 \end{cases}
\end{equation}
for case $\text{I}$ and
\begin{equation}\label{eq:HSV_S2_power}
  \delta S_{A}^{\text{(II)}}(t)=
 \begin{cases}
  \displaystyle 0\,, & \quad t<0 \ ,\\[0.0ex]
  \displaystyle \sigma\hspace{0.1em}\mathcal{I}^{(p)}(t,\tau)|_0^t\,, & \quad 0<t<t_* \ ,\\[0.0ex]
  \displaystyle \sigma\hspace{0.1em}\mathcal{I}^{(p)}(t,\tau)|_0^{t_*}\,,   & \quad t_*<t<t_q\ ,\\[0.0ex]
  \displaystyle \epsilon_0\hspace{0.1em}\mathcal{I}^{(0)}(t,\tau)|_0^{t-t_q}+\sigma\hspace{0.1em}\mathcal{I}^{(p)}(t,\tau)|_{t-t_q}^{t_*}\,, & \quad t_q<t<t_{\text{sat}}\ ,\\[0.0ex]
  \displaystyle  \epsilon_0\hspace{0.1em}\mathcal{I}^{(0)}(t,\tau)|_0^{t_*}\,, & \quad t>t_{\text{sat}}\ ,
 \end{cases}
\end{equation}
for case $\text{II}$ respectively, where all the evaluations are for the integration variable $\tau$. Note that $\mathcal{I}^{(0)}(t,\tau)$ is particularly simple and has the closed-form expression
\begin{align}
 \begin{split}
 \mathcal{I}^{(0)}(t,\tau) &= \frac{ \mathcal{A}_{\Sigma} \, (z \, \tau)^{\frac{1}{z}} \, }{8 \, G_N \, u_H^{d_\theta+z} \, (z+1)} \, \,  {}_2F_{1}\bigg[-\frac{1}{2}, \frac{z+1}{2 d_\theta}, \frac{2 d_\theta+z+1}{2 d_\theta};\big[\frac{\tau}{t_*} \big]^{\frac{2 d_\theta}{z}}  \bigg] .
 \end{split}
\end{align}
These expressions can be easily understood graphically. In the first row of Figure \ref{fig:QuantitiesvsP}, we plot the time-dependent entanglement entropy after a power-law quench for different powers. Different regimes of entanglement growth can be identified from these plots. Nonetheless, they can be made manifest by plotting the relative entropy as a function of time, as shown in the second row of Figure \ref{fig:QuantitiesvsP}. The figure shows a conspicuous cusp, which denotes the end of the driven regime. However, there is no singular behavior at the location of the cusp, as can be verified by studying the instantaneous rate of entanglement growth. We show this in the third row of Figure \ref{fig:QuantitiesvsP}. As in the case of the instantaneous quench, we observe that the instantaneous entanglement velocity need not be bounded by 1. \\

\begin{figure}[!htb]
\begin{tabular}{ccc}
    \includegraphics[width=0.3\textwidth]{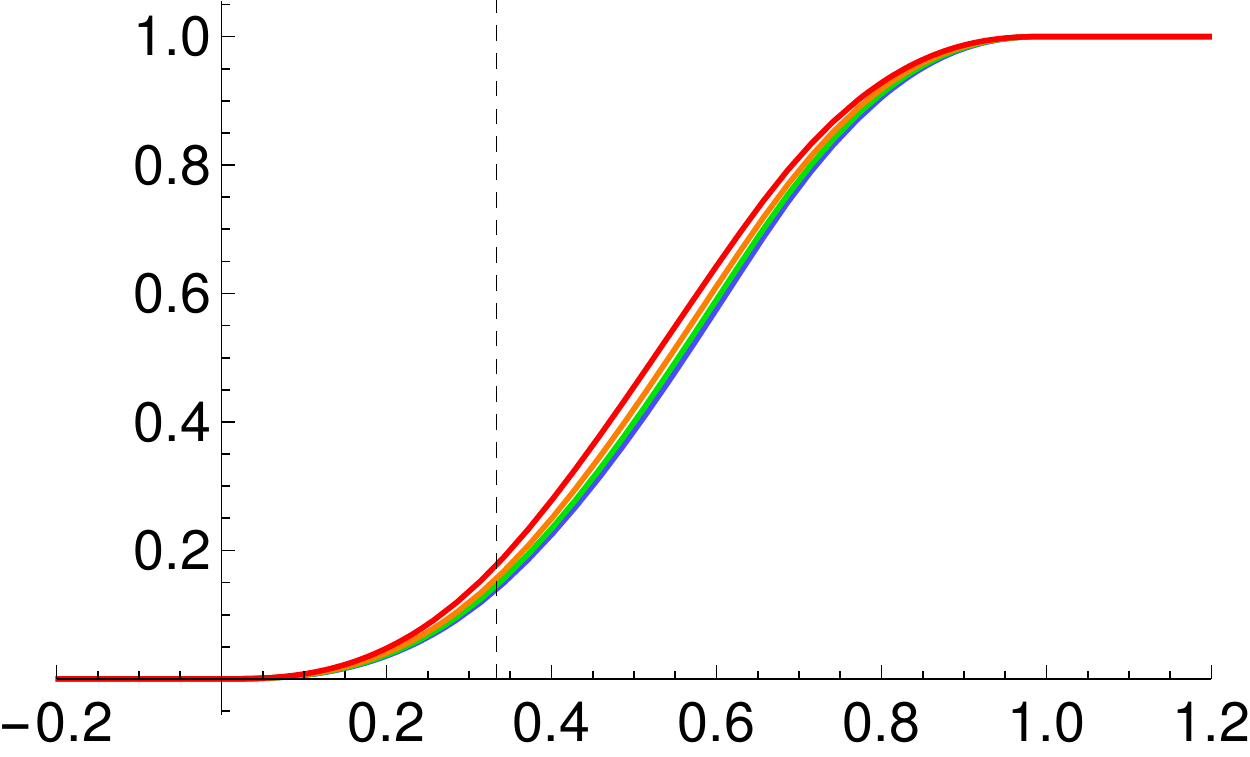} &$\,$ 
     \includegraphics[width=0.3\textwidth]{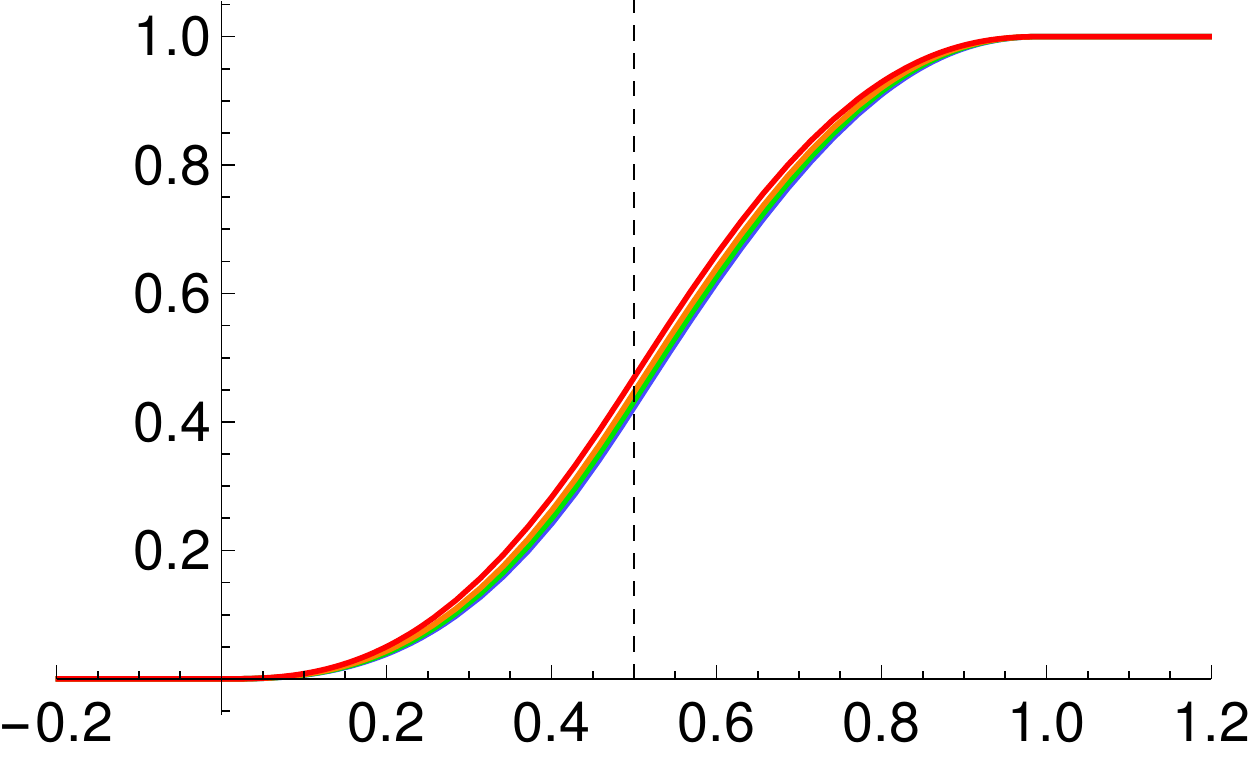} &$\,$ 
     \includegraphics[width=0.3\textwidth]{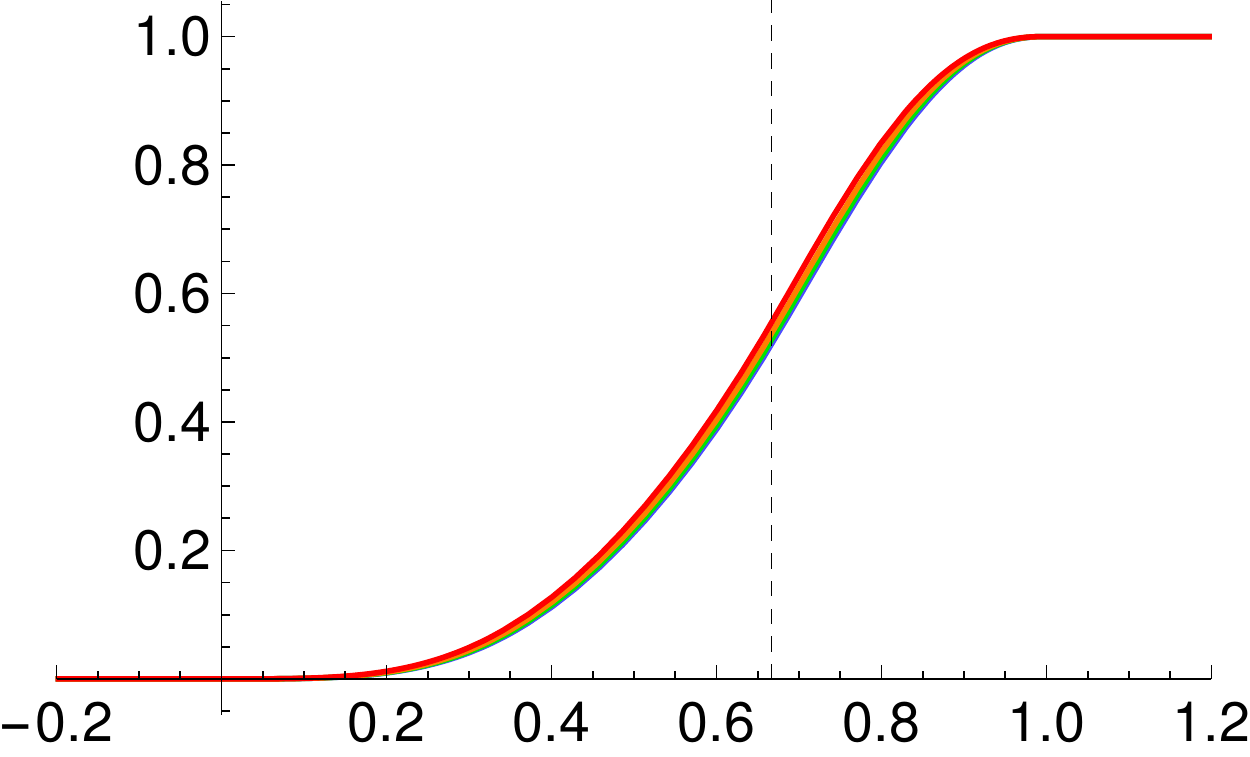} \vspace{5mm} \\
     \includegraphics[width=0.3\textwidth]{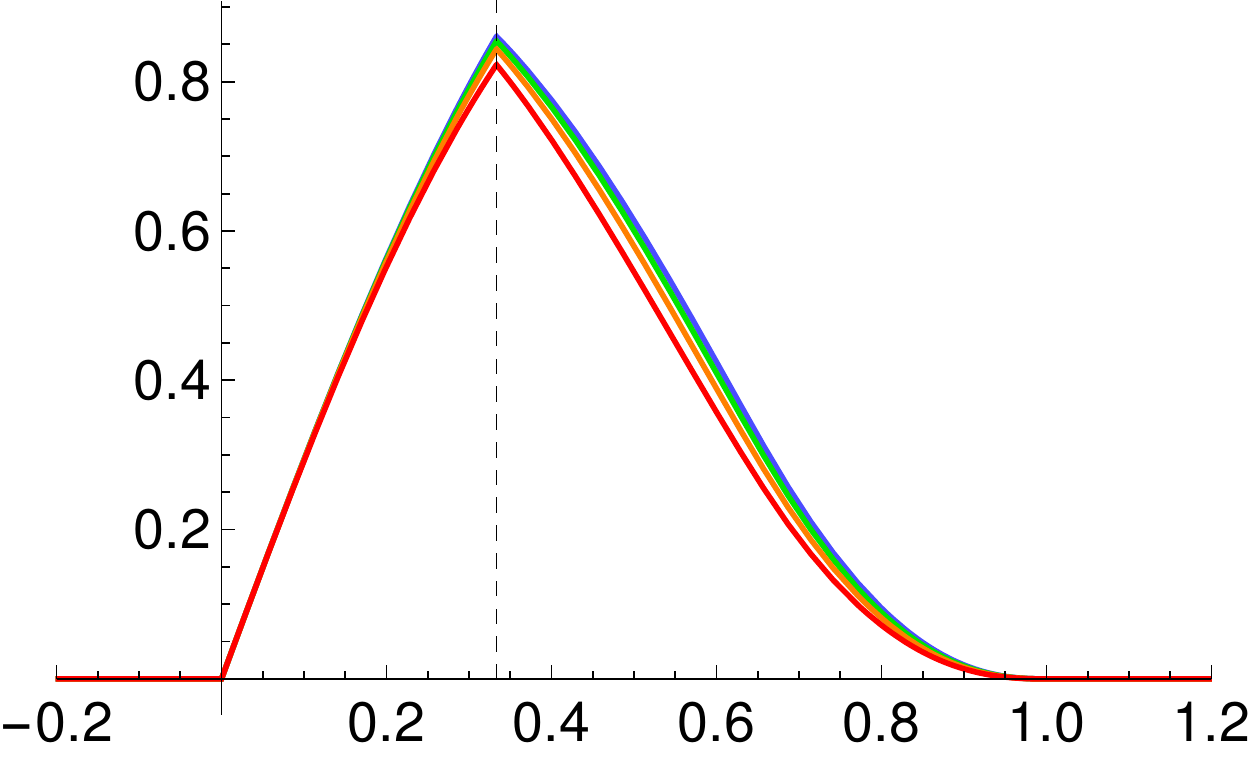} &$\,$ 
     \includegraphics[width=0.3\textwidth]{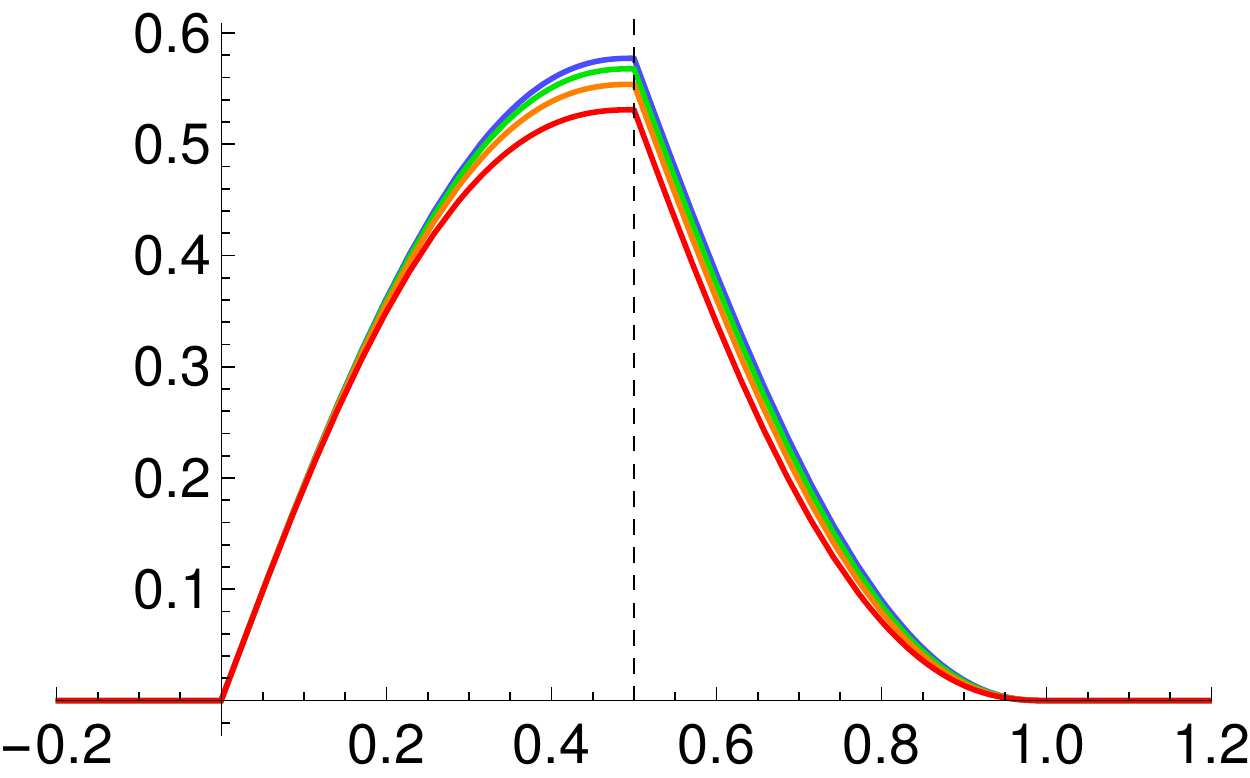} &$\,$ 
     \includegraphics[width=0.3\textwidth]{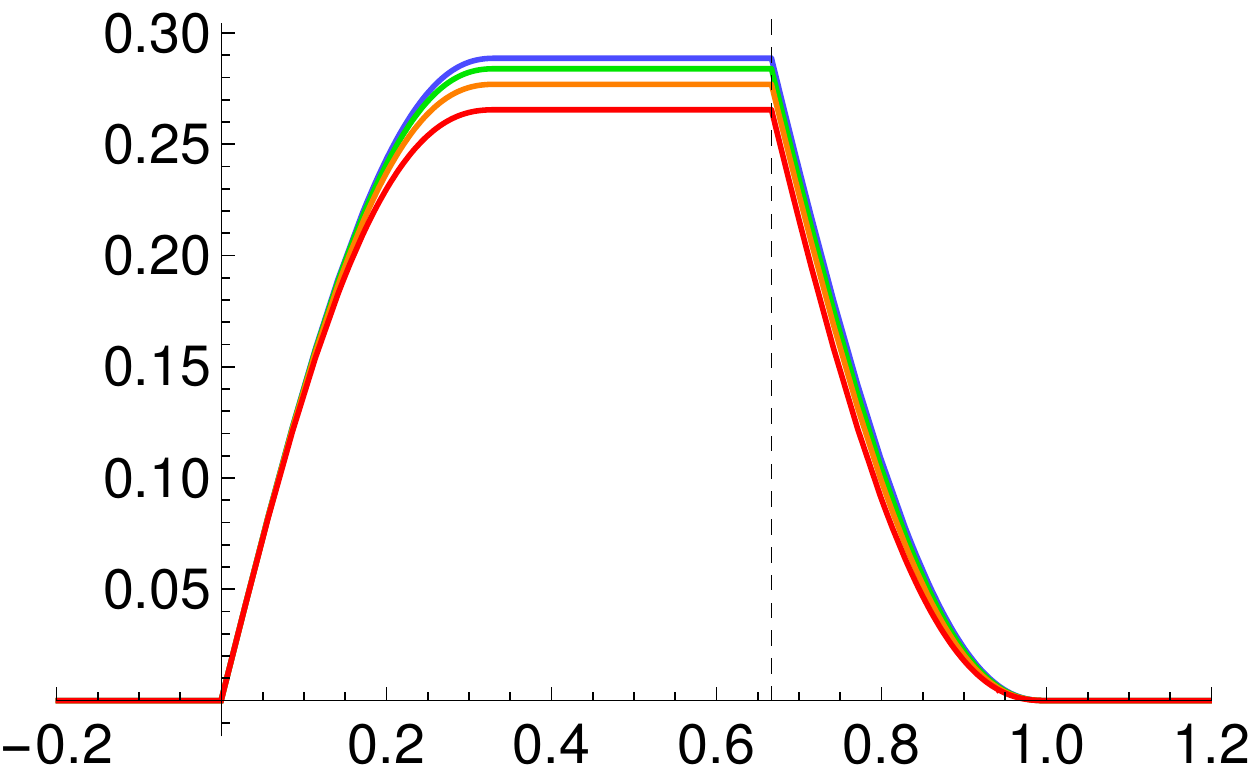} \vspace{5mm}\\
    \includegraphics[width=0.3\textwidth]{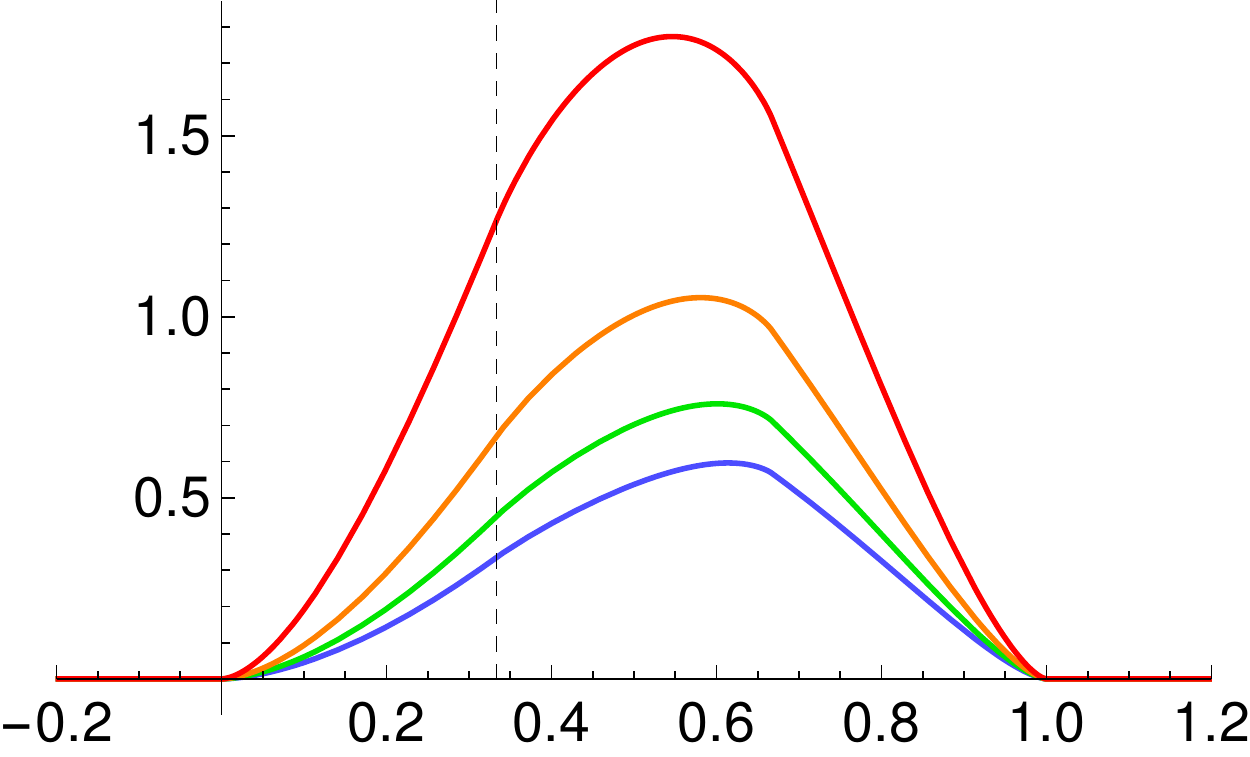} &$\,$ 
    \includegraphics[width=0.3\textwidth]{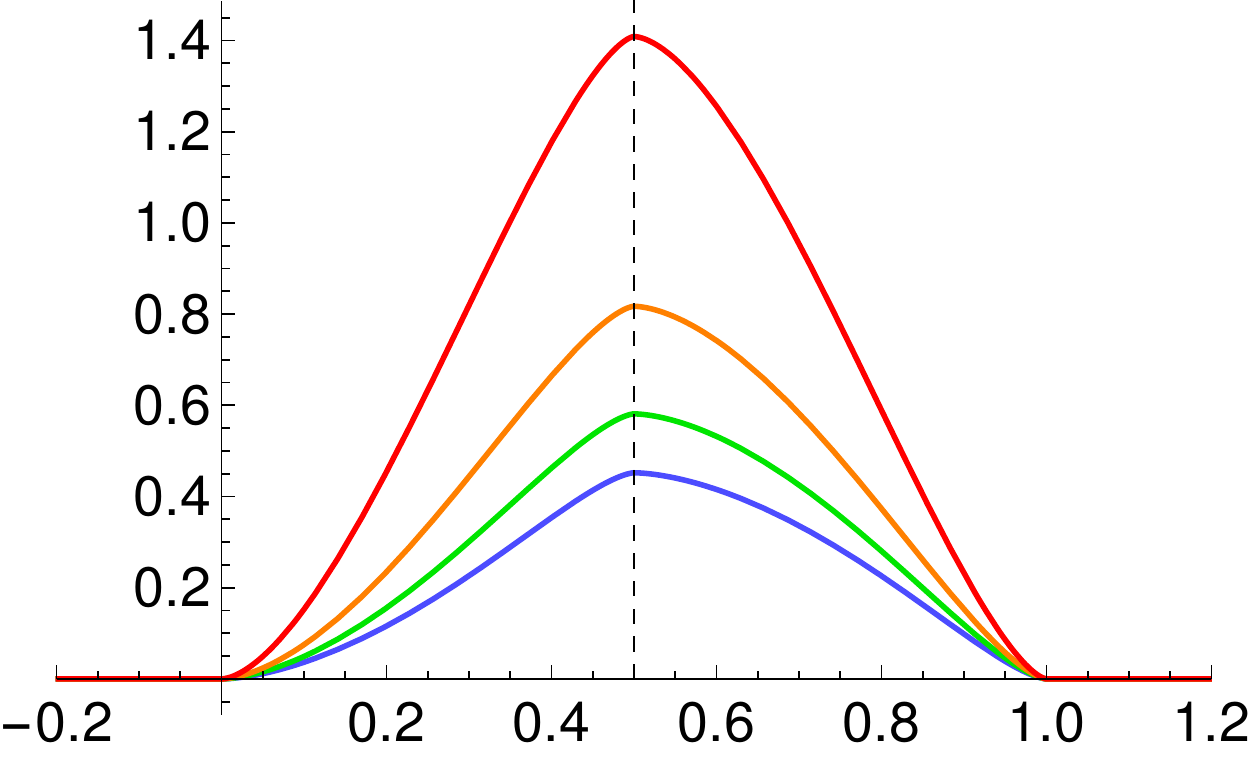} &$\,$ 
    \includegraphics[width=0.3\textwidth]{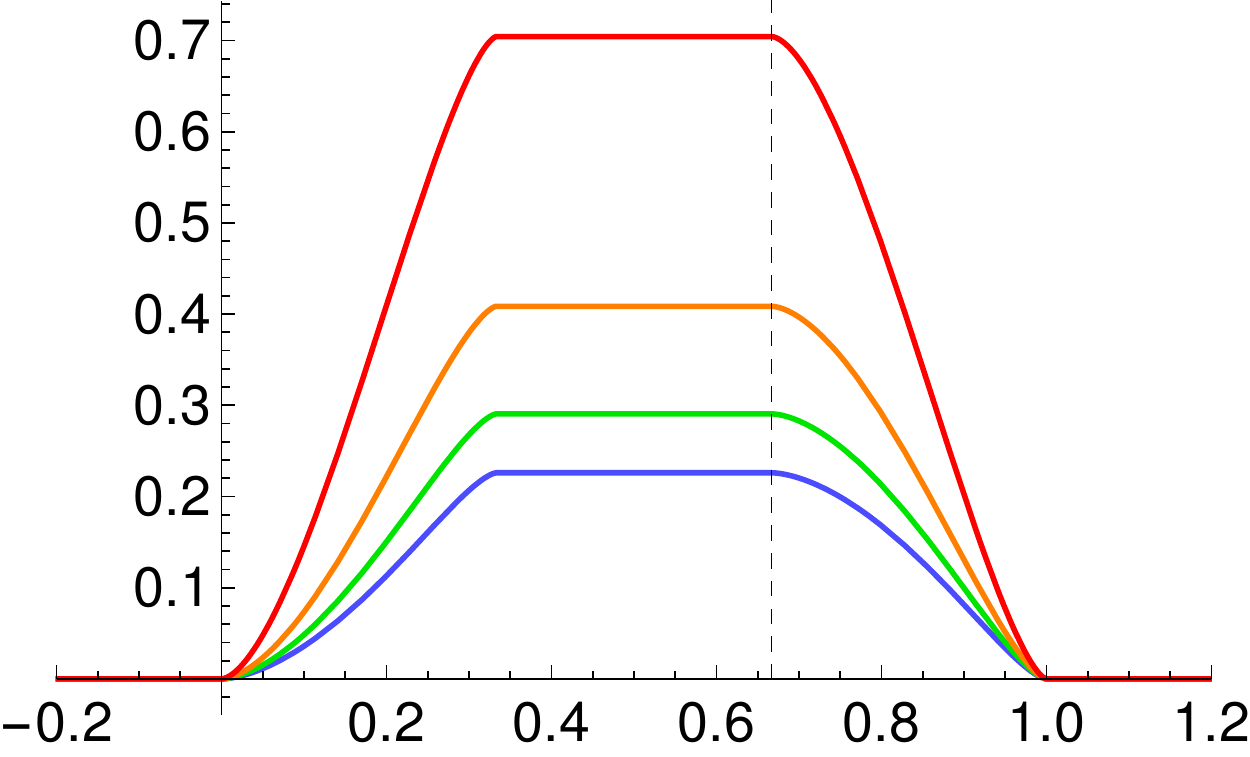} \vspace{5mm}
\end{tabular}
\begin{picture}(0,0)(-460,-145)
\put(-448,157){{\tiny $\delta S/\delta S_{\text{eq}}$}}
\put(-299,157){{\tiny $\delta S/\delta S_{\text{eq}}$}}
\put(-150,157){{\tiny $\delta S/\delta S_{\text{eq}}$}}
\put(-325,85){{\tiny $t$}}
\put(-175,85){{\tiny $t$}}
\put(-26,85){{\tiny $t$}}
\put(-448,57){{\tiny $\delta S_{\text{rel}}/\delta S_{\text{eq}}$}}
\put(-299,57){{\tiny $\delta S_{\text{rel}}/\delta S_{\text{eq}}$}}
\put(-150,57){{\tiny $\delta S_{\text{rel}}/\delta S_{\text{eq}}$}}
\put(-325,-15){{\tiny $t$}}
\put(-175,-15){{\tiny $t$}}
\put(-26,-15){{\tiny $t$}}
\put(-445,-43){{\tiny $\mathfrak{R}_{\text{HSV}}(t)$}}
\put(-295,-43){{\tiny $\mathfrak{R}_{\text{HSV}}(t)$}}
\put(-145,-43){{\tiny $\mathfrak{R}_{\text{HSV}}(t)$}}
\put(-325,-115){{\tiny $t$}}
\put(-175,-115){{\tiny $t$}}
\put(-26,-115){{\tiny $t$}}
\end{picture}
\vspace{-0.6cm}
\caption{Entanglement entropy, relative entropy and entanglement velocity for different dimensions. Colors blue, green, orange and red represent $d=\{5,4,3,2\}$ respectively. Plots from left to right have $t_q/t_*=\{0.5,1,2\}$. In all plots, the dashed vertical line denotes the end of the driven phase $t=t_q$. We have set $p=1, \theta=0.1, z=1.5$.
\label{fig:StuffvsD}}
\end{figure}
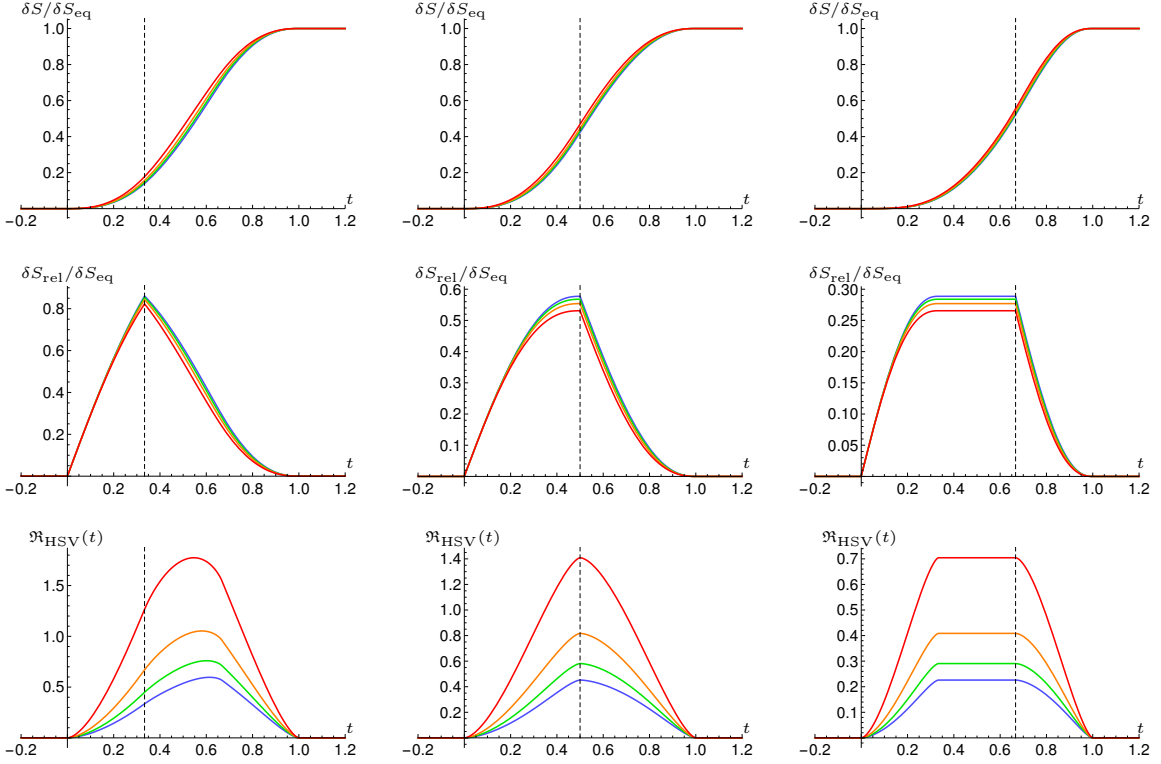

\noindent It is interesting to ask how does the entanglement grow in different dimensions. As shown in Figure \ref{fig:StuffvsD}, we see that for different dimensions the time dependence of entanglement entropy, relative entropy and entanglement velocity is qualitatively uniform. \\

\noindent Finally, in Figures \ref{fig:StuffvsZ} and \ref{fig:StuffvsTheta}, we plot the dependence of entanglement entropy, relative entropy and entanglement velocity as a function of $z$ and $\theta$ respectively. We see a qualitatively similar growth.
\begin{figure}[!htb]
\begin{tabular}{ccc}
    \includegraphics[width=0.3\textwidth]{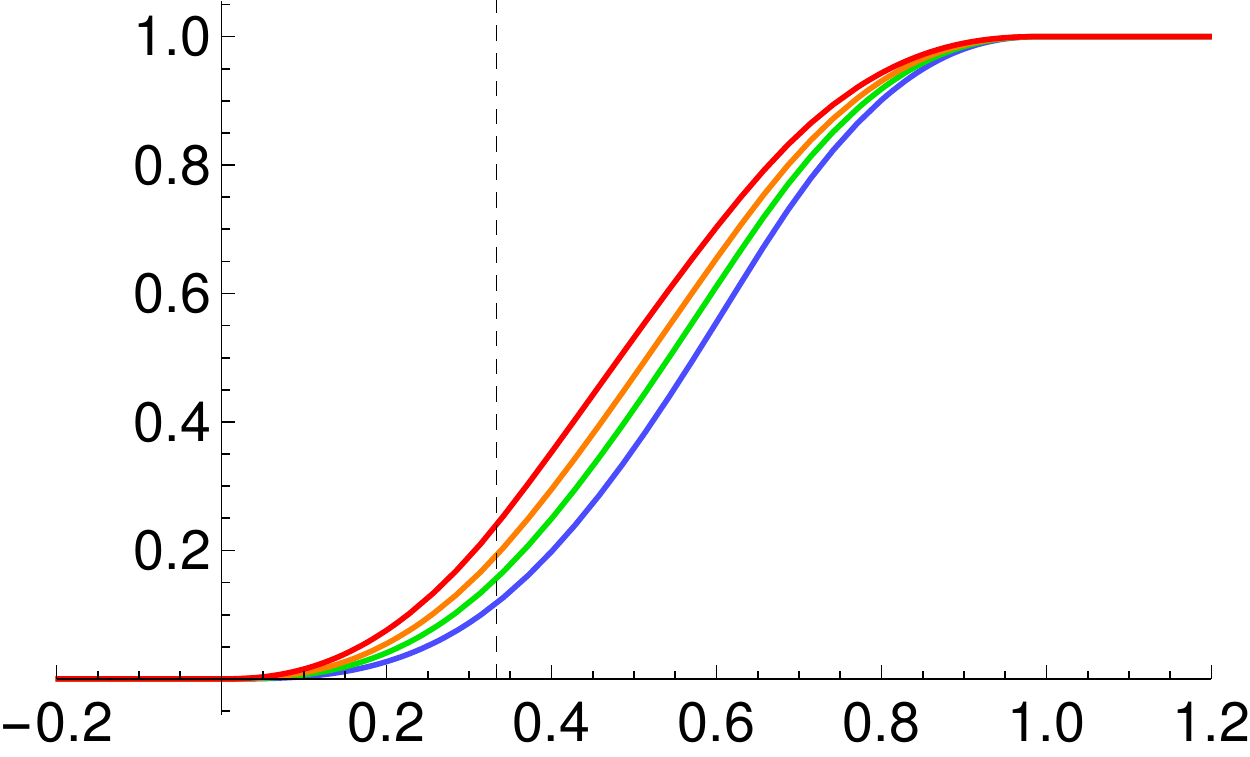} &$\,$ 
    \includegraphics[width=0.3\textwidth]{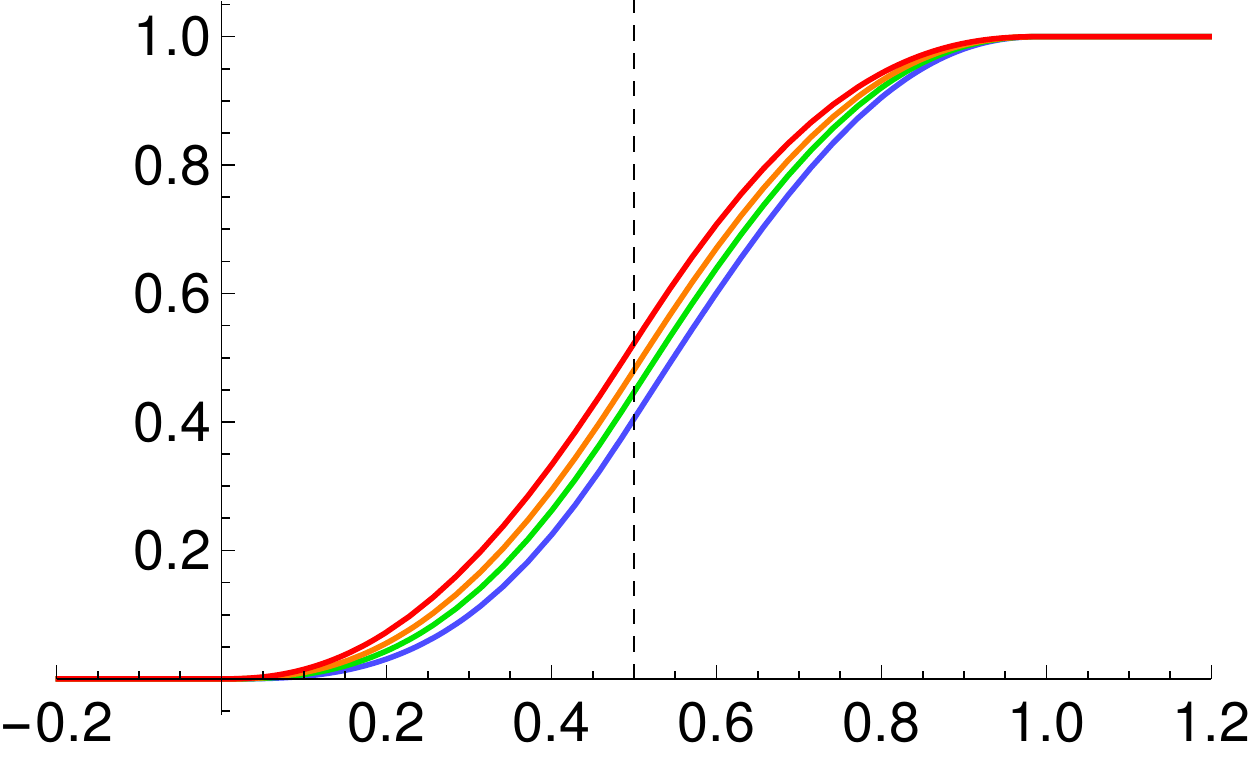} &$\,$ 
    \includegraphics[width=0.3\textwidth]{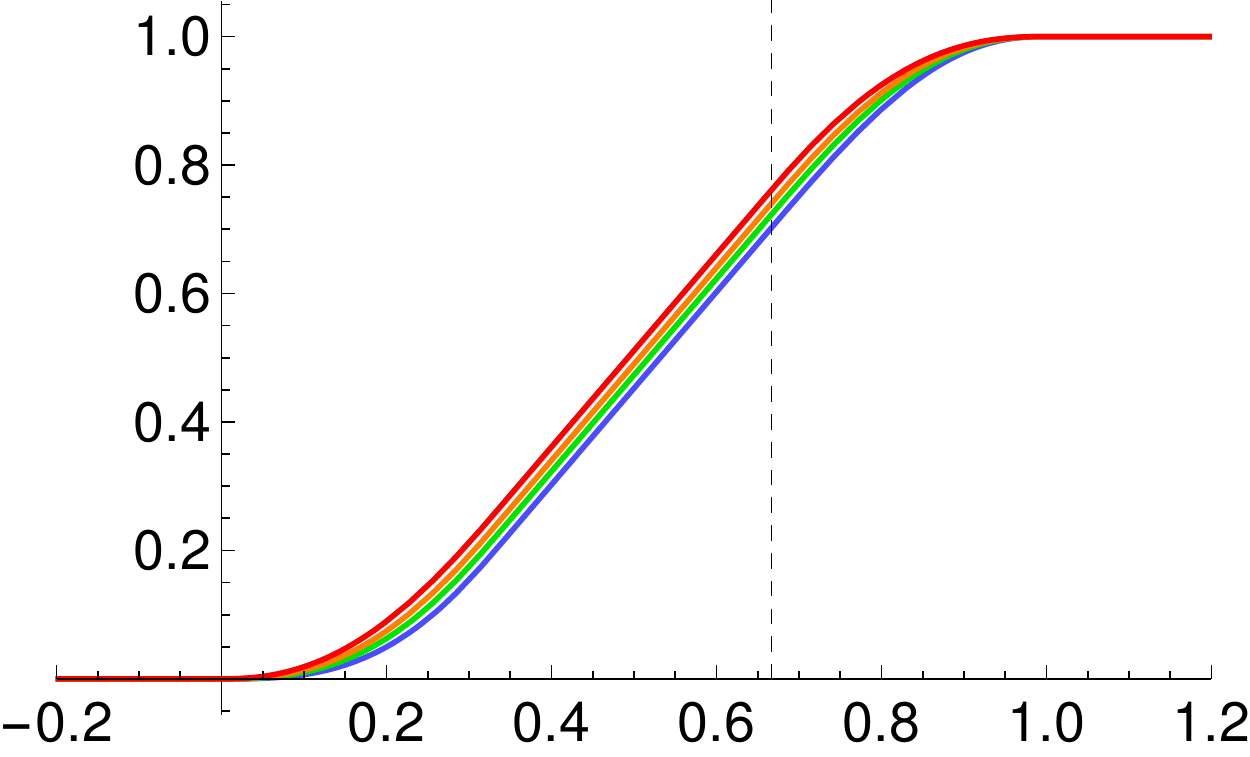} \vspace{5mm} \\
     \includegraphics[width=0.3\textwidth]{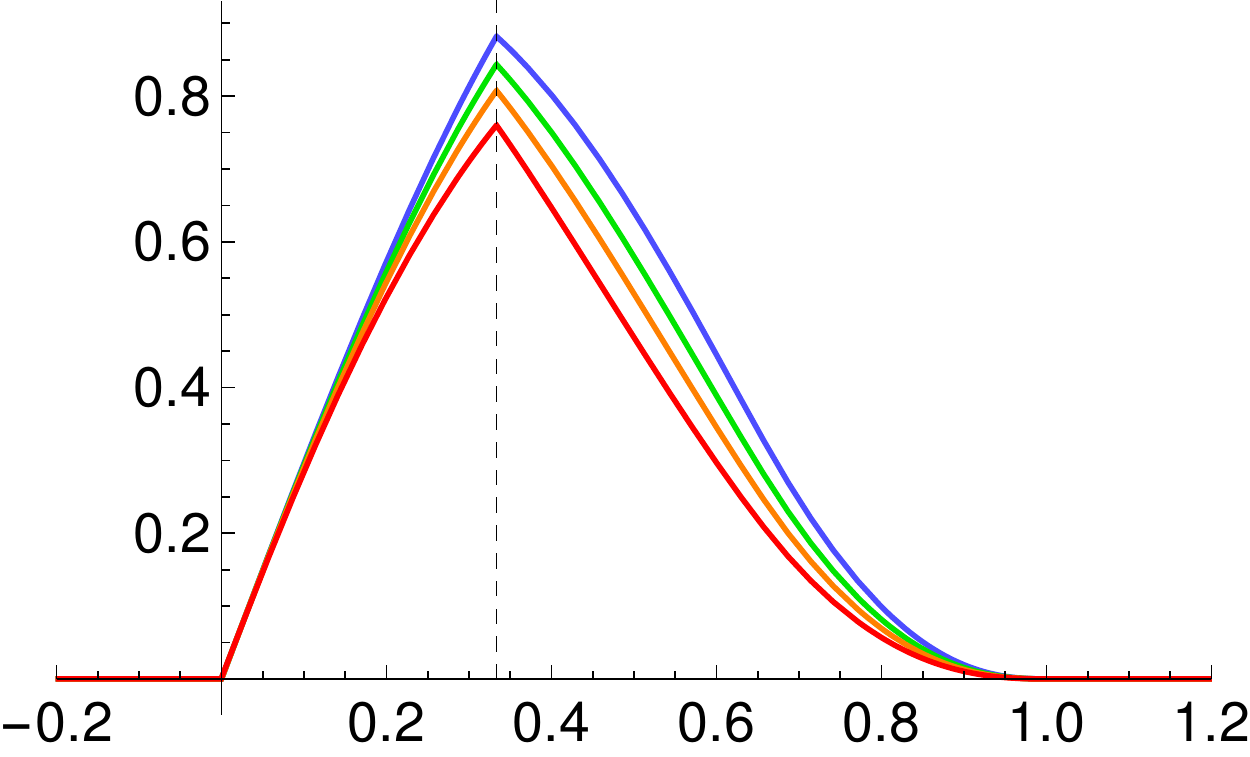} &$\,$ 
    \includegraphics[width=0.3\textwidth]{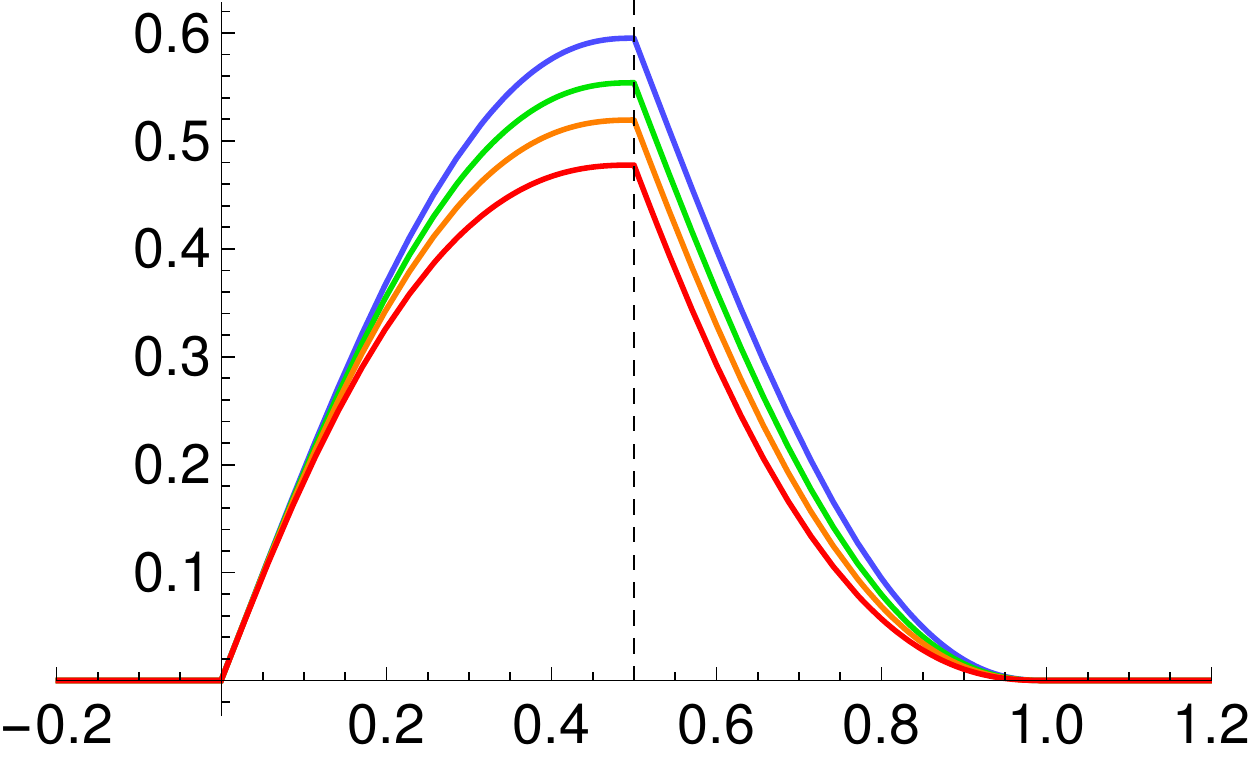} &$\,$ 
    \includegraphics[width=0.3\textwidth]{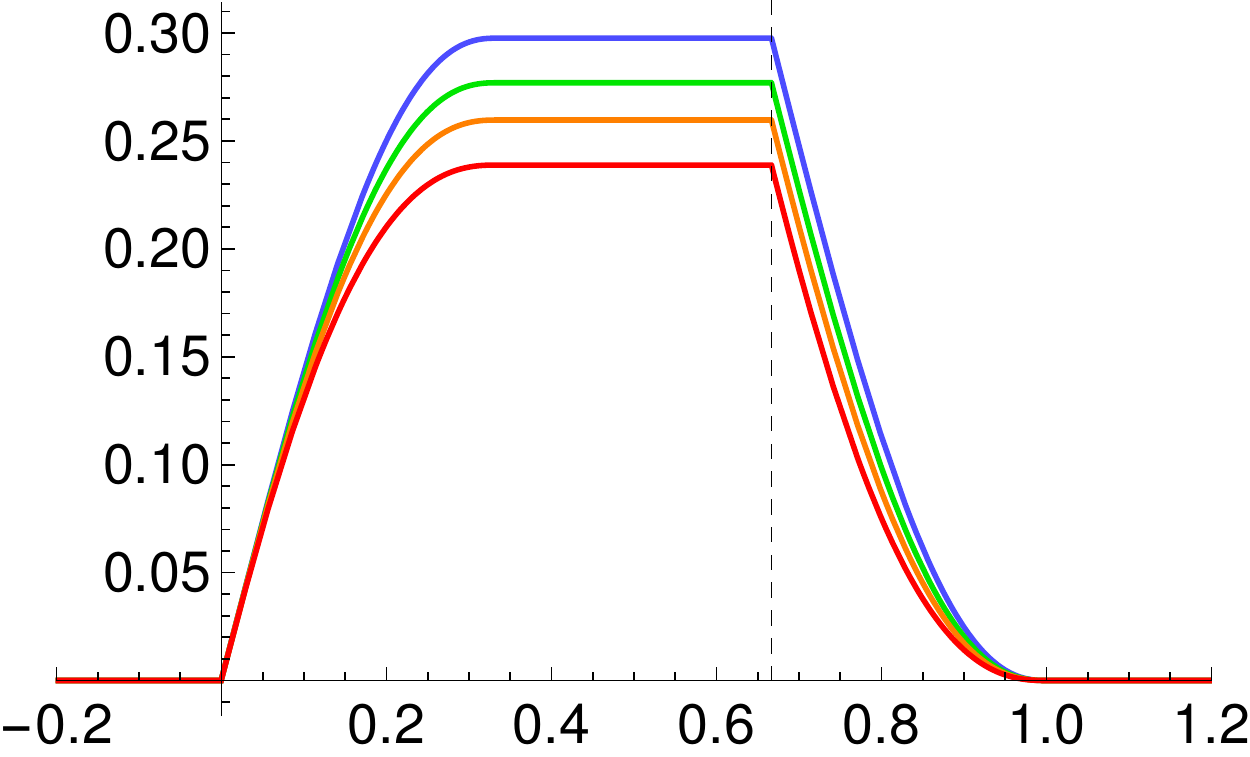} \vspace{5mm}\\
    \includegraphics[width=0.3\textwidth]{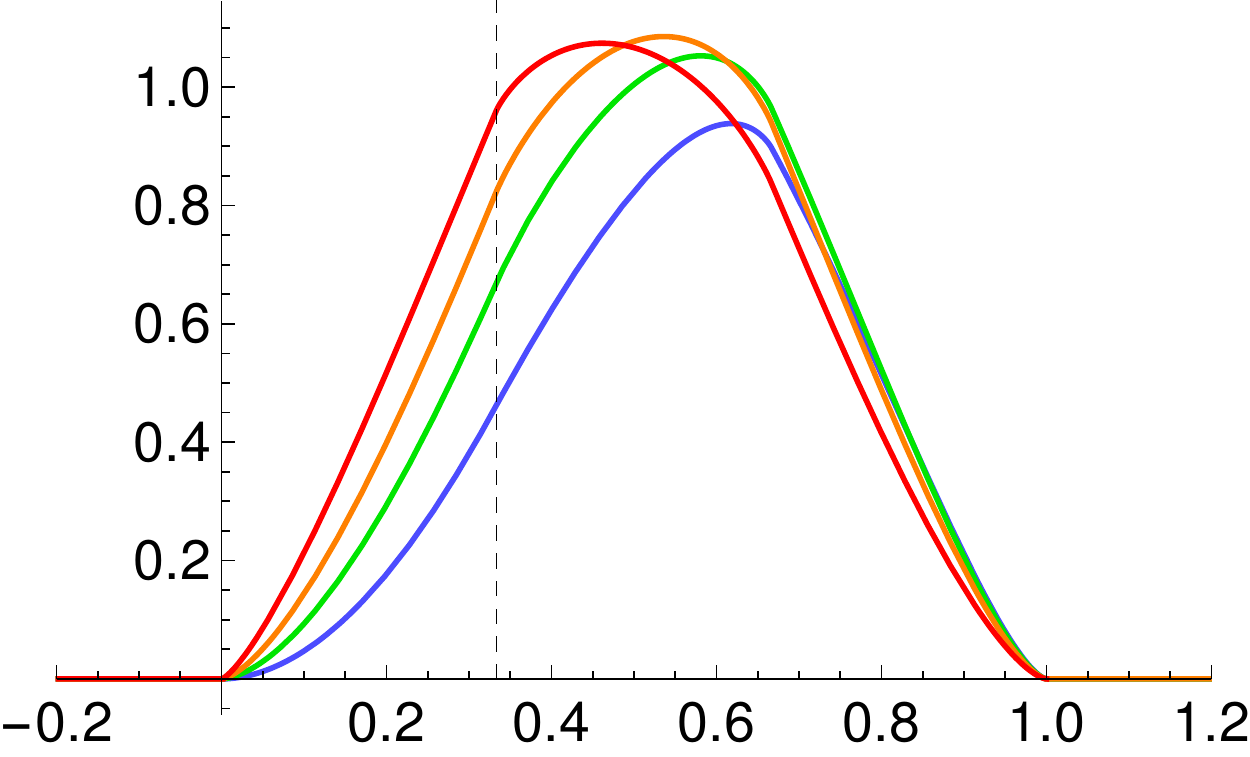} &$\,$ 
   \includegraphics[width=0.3\textwidth]{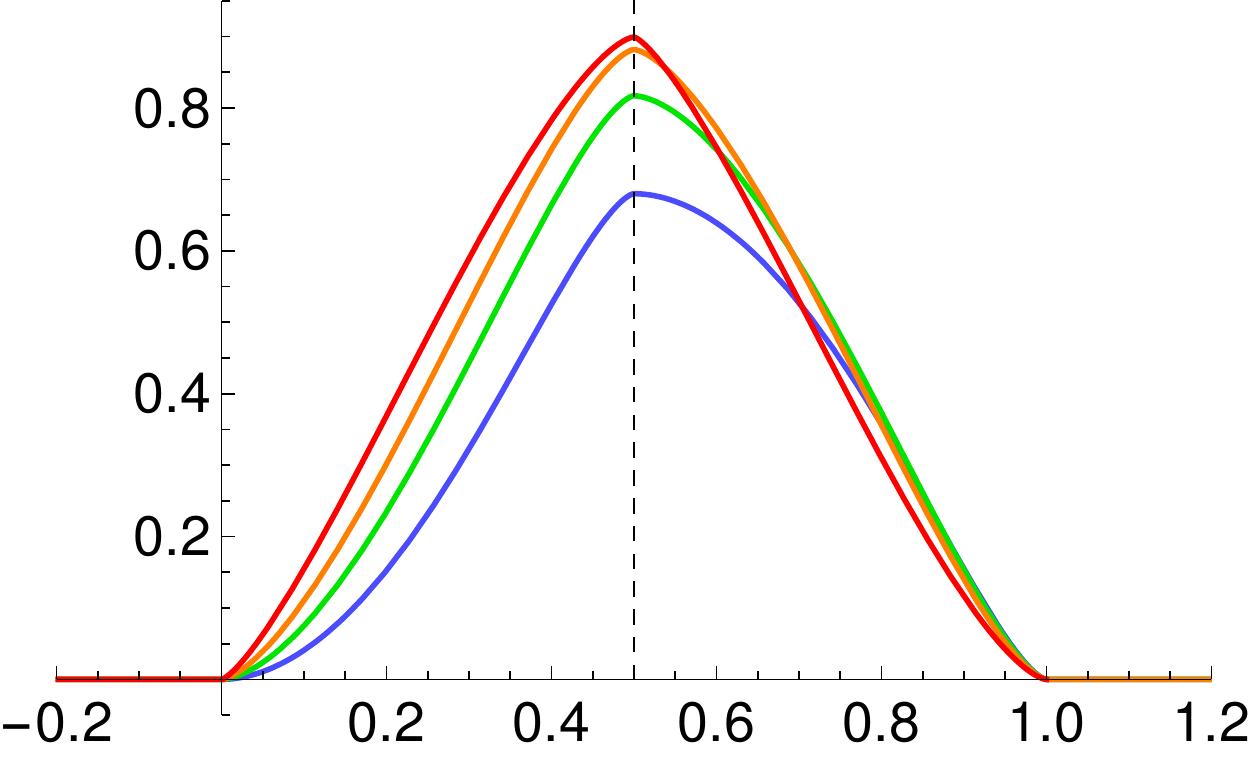} &$\,$ 
   \includegraphics[width=0.3\textwidth]{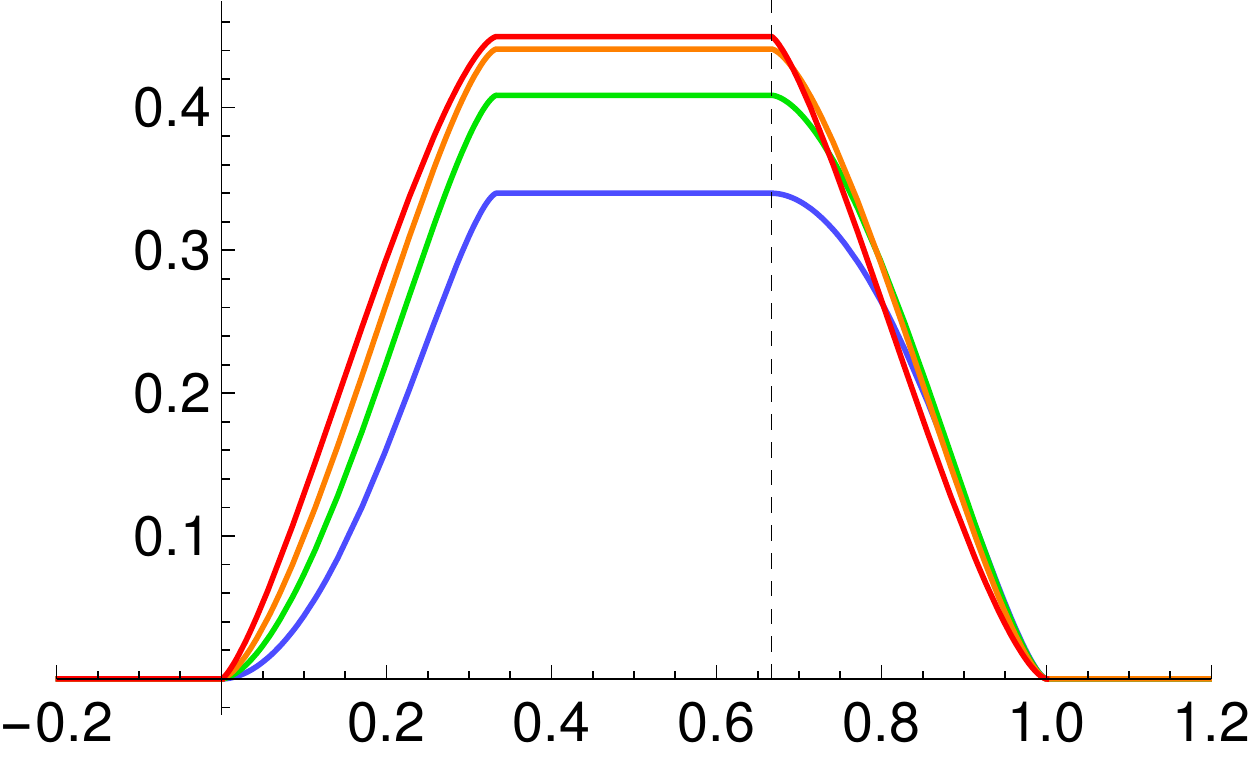} \vspace{5mm}
\end{tabular}
\begin{picture}(0,0)(-460,-145)
\put(-448,157){{\tiny $\delta S/\delta S_{\text{eq}}$}}
\put(-299,157){{\tiny $\delta S/\delta S_{\text{eq}}$}}
\put(-150,157){{\tiny $\delta S/\delta S_{\text{eq}}$}}
\put(-325,85){{\tiny $t$}}
\put(-175,85){{\tiny $t$}}
\put(-26,85){{\tiny $t$}}
\put(-448,57){{\tiny $\delta S_{\text{rel}}/\delta S_{\text{eq}}$}}
\put(-299,57){{\tiny $\delta S_{\text{rel}}/\delta S_{\text{eq}}$}}
\put(-150,57){{\tiny $\delta S_{\text{rel}}/\delta S_{\text{eq}}$}}
\put(-325,-15){{\tiny $t$}}
\put(-175,-15){{\tiny $t$}}
\put(-26,-15){{\tiny $t$}}
\put(-445,-43){{\tiny $\mathfrak{R}_{\text{HSV}}(t)$}}
\put(-295,-43){{\tiny $\mathfrak{R}_{\text{HSV}}(t)$}}
\put(-145,-43){{\tiny $\mathfrak{R}_{\text{HSV}}(t)$}}
\put(-325,-115){{\tiny $t$}}
\put(-175,-115){{\tiny $t$}}
\put(-26,-115){{\tiny $t$}}
\end{picture}
\vspace{-0.6cm}
\caption{Entanglement entropy, relative entropy and entanglement velocity for different values of $z$. Colors blue, green, orange and red represent $z=\{1.1,1.5,2,3\}$ respectively. Plots from left to right have $t_q/t_*=\{0.5,1,2\}$. In all plots, the dashed vertical line denotes the end of the driven phase $t=t_q$. We have set $d=3, p=1, \theta=0.1$.
\label{fig:StuffvsZ}}
\end{figure}
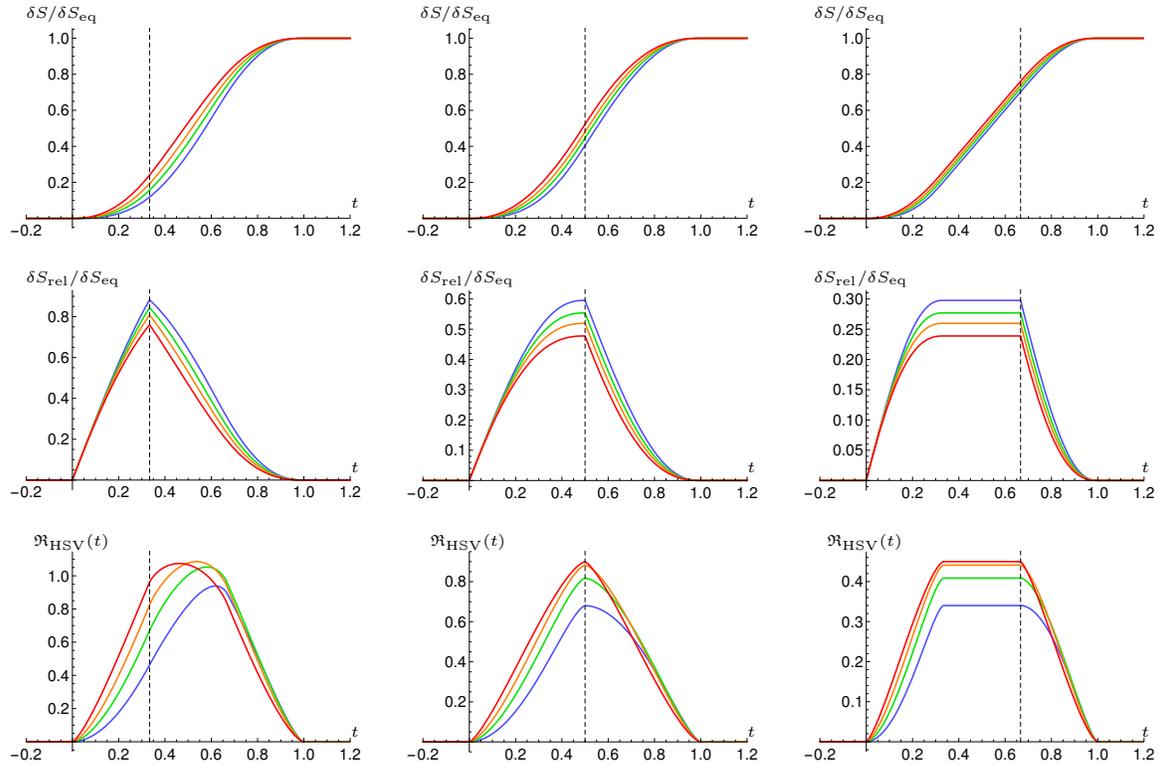
\begin{figure}[!htb]
\begin{tabular}{ccc}
    \includegraphics[width=0.3\textwidth]{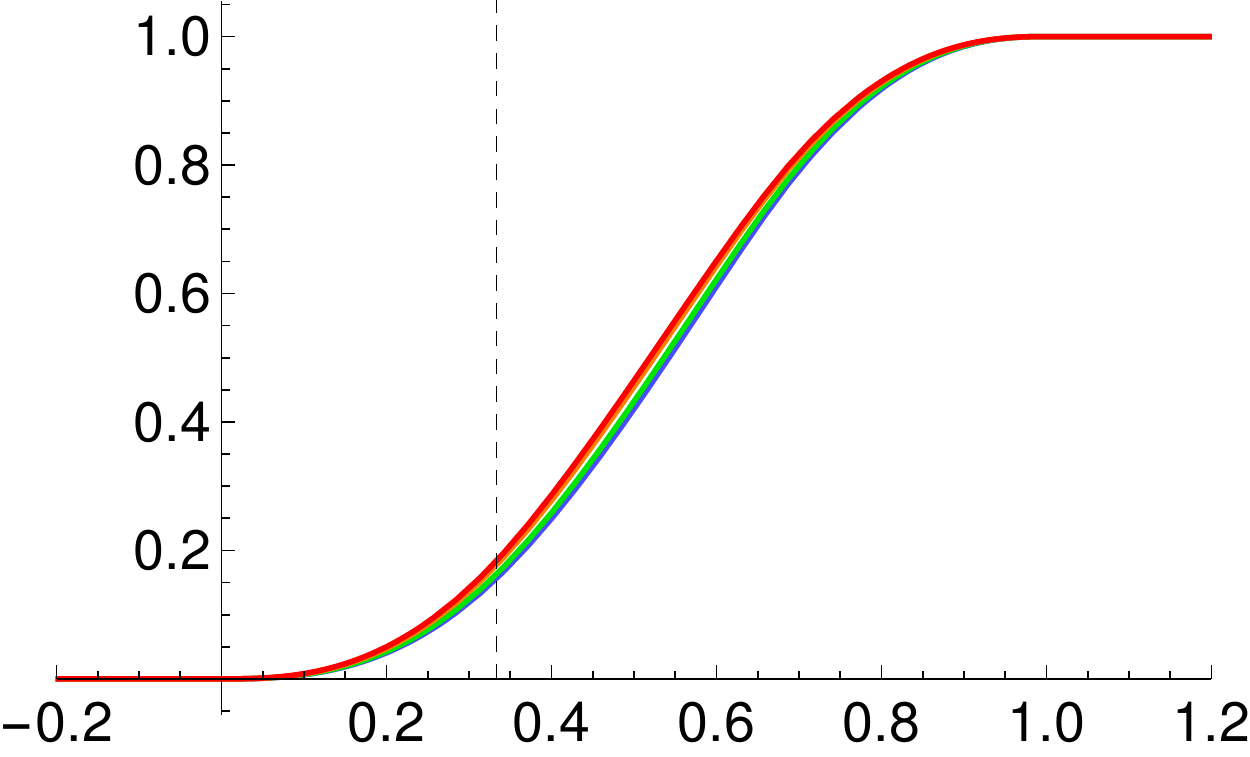} &$\,$ \includegraphics[width=0.3\textwidth]{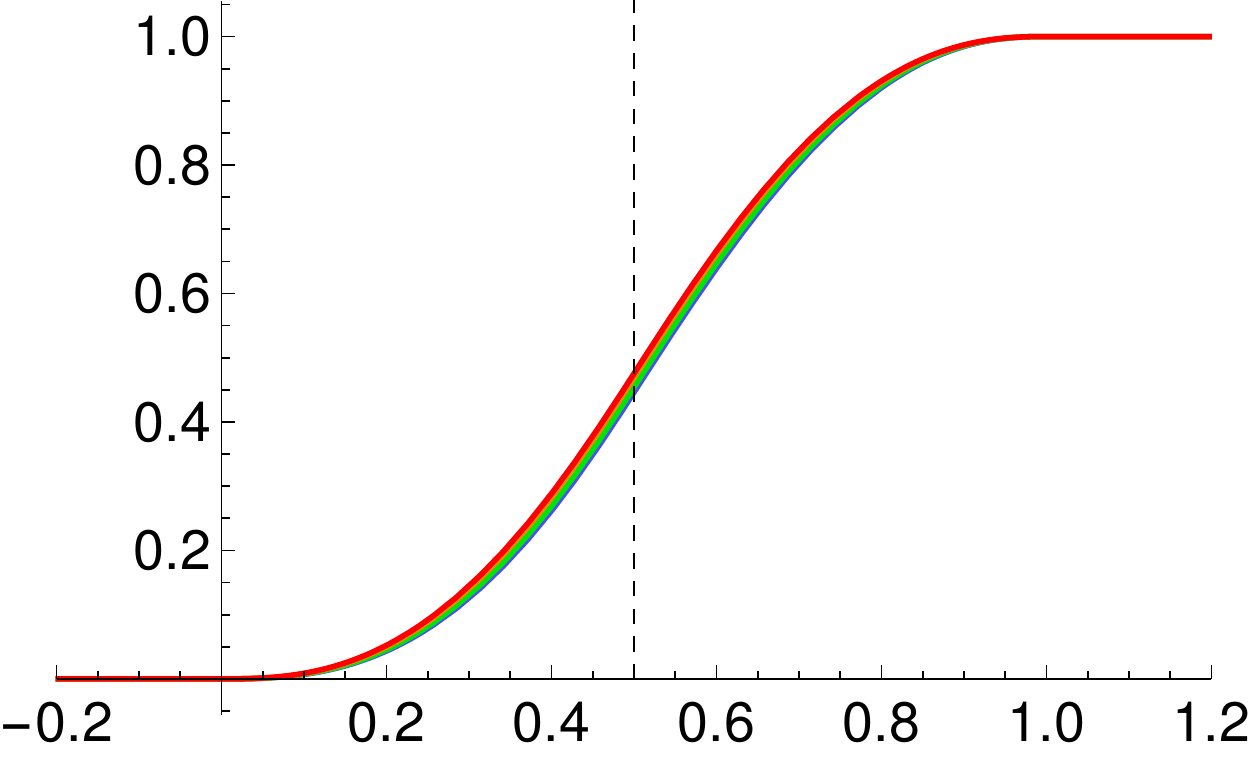} &$\,$ \includegraphics[width=0.3\textwidth]{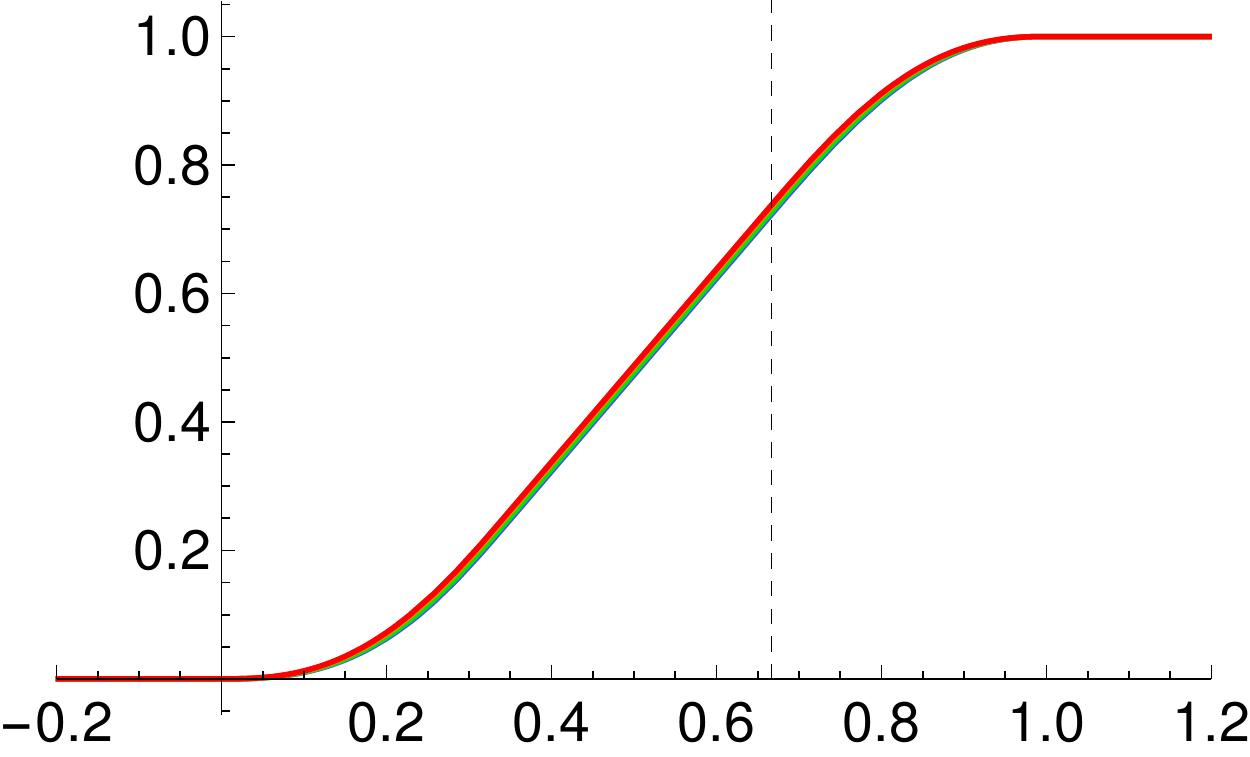} \vspace{5mm} \\
     \includegraphics[width=0.3\textwidth]{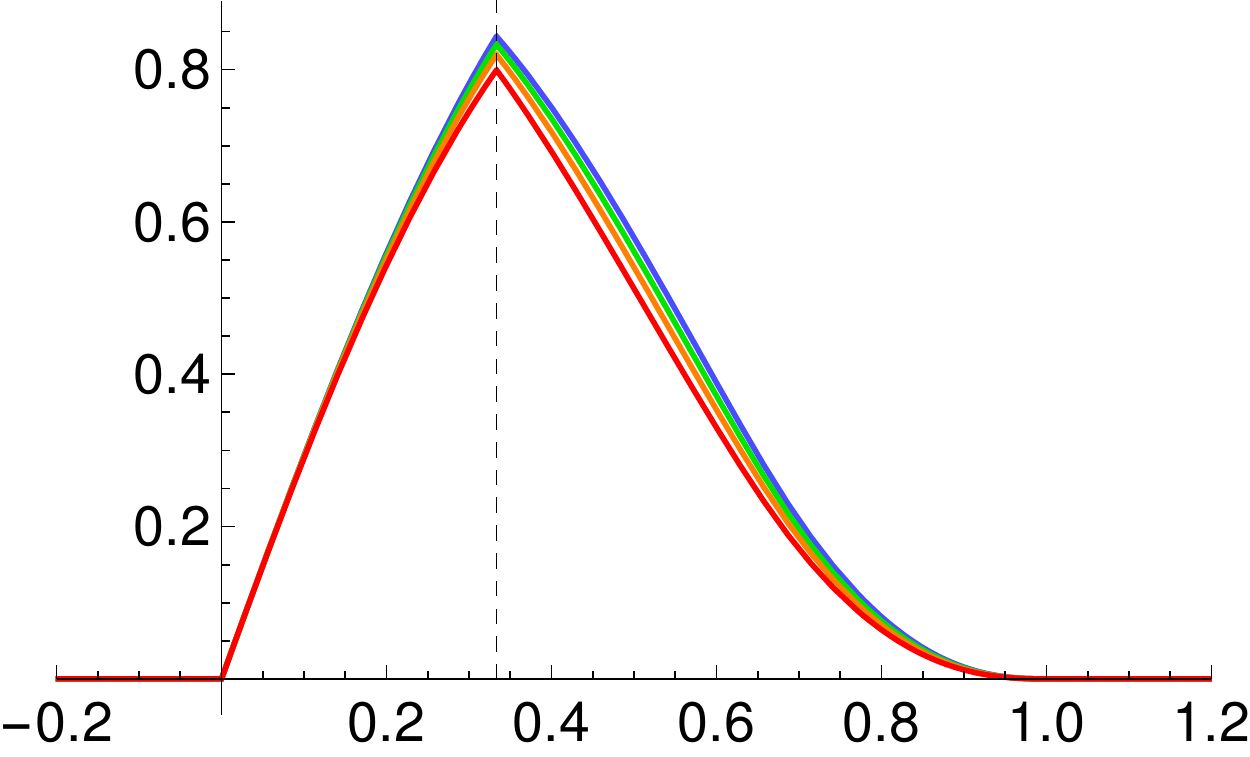} &$\,$ \includegraphics[width=0.3\textwidth]{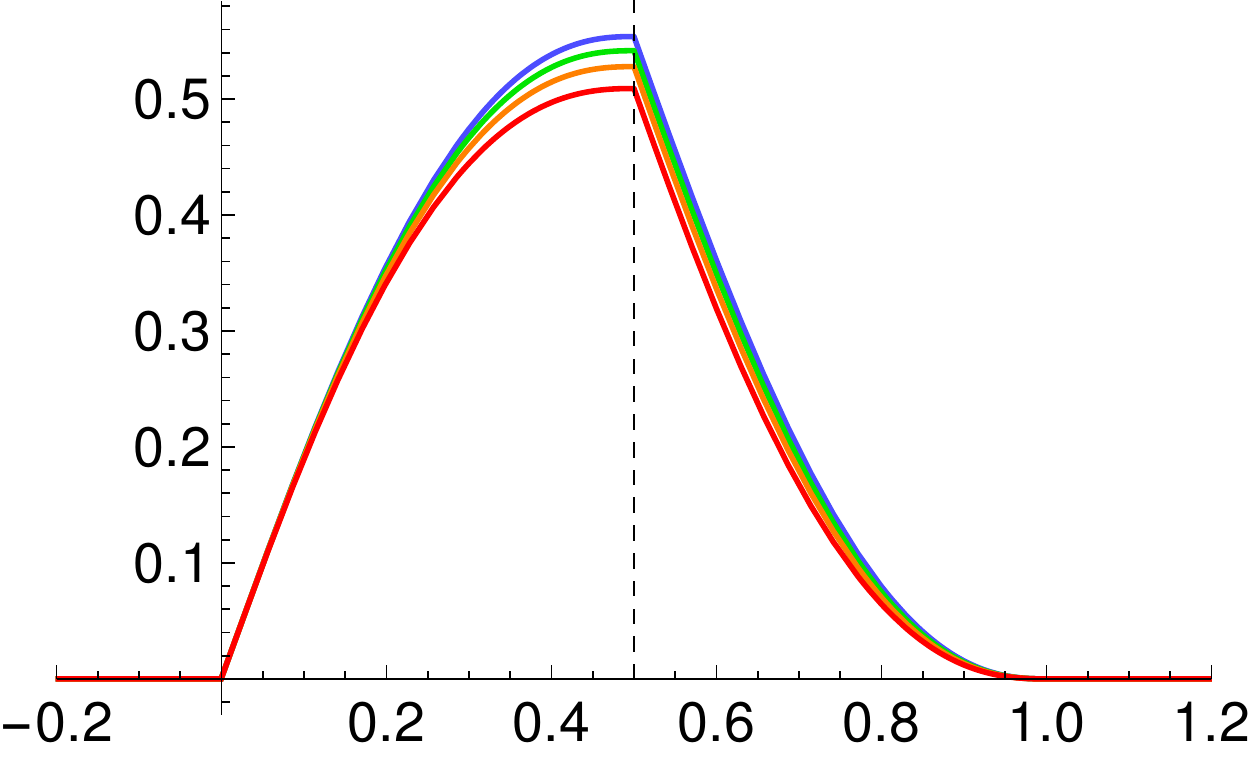} &$\,$ \includegraphics[width=0.3\textwidth]{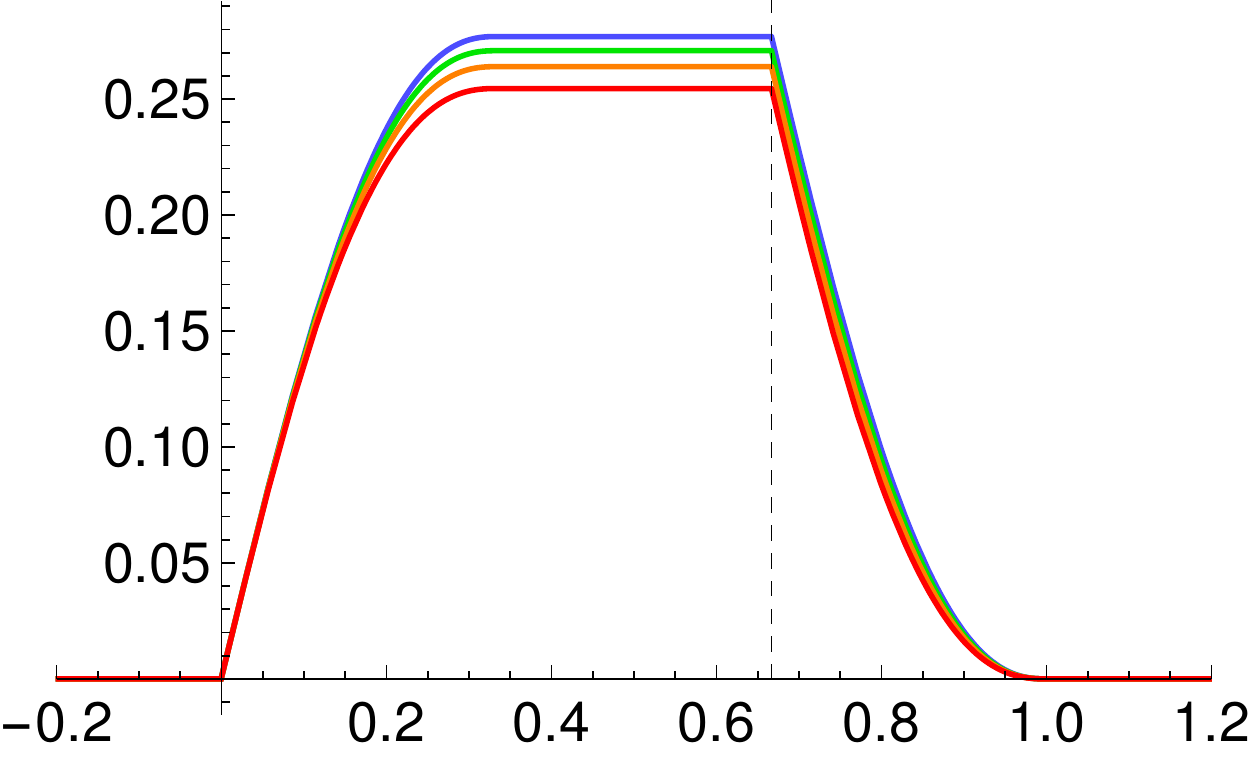} \vspace{5mm}\\
    \includegraphics[width=0.3\textwidth]{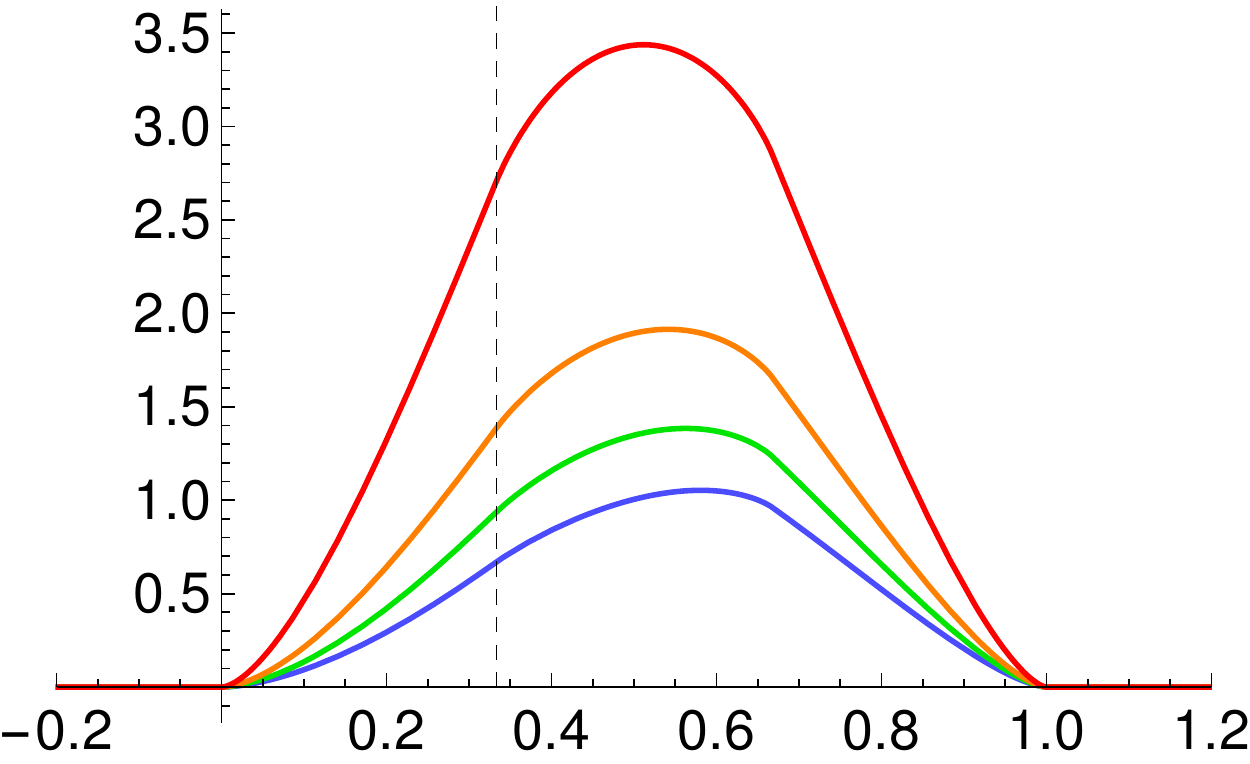} &$\,$ \includegraphics[width=0.3\textwidth]{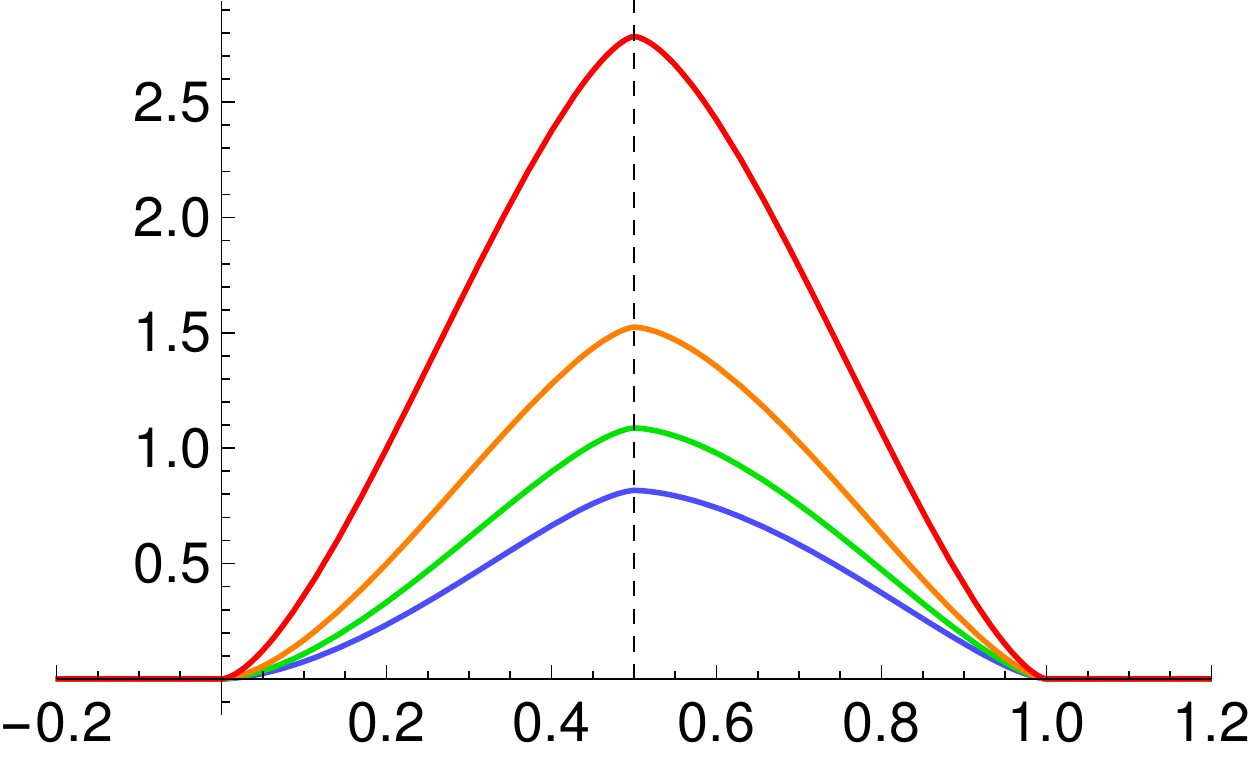} &$\,$ \includegraphics[width=0.3\textwidth]{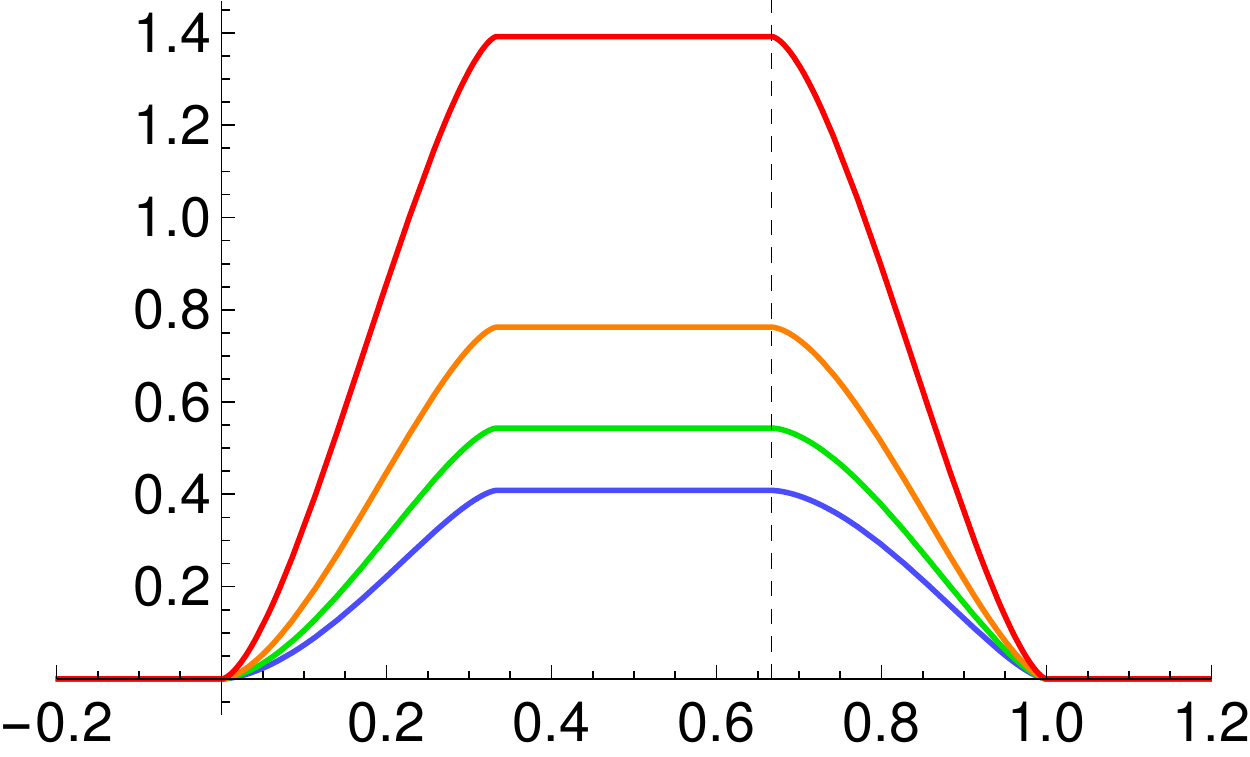} \vspace{5mm}
\end{tabular}
\begin{picture}(0,0)(-460,-145)
\put(-448,157){{\tiny $\delta S/\delta S_{\text{eq}}$}}
\put(-299,157){{\tiny $\delta S/\delta S_{\text{eq}}$}}
\put(-150,157){{\tiny $\delta S/\delta S_{\text{eq}}$}}
\put(-325,85){{\tiny $t$}}
\put(-175,85){{\tiny $t$}}
\put(-26,85){{\tiny $t$}}
\put(-448,57){{\tiny $\delta S_{\text{rel}}/\delta S_{\text{eq}}$}}
\put(-299,57){{\tiny $\delta S_{\text{rel}}/\delta S_{\text{eq}}$}}
\put(-150,57){{\tiny $\delta S_{\text{rel}}/\delta S_{\text{eq}}$}}
\put(-325,-15){{\tiny $t$}}
\put(-175,-15){{\tiny $t$}}
\put(-26,-15){{\tiny $t$}}
\put(-445,-43){{\tiny $\mathfrak{R}_{\text{HSV}}(t)$}}
\put(-295,-43){{\tiny $\mathfrak{R}_{\text{HSV}}(t)$}}
\put(-145,-43){{\tiny $\mathfrak{R}_{\text{HSV}}(t)$}}
\put(-325,-115){{\tiny $t$}}
\put(-175,-115){{\tiny $t$}}
\put(-26,-115){{\tiny $t$}}
\end{picture}
\vspace{-0.6cm}
\caption{Entanglement entropy, relative entropy and entanglement velocity for different values of $\theta$. Colors blue, green, orange and red represent $\theta=\{0.1,0.7,1.2,1.7\}$ respectively. Plots from left to right have $t_q/t_*=\{0.5,1,2\}$. In all plots, the dashed vertical line denotes the end of the driven phase $t=t_q$. We have set $d=3, p=1, z=1.5$.}
\label{fig:StuffvsTheta}
\end{figure}

\subsection{Linear Pump}
\label{linear-pump}

In this subsection we will comment on the case of a linear quench $g(t) \sim t$. We call this the Linear Pump. For a CFT, it was studied in detail in \cite{OBannon:2016exv, Lokhande:2017jik}. \cite{OBannon:2016exv} in particular proposed a First Law of Entanglement Rates for small subsystems given by
\begin{equation}
\frac{d \delta S_A(t)}{d t} = \frac{d \epsilon(t)}{dt} \, \frac{V_A}{T_A} ,
\end{equation}
where $T_A$ is the entanglement temperature for a CFT calculated by the CFT limit of equation \eqref{eq:const_ent_Temp}. This law can be interpreted as a derivative form of our general time-dependent first law of entanglement entropy \eqref{eq:general_first_law} in the CFT limit. We now use our detailed discussion of the general power law quench from Subsection \ref{power-law} to discuss the spread of entanglement entropy after a linear pump. In fact the cases $p=1, 2$ can be solved completely analytically. The exact integrals for the linear quench can be derived from the general equations \eqref{eq:HSV_S1_power}, \eqref{eq:HSV_S2_power}. Here we only discuss the plots in brief. In the first row of Figure \ref{fig:StuffvsD}, we can see the growth of entanglement entropy after a linear pump. The dashed black line indicates the end of the driven phase of entanglement growth. The four different colors indicate different dimensions. In the second row of the same figure, we have relative entropy as a function of time for different dimensions. It first increases and then decreases, unlike the case of the instantaneous quench. This behavior can be understood by observing that until time $t_q$, the system is forced. As a result of this, it keeps going farther away from equilibrium. Keeping in line with our philosophy that the relative entropy measures how far the system is from equilibrium, it makes sense that it increases initially. The initial and the final states are both equilibrium states, hence the relative entropy is zero in these states. If it increases initially, it must decrease later to account for this fact.

\subsection{Floquet Quench}
\label{Floquet}

Time periodic forces are commonly used in laboratory situations. The study of differential equations with a periodic function in the differential operator is called Floquet Theory. Adopting this name, we refer to a quench that is periodic in time as Floquet Quench. Thus, the time-dependent function $g(t)$ in the source function $\mathfrak{m}(t)$ (equation \eqref{eq:source_func}) looks like 
\begin{equation}
\label{eq:periodic_energy_density}
g(t)=\sin(\Omega \, t ) \,  \Theta(t)   \, ,
\end{equation}
where $\Omega$ is frequency of the external source and we have set the amplitude of the source to 1. Also observe that the quench does not have a finite duration ($t_q \to \infty$). This case is more interesting because a Floquet Quench that acts for finite duration can be very-well approximated by a finite combination of power-law quenches, whose exact description we know. \\

\noindent From the convolution equation \eqref{eq:EE_lin_resp}, at late times we expect the entanglement growth to also be periodic in time. In particular, there is no saturation of entanglement. This is because the parameter $t$ only appears in the source function, and there it is periodic with a trigonometric expression. Thus, we expect the entanglement growth to have the form
\begin{equation}\label{eq:perEEexpect}
\delta S(t) = P(t) +\Psi  \, \sin(\Omega \, t + \Phi) ,
\end{equation}
where $P(t)$ is some smooth polynomial in $t$ that interpolates to the $\sin$ function around $t=t_*$. The function $P(t)$, the amplitude $\Psi$ and the phase $\Phi$ are all functions of the parameters $d, \theta, z, \Omega$ and the subsystem size $\ell$. In fact, We can argue for the the existence of $P(t)$ on general grounds. The entanglement entropy is zero for $t<0$ and then it starts growing at $t=0$. However, the initial growth of entanglement is dictated solely by the symmetries of the theory to be $t^{1+z}$. Since at late times, the entanglement growth is entirely driven by the periodic source, there must exist a polynomial $P(t)$ such that it interpolates smoothly  between the early and the late time growth.\\

\noindent Expanding the $\sin$ function in a Taylor series, the exact expression for the growth of entanglement entropy is becomes
\begin{align}
 \begin{split}
\delta S(t) &= \sum\limits_{p=0}^\infty \frac{-(-1)^{(p+1)^2} \, \mathcal{A}_{\Sigma} \, \Omega^{2p+1}}{8 \, G_N \, u_H^{d_\theta+z} \, (2p+1)!} \, \int_{0}^{t_*} \, d \tau \, (t-\tau)^{2p+1} \, (z \, \tau)^{\frac{1}{z}} \, \sqrt{1-\big[\frac{\tau}{t_*}\big]^{\frac{2 d_\theta}{z}}} .
 \end{split}
\end{align}
Referring to the indefinite integral \eqref{eq:p-indef-int1}, we can write this as,
\begin{align}
 \begin{split}
 \label{eq:floquet_EE}
\delta S(t) &=    \sum\limits_{p=0}^\infty \frac{-(-1)^{(p+1)^2}  \, \Omega^{2p+1}}{ (2p+1)!}
 \end{split} \, \mathcal{I}^{(2p+1)}(t,\tau)\bigg|_{\tau=0}^{\tau=t_*}
\end{align}
where the exact expression for $\mathcal{I}^{(p)}(t,\tau)$ is given in equation \eqref{eq:p-indef-int2}. This gives us the entanglement growth as an infinite series of hypergeometric functions. This description of the growth is most useful when the driving frequency $\Omega$ is small compared to 1. Then, it is possible to get a good approximation to the entanglement growth by truncating the above series at some appropriate power of $\Omega$. In the cases when the driving frequency is not small, it is not possible to get a closed-form analytic expression for the entanglement growth with the usual methods. Thus, we will study the general case numerically instead.  This interesting problem was also studied numerically in \cite{Rangamani:2015sha}, without a discussion of the underlying analytic form of entanglement entropy. For the limit of small subsystems as defined in this paper, our expression for entanglement entropy as a convolution provides an explanation and the underlying analytics for their numerical results.   \\

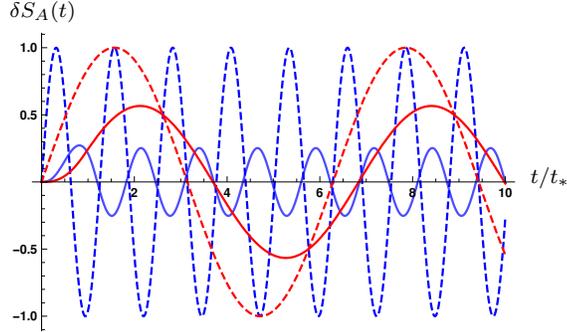
\begin{figure}[!h]
$$
  \includegraphics[angle=0,width=0.43\textwidth]{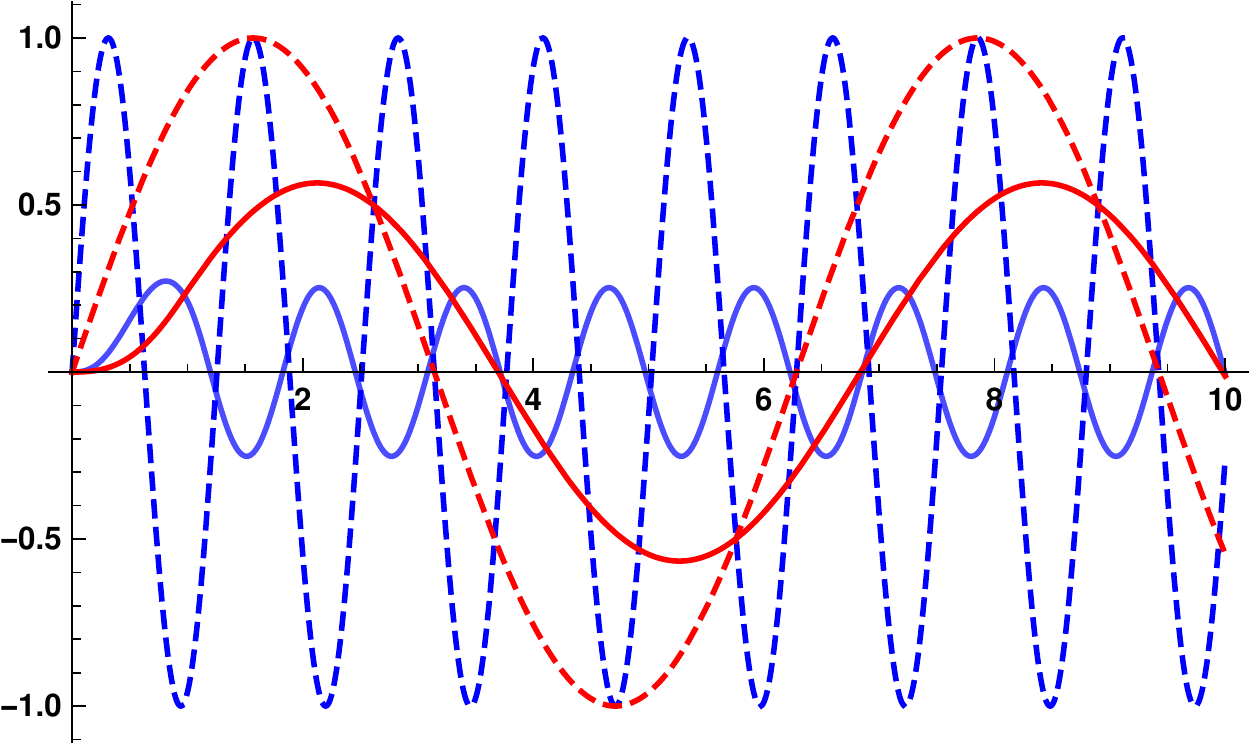}
$$
\begin{picture}(0,0)
\put(125,141){{\scriptsize $\delta S_A(t)$}}
\put(322,78){{\scriptsize $t/t_*$}}
\end{picture}
\vspace{-0.7cm}
\caption{\small Entanglement entropy (solid line) as a function of time after a Floquet Quench, plotted against the periodic source (dashed line). The colors blue and red denote frequencies $\Omega=\{5,1\}$ respectively. We have set $d=3, \theta=0.1, z=1.5$. }
\label{fig:perEEvsSource}
\end{figure}

\noindent In Figure \ref{fig:perEEvsSource}, we verify our expectation that the entanglement grows with time according to equation \eqref{eq:perEEexpect}.In fact, after a time $t_*$ it grows as a $\sin$ function of the same frequency as the source $g(t)$.  One can also observe the relative amplitude and phase between the entanglement entropy and the energy density. The entanglement entropy has a smaller amplitude, which can be understood from the fact that to leading order, it is the convolution of a $\sin$ function and $\tau^{\frac{1}{z}}$, both smaller than 1 (since $t_*$ is effectively 1). The plots moreover show that the entanglement growth at early times interpolates to the $\sin$ growth at late time smoothly. It is possible to model this smooth interpolation using spline theory. But we will not do so in this paper. 

\noindent As the dashed red line in Figure \ref{fig:perEEvsSource} indicates, the energy density oscillates. This is not physical because if the Null Energy Condition (NEC) is satisfied in the bulk, energy density should not decrease. There are two ways one can rectify this situation. The Floquet quench involves continuously driving the system externally and if NEC is applied to the whole system of the external apparatus plus \textbf{hvLif} theory together, we expect the NEC to be satisfied. Secondly, one could modify the quench by adding a linear pump to the Floquet quench. In this case, we have verified that the NEC is satisfied. \\ 

\noindent In Figure \ref{fig:perEEvsDW}, we show how the entanglement growth changes as we change the number of dimensions and also its dependence on the frequency of the source.
\begin{figure}[!htb]
$$
\begin{array}{cc}
  \includegraphics[angle=0,width=0.43\textwidth]{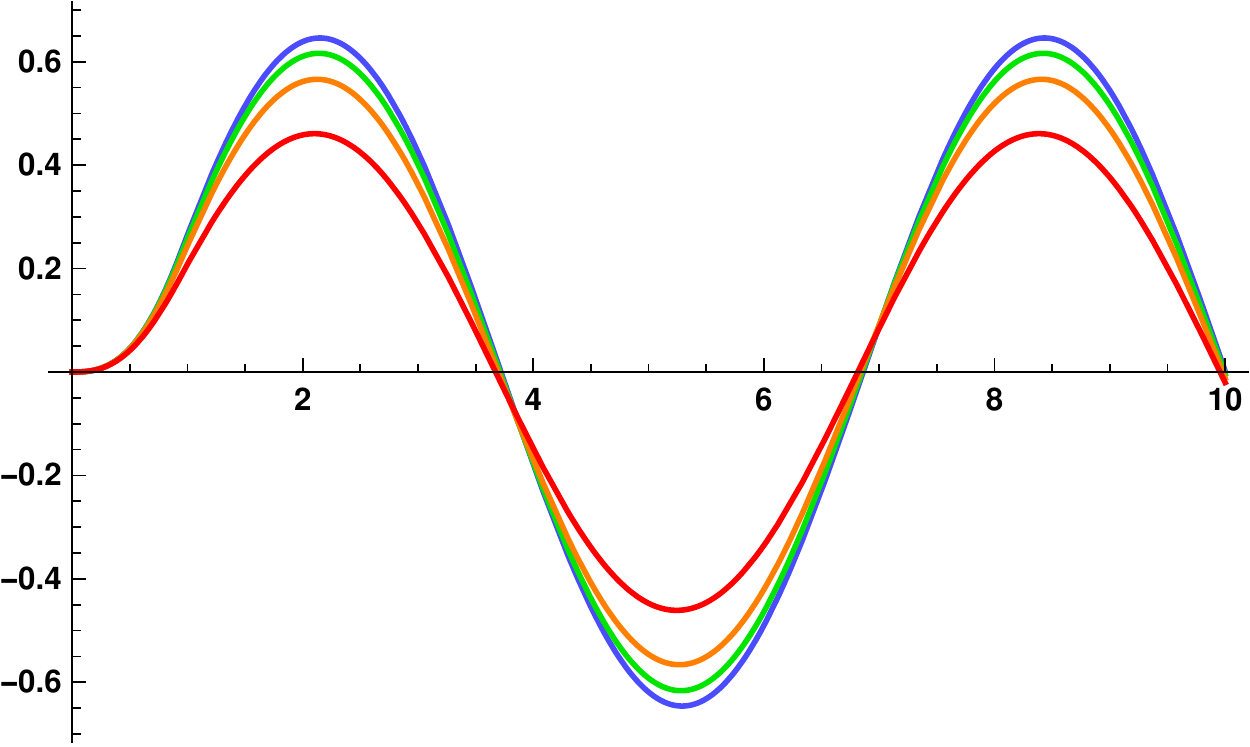} \qquad \qquad & \includegraphics[angle=0,width=0.43\textwidth]{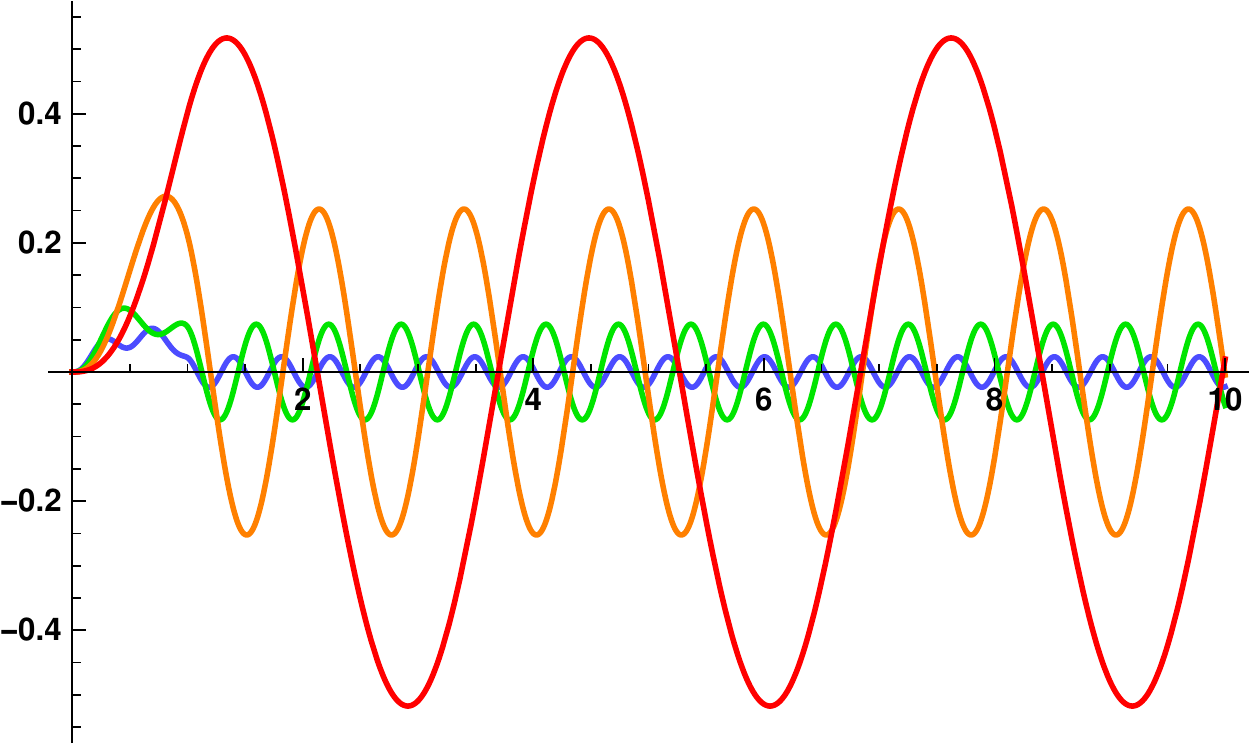}\\
  (a) \qquad \qquad & (b)
\end{array}
$$
\begin{picture}(0,0)
\put(10,152){{\scriptsize $\delta S_A(t)$}}
\put(245,152){{\scriptsize $\delta S_A(t)$}}
\put(200,92){{\scriptsize $t/t_*$}}
\put(435,92){{\scriptsize $t/t_*$}}
\end{picture}
\vspace{-0.7cm}
\caption{\small Entanglement entropy after a Floquet Quench as a function of (a) dimension and (b) frequency. The colors blue, green, orange and red denote $d=\{5,4,3,2\}$ in part (a) and $\Omega=\{15, 10, 5, 2\}$ in part (b), respectively. We have set $\theta=0.1, z=1.5$.}
\label{fig:perEEvsDW}
\end{figure}
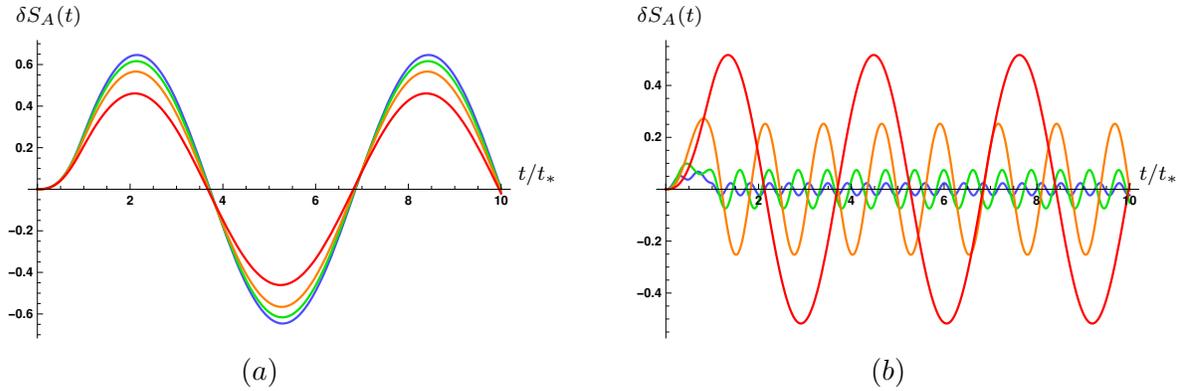\\

\noindent  We see that at higher dimensions, the amplitude of the entanglement entropy is higher but it decreases as the frequency increases. \\

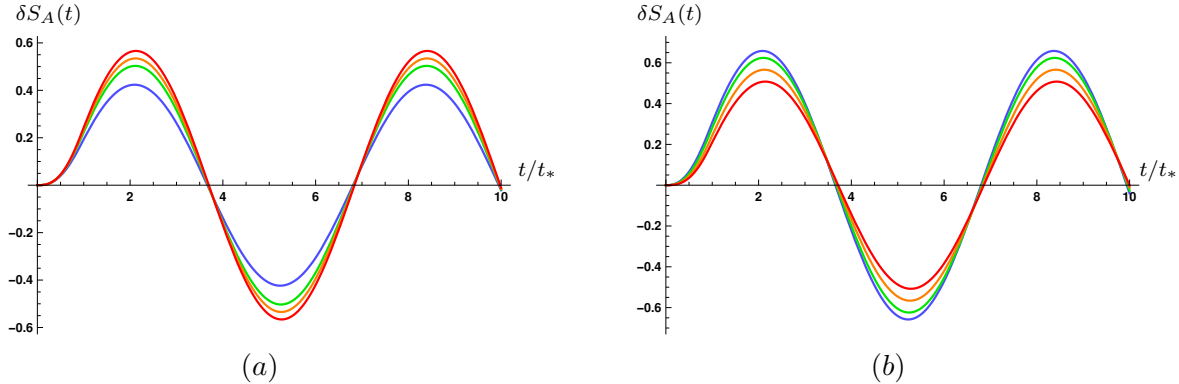
\begin{figure}[!htb]
$$
\begin{array}{cc}
  \includegraphics[angle=0,width=0.43\textwidth]{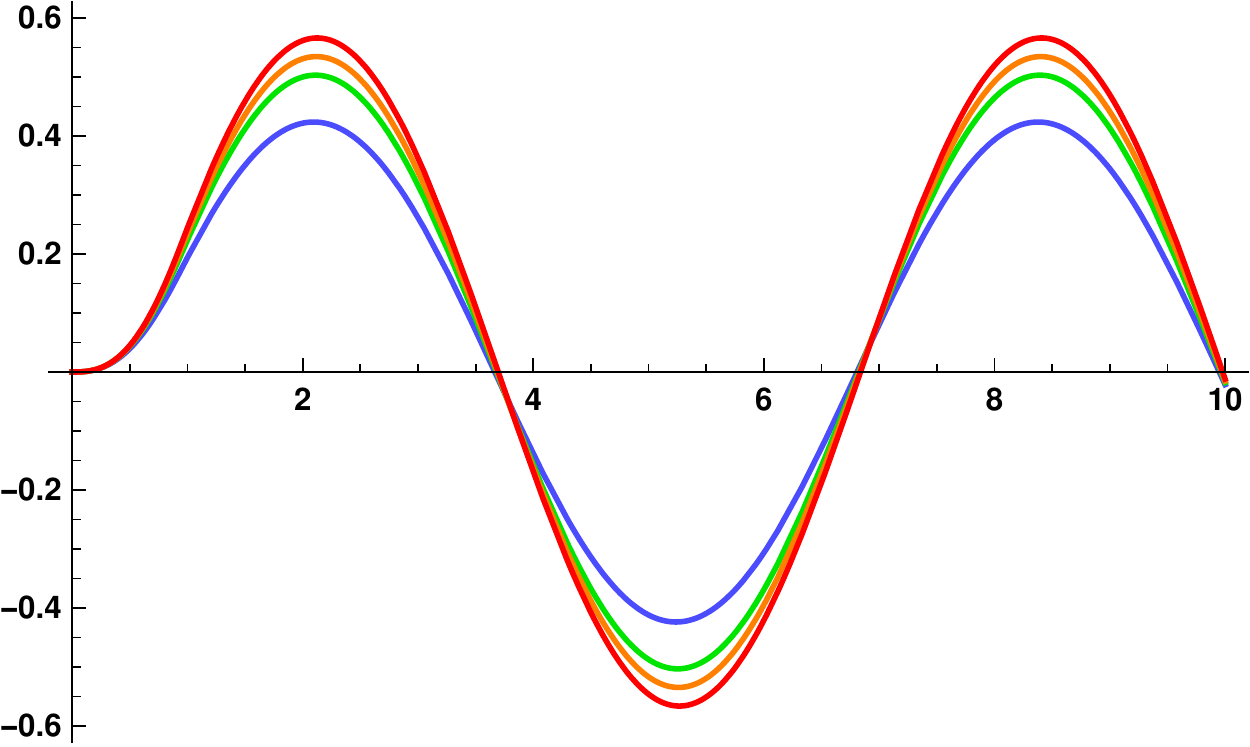} \qquad \qquad & \includegraphics[angle=0,width=0.43\textwidth]{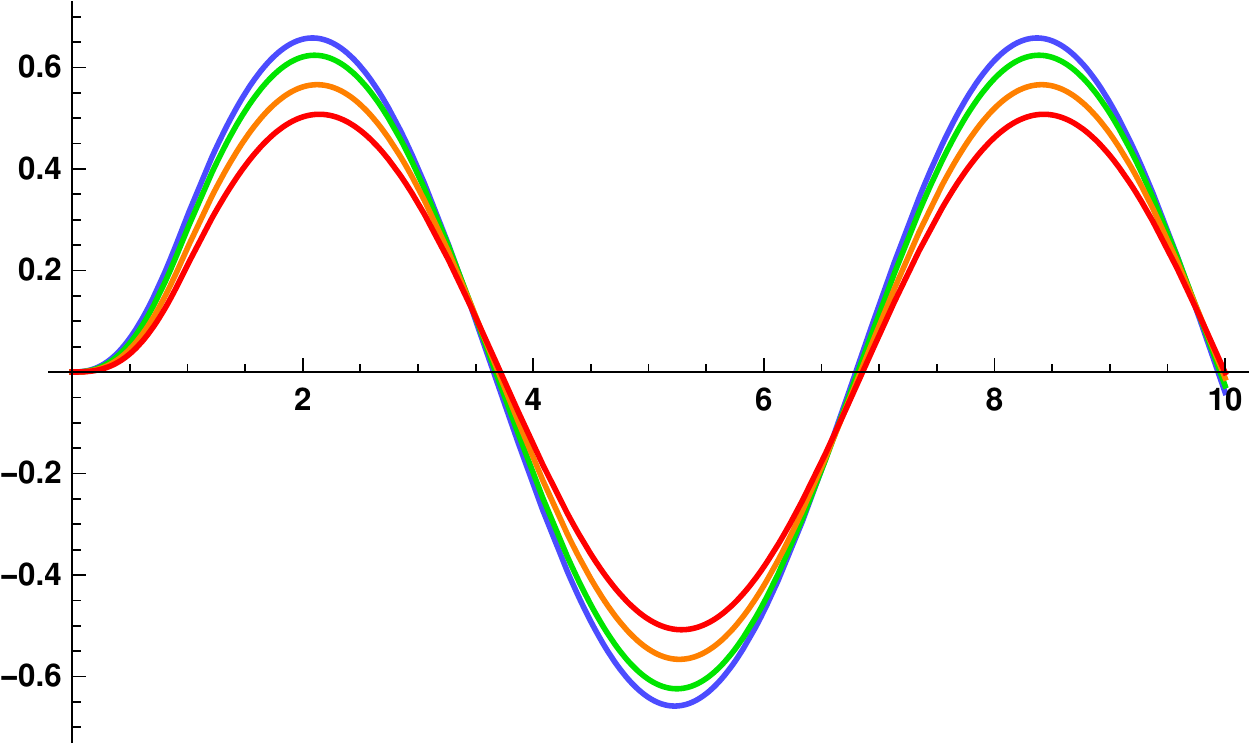}\\
  (a) \qquad \qquad & (b)
\end{array}
$$
\begin{picture}(0,0)
\put(10,151){{\scriptsize $\delta S_A(t)$}}
\put(245,151){{\scriptsize $\delta S_A(t)$}}
\put(200,92){{\scriptsize $t/t_*$}}
\put(435,92){{\scriptsize $t/t_*$}}
\end{picture}
\vspace{-0.7cm}
\caption{\small Entanglement entropy after a Floquet Quench as a function of (a) $\theta$ and (b) $z$. The colors blue, green, orange and red denote $\theta=\{1.3, 0.8, 0.5, 0.1\}$ in part (a) and $z=\{2.1, 1.8, 1.5, 1.3\}$ in part (b), respectively. We have set $d=3, \Omega=1$.}
\label{fig:perEEvsNRparams}
\end{figure}

\noindent In Figure \ref{fig:perEEvsNRparams}, we plot the dependence of entanglement entropy on the non-relativistic parameters $\theta$ and $z$. We see an increase in the amplitude upon increasing $z$ but a decrease as $\theta$ is increased. 
\noindent The above graphs are indicative of how the amplitude of the entanglement entropy depends on various parameters but it is not very clear what is happening to the phase. Hence, we can now discuss both the amplitude and the phase in some more detail. In Figure \ref{fig:perAmps}, we plot the amplitude of entanglement entropy as a function of dimension, $z$ and $\theta$. Observe that the amplitude vanishes for $\theta>d-1$ because the Vaidya solution is defined only for these values.
\begin{figure}[!htb]
$$
\begin{tabular}{ccc}
    \includegraphics[width=0.3\textwidth]{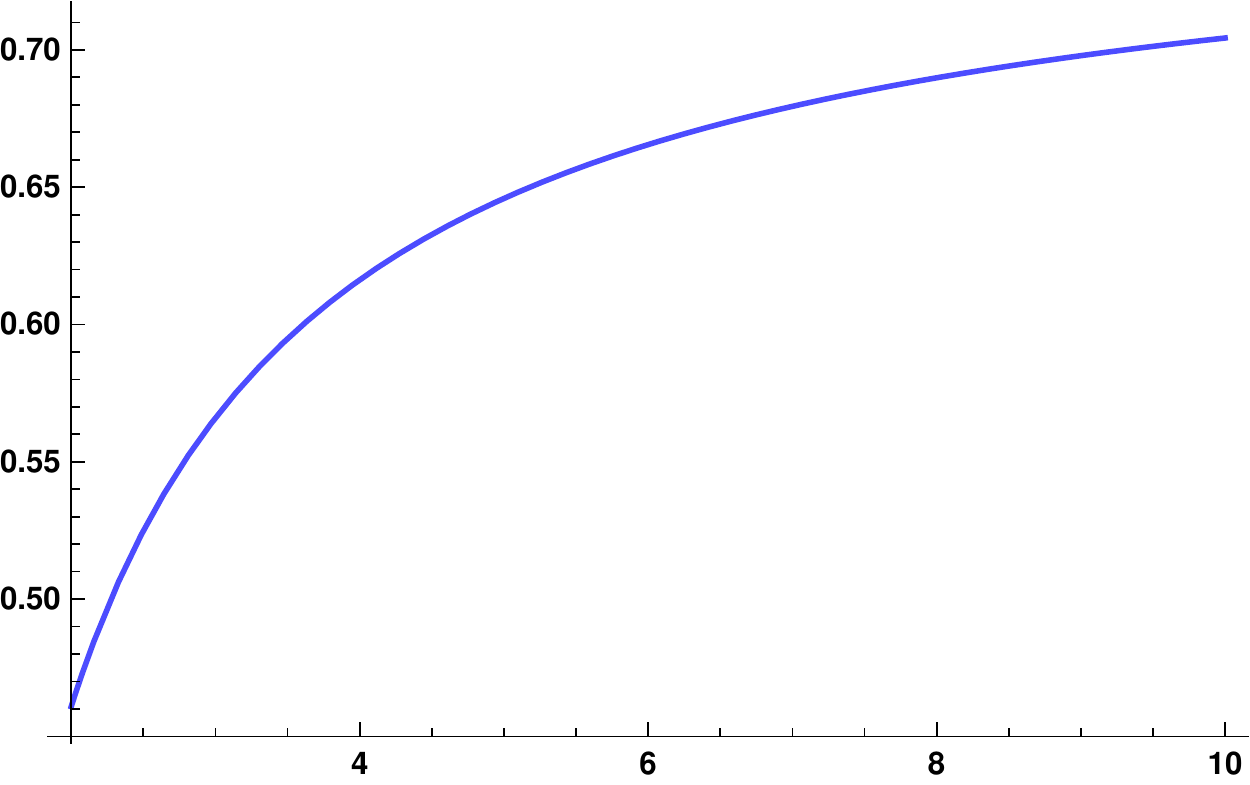} & \hspace{2mm} \includegraphics[width=0.3\textwidth]{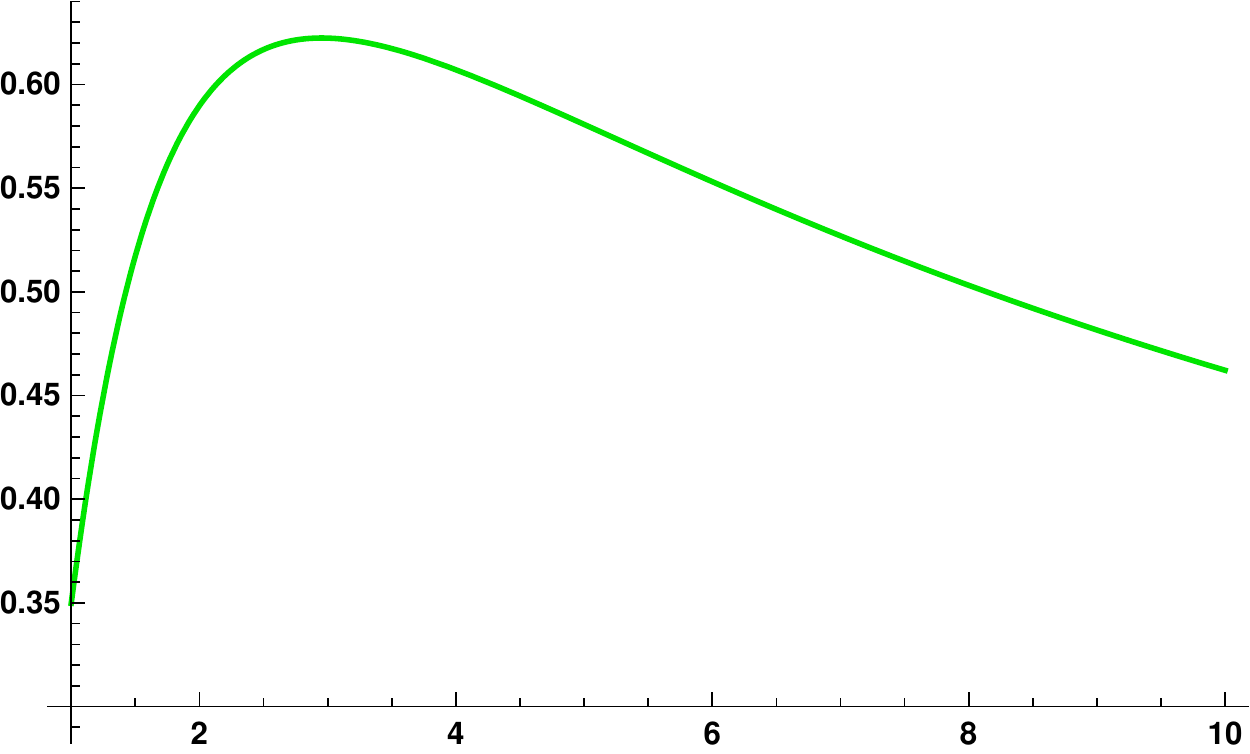} & \hspace{2mm} \includegraphics[width=0.3\textwidth]{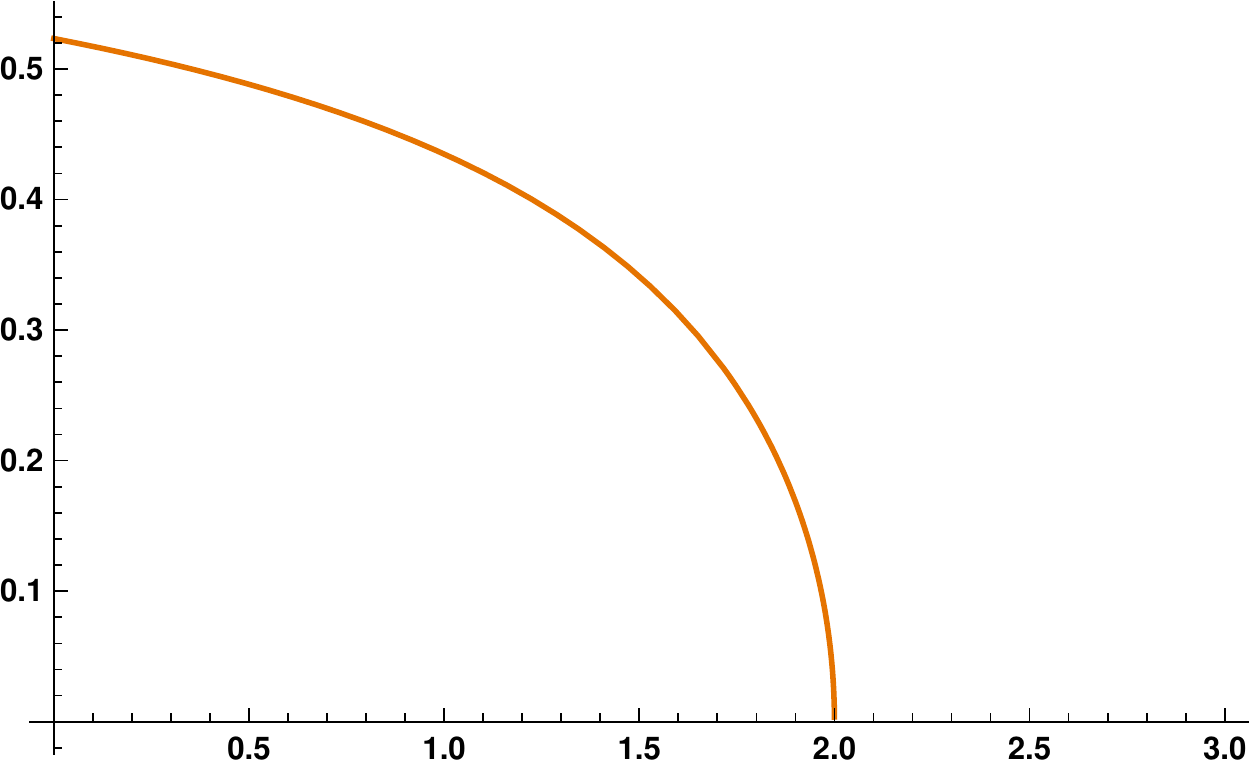} \vspace{5mm} \\
\end{tabular}
$$
\begin{picture}(0,0)(-460,-145)
\put(-453,-28){ {\small $\Psi$}}
\put(-299,-28){ {\small $\Psi$}}
\put(-150,-28){ { \small $\Psi$}}
\put(-320,-106){{ \small $d$}}
\put(-170,-106){ {\small $z$}}
\put(-20,-106){ {\small $\theta$}}
\end{picture}
\vspace{-0.7cm}
\caption{Amplitude $\Psi$ of entanglement entropy as a function of (a) dimension, (b) $z$ and (c) $\theta$. In (a), we have $\{z=1.5, \theta=0.1\}$. In (b), we have $\{d=3,\theta=0.1\}$ and in (c), $\{d=3, z=1.5\}$. We have set $\Omega=1$ everywhere.}
\label{fig:perAmps}
\end{figure}
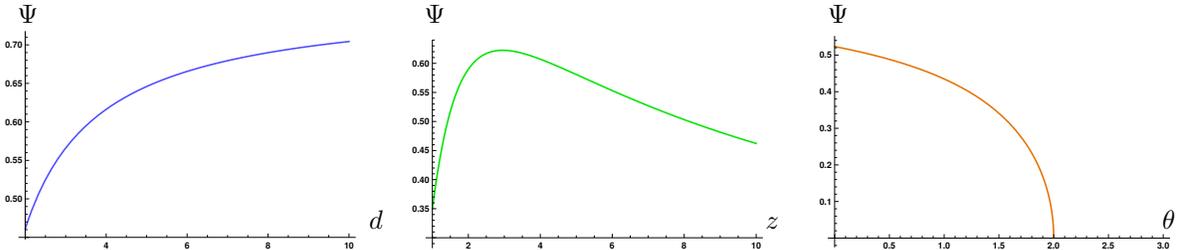 \\

\begin{figure}[!ht]
$$
\begin{tabular}{ccc}
    \includegraphics[width=0.3\textwidth]{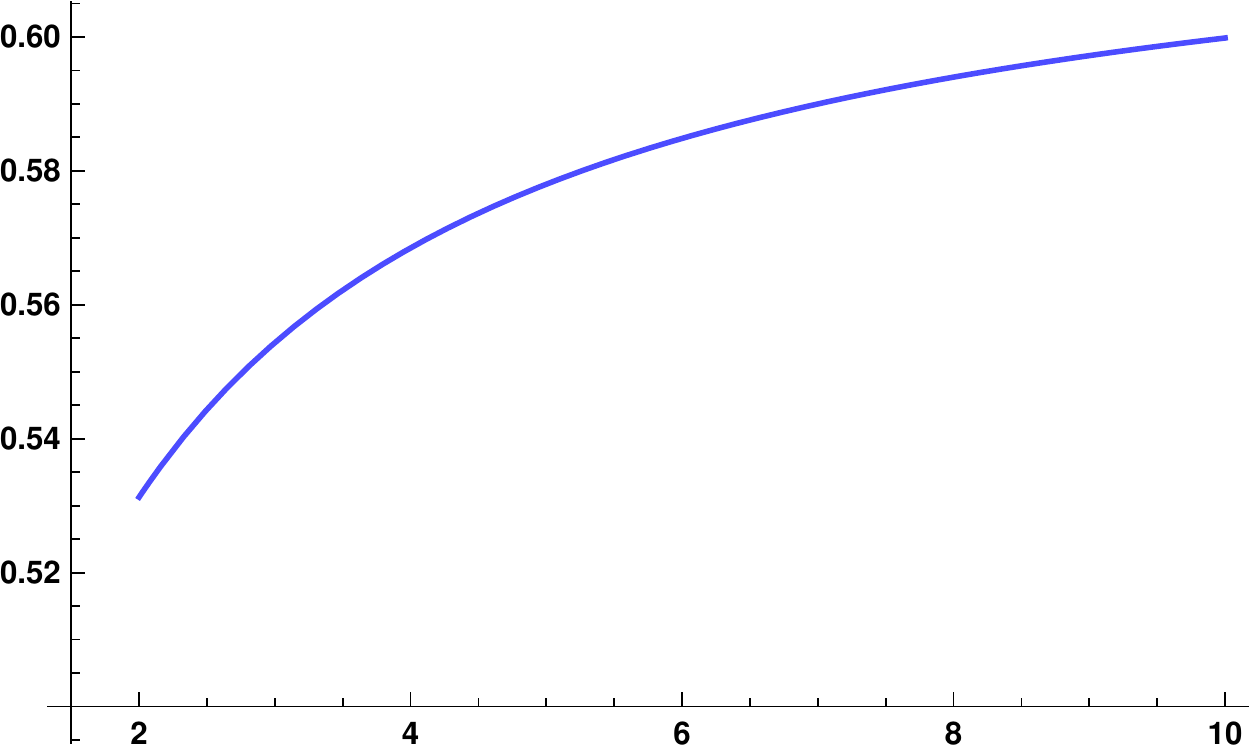} & \hspace{2mm} \includegraphics[width=0.3\textwidth]{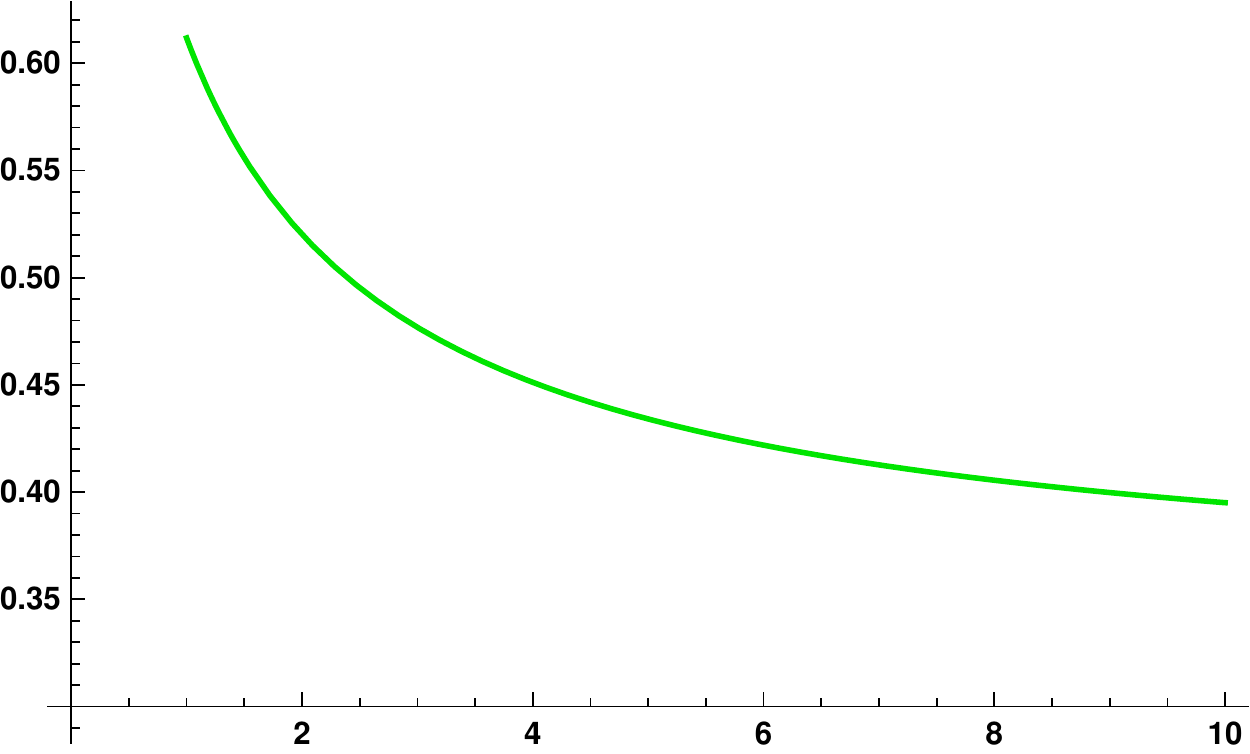} & \hspace{2mm} \includegraphics[width=0.3\textwidth]{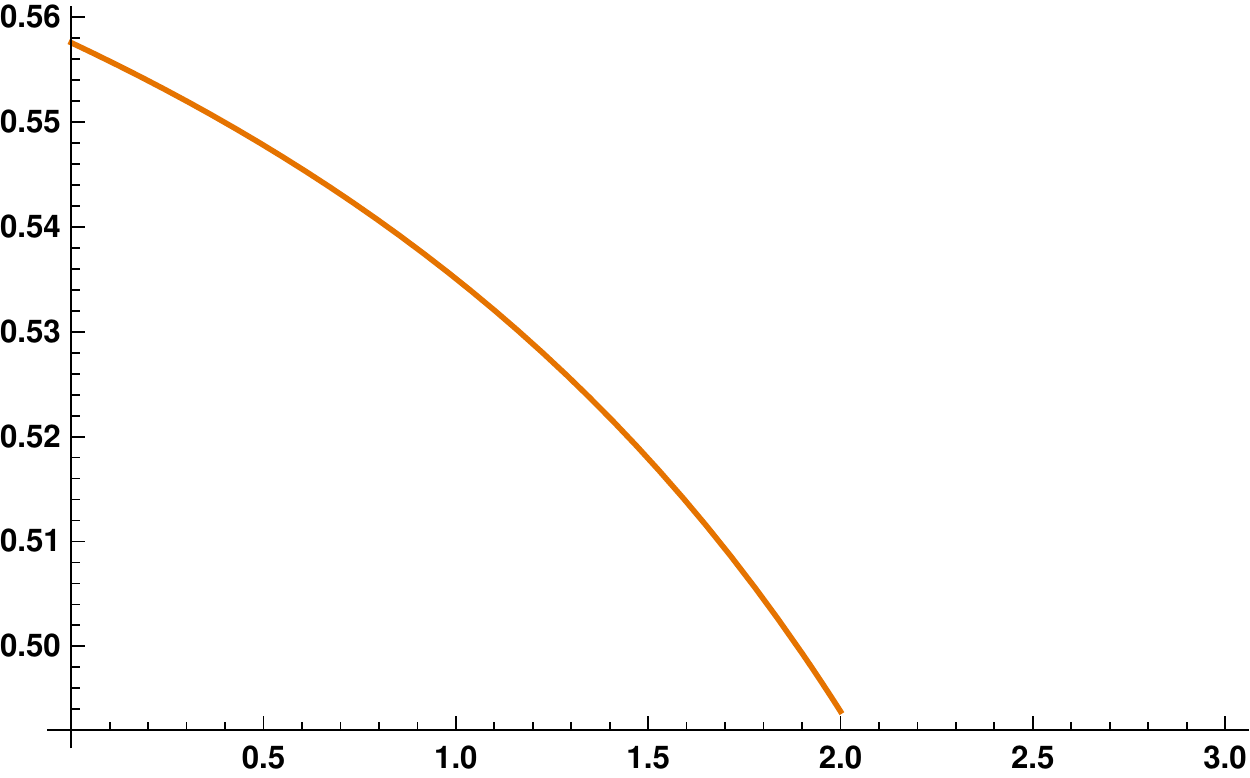} \vspace{5mm} \\
\end{tabular}
$$
\begin{picture}(0,0)(-460,-145)
\put(-450,-28){{\small $\Phi$}}
\put(-295,-28){{\small $\Phi$}}
\put(-142,-28){{\small $\Phi$}}
\put(-320,-105){{\small $d$}}
\put(-166,-105){{\small $z$}}
\put(-18,-105){{\small $\theta$}}
\end{picture}
\vspace{-0.7cm}
\caption{Relative phase $\Phi$ of entanglement entropy as a function of (a) dimension, (b) $z$ and (c) $\theta$. In (a), we have $\{z=1.5, \theta=0.1\}$. In (b), we have $\{d=3,\theta=0.1\}$ and in (c), $\{d=3, z=1.5\}$. We have set $\Omega=1$ everywhere.}
\label{fig:perPhases}
\end{figure}
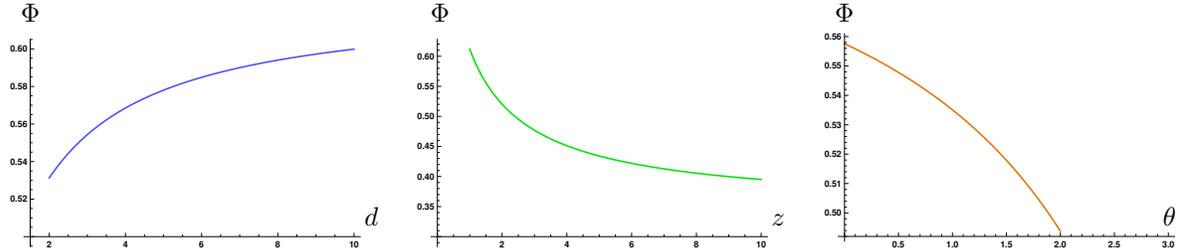

\noindent In Figure \ref{fig:perPhases}, we plot the phase of the entanglement entropy relative to the energy density as a function of dimension, $z$ and $\theta$. With this we conclude our discussion of Floquet quench.

\section{Summary and Outlook}
\label{summary}

\noindent In this paper, we studied global quantum quenches to holographic hyperscaling-violating-Lifshitz (\textbf{hvLif}) field theories, using entanglement entropy of a subregion as a probe to study thermalization. In the Introduction \ref{intro}, we argued that such a theories appear in the IR description of finite energy and finite charge density excited states in the CFT. Thus, our results describe thermalization in such states approximately. \\

\noindent In Section \ref{holo_renorm}, we specialized our discussion to small subregions ($\ell^z \ll 1/T$) and precisely defined the universal corrections we calculate. Using $\ell^z T$ as a perturbative parameter, we argued in \ref{perturb_expansion} that the holographic entanglement entropy becomes simple. In Section \ref{Vaidya_model}, we proposed the \textbf{hvLif}-Vaidya geometry
\begin{align}
\begin{split}
 ds^2 &= \frac{1}{u^{2 d_\theta/(d-1)}} \,  \bigg(-\frac{2 du \, dv}{u^{2(z-1)}} -\frac{f(u,v) \, dv^2}{u^{2(z-1)}} + dx_i^2 \bigg)  \, ,\\
 f(u,&v)  = 1 - g(v) \, \bigg(\frac{u}{u_H} \bigg)^{z+d_\theta} \, .
\end{split}
\end{align}
as a simple holographic toy model to study the time evolution of entanglement entropy. \\

\noindent Using the perturbative expansion in \textbf{hvLif}-Vaidya geometry, we calculated the holographic entanglement entropy of a small subsystem $A$ in \ref{spread_EE} to be
\begin{equation}
\delta S(t) = \frac{ \ell_p^{d-2}}{4 \, G_N \, u_H^{d_\theta+z}} \, \int_0^{u_*} d u \, u^z \, \sqrt{1 - \left[\frac{u}{u_*}\right]^{2d_\theta}} \, g \Big(t - \frac{u^z}{z} \Big) ,
\end{equation}
for a general quench parameterized by the function $g(v)$. Motivated by the simplicity of this equation, we interpreted this equation in Section \ref{linear_response} as a \textit{Linear Response} 
\begin{equation}
 \delta S(t) = \int_{-\infty}^{\infty} \, dt' \, \mathfrak{m}(t-t') \, \mathfrak{n}(t')
\end{equation}
with the \textit{Source} function $\mathfrak{m}(t)$ given by the energy density of the quench
\begin{equation}
 \mathfrak{m}(t) \equiv \epsilon(t) = \frac{d_\theta \, g(t) }{16 \, \pi \, G_N \, u_H^{d_\theta+z}}  \, .
\end{equation}
and the \textit{Kernel} function $\mathfrak{n}(t)$ given in terms of the shape and the size of the subregion
\begin{equation}
 \mathfrak{n}(t) =\frac{2 \, \pi \, \mathcal{A}_{\Sigma} \,  (z \, t)^{\frac{1}{z}} }{d_\theta} \, \, \bigg[1-\bigg( \frac{t}{t_*} \bigg)^{\frac{2 \, d_\theta}{z}} \bigg]^{\frac{1}{2}} \, \bigg[ \Theta(t) - \Theta(t - t_*) \bigg] \, ,
\end{equation}
This interpretation allowed us to define an adiabatic time-dependent first law of entanglement entropy for small subsystems in \ref{first_law}
\begin{equation}
  \delta S_A(t) =  \frac{\delta E_A(t)}{T_A} -   \int_0^{t_*} dt' \,  \frac{d \epsilon(t-t')}{d t'} \, \mathfrak{B}(t')  \, .
\end{equation}
where the function $\mathfrak{B}(t)$ can be thought of as the \textit{anti-derivative} of the \textit{kernel} function
\begin{equation}
 \mathfrak{B}(t)  \equiv  \frac{ 2 \pi \mathcal{A}_{\Sigma} \, (z \, t)^{1+\frac{1}{z}}}{d_\theta \, (z+1)} \,  \,   {}_2F_1 \bigg[ -\frac{1}{2}, \frac{z+1}{2 d_\theta}, \frac{2 d_\theta+z+1}{2 d_\theta}; \bigg( \frac{t}{t_*} \bigg)^{\frac{2 d_\theta}{z}}  \, \bigg]   - \frac{V_A}{T_A} \, . 
\end{equation}
Here, $-\frac{V_A}{T_A}$ denotes an integration constant that is chosen to reproduce the time-independent first law of entanglement entropy \eqref{eq:first_law_const_energy}. See the discussion above equation \eqref{eq:general_first_law} for details. Moreover, in \ref{relative_entropy}, we used the linear response to study a time-dependent analogue of relative entropy
\begin{equation}
\delta S_{\text{rel}}(t) \equiv \frac{\delta E_A(t)}{T_A} - \delta S_A(t) ,
\end{equation}
which we argued is a good parameter to characterize out-of-equilibrium states at a time $t$ compared to an equilibrium state with the energy density $\epsilon(t)$. \\

\noindent In Section \ref{special_cases}, we started studying special examples of global quenches, in particular
\begin{align}
\begin{split}
g(t) &= \Theta(t) \, ,\\
g(t) &= \sigma \, t^p \, \big[\Theta(t)-\Theta(t-t_q) \big] + \epsilon_0  \, \Theta(t-t_q) \, , \\
g(t) &= \sin(\Omega \, t ) \,  \Theta(t)   \, . 
\end{split}
\end{align}
In \ref{instantaneous}, we studied the first of these cases i.e. the instantaneous quench $g(t) = \Theta(t)$ and showed that the entanglement entropy is
\begin{align}
\begin{split}
\delta S(t) &= \delta S_{eq} \, \mathcal{F}(x) \, ,  \\ 
\mathcal{F}(x) &\equiv  \frac{\Gamma \big[\frac{3d_\theta + z+ 1}{2 d_\theta} \big] }{\Gamma \big[\frac{3}{2} \big] \, \Gamma \big[ \frac{z+1}{2 d_\theta}\big]} \, \cdot \beta \bigg[x^{2 d_\theta} , \frac{z+1}{2 d_\theta}, \frac{3}{2} \bigg] \, , \\
\delta S_{eq} &= \frac
{\sqrt{\pi} \ell_p^{d-2} \, \, u_*^{1+z} \, \Gamma \big[\frac{2d_\theta+z+1}{2 d_\theta} \big]}
{8 \, G_N \, (1+z) \, u_H^{d_\theta+z} \, \Gamma \big[\frac{3 d_\theta+z+1}{2d_\theta}\big]} \, .
\end{split}
\end{align}
We then studied this example in great detail: deriving the early time universal growth given in equation \eqref{eq:deltaS_quadratic}, the quasilinear growth with a slope (the \textit{entanglement velocity}) given by equation \eqref{eq:EE_velocity_HSV} and the near-saturation regime of the evolution characterized using equations \eqref{eq:near_sat}. \\

\noindent For the case of the power law quench, we studied the entanglement entropy in detail in \ref{power-law}. For arbitrary real power $p \in \mathbb{R}$, we obtained a formal expression for the entanglement entropy as an infinite series using the explicit function
\begin{align}
 \begin{split}
  \mathcal{I}^{(p)}(t,\tau) &=  \frac{  \mathcal{A}_{\Sigma} \, }{8 \, G_N \,  \,  u_H^{d_\theta+z}} \, \sum\limits_{k=0}^\infty \begin{pmatrix} 
                                       p \\
                                       k
                                      \end{pmatrix} \, t^{p-k} \,  \,\frac{ (z \, \tau)^{1+\frac{1}{z}} \, (-\tau)^{k}}{(z+1+kz)} \,   \\
&\qquad \times {}_2F_{1}\bigg[-\frac{1}{2}, \frac{z+1+k z}{2 d_\theta},  \frac{2 d_\theta + z+1+k z}{2 d_\theta}; \big[\frac{\tau}{t_*} \big]^{\frac{2 d_\theta}{z}} \bigg]  .
 \end{split}
\end{align}
For an integral power $p \in \mathbb{Z}$, we derived an explicit closed form expression for the entanglement entropy in equations \eqref{eq:HSV_S1_power} and \eqref{eq:HSV_S2_power}. We plotted the evolution of (normalized) entanglement entropy $\delta S(t)/\delta S_{\text{eq}}$, relative entropy $\delta S_{\text{rel}}(t)/\delta S_{\text{eq}}$ and entanglement velocity $\mathfrak{R}_{\text{HSV}}(t)$ in Figures \ref{fig:QuantitiesvsP}, \ref{fig:StuffvsD}, \ref{fig:StuffvsZ}, and \ref{fig:StuffvsTheta}. In \ref{linear-pump}, we made some comments about the interesting particular case of the linear in time quench. \\

\noindent Finally, in \ref{Floquet}, we studied the periodic in time quench, also called Floquet quench. We derived a formal expression for entanglement entropy in equation \eqref{eq:floquet_EE} as an infinite series. This expression is not illuminating, so we plotted entanglement entropy in Figure \ref{fig:perEEvsSource} as a function of time and in Figure \ref{fig:perEEvsDW} as a function of dimension and frequency. Figures \ref{fig:perAmps} and \ref{fig:perPhases} show the amplitude and the phase of the entanglement entropy as a function of $d$, $\theta$ and $z$. \\

\noindent In this paper, we could not study the problem of holographic renormalization in detail. That is an obvious direction to extend our work in. Further, notwithstanding the expectation that our simple holographic toy model captures gross universal features of thermalization, it is not very realistic. It would be interesting to have a more realistic holographic model for studying quenches in \textbf{hvLif} field theories. We wish to come back to these issues in future.

\section*{Acknowledgements}

I would like to thank Juan F. Pedraza for several useful discussions. I am grateful to Juan F. Pedraza, Jay Armas, Rik van Bruekelen, Gerben W.J. Oling and an anonymous referee for suggestions to improve the draft. This work is supported by the Netherlands Organisation for Scientific Research (NWO-I), which is funded by the Dutch Ministry of Education, Culture and Science (OCW).

\section*{Data Availability Statement}

All data used in this study have been described in the article.

\appendix

\section{Stress Tensor for hvLif Theories}
 \label{stress_tensor_HSV}

In this Appendix, we use the so-called Minimal Approach to find the stress-energy tensor for an asymptotically \textbf{hvLif} black hole in general dimensions. This method is essentially an application of the first law of thermodynamics. Let us consider the following metric of a static black hole in a \textbf{hvLif} background \cite{Gouteraux:2011ce, Huijse:2011ef, Dong:2012se, Alishahiha:2012qu, Pedraza:2018eey}
\begin{equation}
\label{eq:static_hvLif_BH}
  ds^2 = \frac{1}{u^{2d_\theta/(d-1)}} \bigg(\frac{du^2}{f(u)} -\frac{f(u)}{u^{2(z-1)}} \, dt^2 +   dx_i^2 \bigg) \, ,
\end{equation}
where the blackening function is
\begin{equation}
 f(u) = 1 - \bigg(\frac{u}{u_H}\bigg)^{(d_\theta+z)} \, ,
\end{equation}
and $u_H$ is as usual the horizon radius. The entropy of this black hole will be given by the area of the horizon in units of the $(d+1)$ dimensional Planck area $4 \, G_N$. Using the above metric we get
\begin{equation}
 S_{BH}=\frac{1}{4 \, G_N} \, \bigg( \frac{1}{u_H^{2d_\theta/(d-1)}} \bigg)^{\frac{d-1}{2}} \times \text{Volume} \, .
\end{equation}
Hence the entropy density becomes
\begin{align}
\begin{split}
s_{BH} &= \frac{1}{4 \, G_N} \,  \frac{1}{u_H^{d_\theta}} \, .
\end{split}
\end{align}
Now let us calculate the temperature of this black hole. First we Eulideanize the metric as follows
\begin{align}
\begin{split}
ds_E^2&=  \frac{1}{u^{2d_\theta/(d-1)}} \bigg( \frac{f(u)}{u^{2(z-1)}} \, d\tau^2 + \frac{du^2}{f(u)} + dx_i^2 \bigg) \, .
\end{split}
\end{align}
Then, we evaluate the blackening function near the horizon.
\begin{equation}
f(u) \sim \frac{(u_H - u) \, (d_\theta +z)}{u_H} \, .
\end{equation}
Taking into account the function multiplying the paranthesis, the $x^i$ directions form a conformal orthogonal sphere and decouple. So we focus only on the $(\tau,r)$ direction. Then the near-horizon metric becomes
\begin{align}
\begin{split}
ds^2_E&= \Omega^2(u) \, \bigg( \frac{(u_H-u) \, (d_\theta+z)}{u_H \, u_H^{2(z-1)}} \, d\tau^2 + \frac{u_H \, du^2}{(u_H-u) \, (d_\theta+z)} \bigg) \, ,
\end{split}
\end{align}
where the conformal pre-factor
\begin{equation}
 \Omega^2(u) =  u^{-\frac{2d_\theta}{(d-1)}} \, 
\end{equation}
plays no role in defining the black hole temperature. We also note that we have evaluated the denominator of $d \tau^2$ at the horizon radius $u_H$, being consistent with the fact that we are evaluating the metric near the horizon. We want to find the periodicity of the coordinate $\tau$. To do this, let us define the following coordinates
\begin{equation}
\rho^2 \equiv \, (u_H - u) \, , \quad \tilde{\tau} \equiv  \frac{(d_\theta+z)}{\, u_H^{z}} \, \tau \, .
\end{equation}
In terms of these new coordinates the metric becomes:
\begin{equation}
 ds^2_E = \Omega^2(u) \, \bigg(\frac{\rho^2 \, u_H \, d\tilde{\tau}^2}{(d_\theta+z)} + \frac{4 \, u_H \, \rho^2 \, d \rho^2}{\rho^2 \, (d_\theta+z) }  \bigg) \, .
\end{equation}
Absorbing the factor $\frac{4 \,u_H}{(d_\theta+z)}$ into $\Omega^2(u)$, we get
\begin{equation}
 ds^2_E= \Omega'^2(u) \bigg( \frac{\rho^2 \, d\tilde{\tau}^2 }{4} + d \rho^2  \bigg) \, .
\end{equation}
Here the coordinate $\tilde{\tau}/2$ has periodicity $2 \pi$. Thus, denoting the periodicity of the coordinate $\tau$ by $\beta$, we obtain
\begin{equation}
\label{eq:temp_explicit}
 \beta = 4 \, \pi \, \bigg( \frac{d_\theta+z}{ u_H^{z}} \bigg)^{-1} \implies T=\frac{(d_\theta+z)}{4 \, \pi \, u_H^{z}} \, .
\end{equation}
As one can check, we get the same expression for the temperature if we define it using the equation
\begin{equation}
 T = \frac{1}{4 \pi} \bigg| \frac{df(u)}{du} \bigg|_{u_H} \, .
\end{equation}
Now we can proceed to obtain an expression for energy density using the first law
\begin{equation}
 d \epsilon = T \, d s_{BH} \, .
\end{equation}
The differential of the entropy density is
\begin{equation}
 ds_{BH} =  \frac{1}{4 \, G_N} \,  \frac{-d_\theta}{u_H^{(d-2-\theta)}} \, .
\end{equation}
Therefore the energy density is given by
\begin{align}
 \begin{split}
d \epsilon &= \frac{(d_\theta+z)}{4 \, \pi \, u_H^{z}}  \,   \frac{1}{4 \, G_N} \,  \frac{-d_\theta}{u_H^{(d-2-\theta)}} \, ,\\
&= - \frac{d_\theta}{16 \, \pi G_N} \, \frac{(d_\theta+z)}{u_H^{d-2-\theta+z}} \, .
 \end{split}
\end{align}
Integrating and setting the integration constant to zero, this becomes
\begin{equation}
 \epsilon = \frac{d_\theta}{16 \, \pi G_N} \, \frac{1}{u_H^{d_\theta+z}} \, .
\end{equation}
In the case of small subsystems, the black hole entropy becomes the final entanglement entropy after the perturbation. \\

\noindent We could now consider making the bulk solution time-dependent in an adiabatic way, such that the blackening function is given by
\begin{equation}
f(t,u) = 1- g(t) \, \bigg( \frac{u}{u_H} \bigg)^{d_\theta+z} 
\end{equation}
with $g(t)$ being an adiabatic function in time. This property implies thermodynamics can still be defined as before and we can repeat the above derivation to obtain an energy density
\begin{equation}
\epsilon(t) =   \frac{d_\theta}{16 \, \pi G_N} \, \frac{g(t)}{u_H^{d_\theta+z}} \, .
\end{equation}
This remains mostly constant as $g(t)$ varies slowly with time. \\

\noindent We may now consider a time-dependent bulk geometry that is not a small perturbation of \eqref{eq:static_hvLif_BH}. For example, consider the scenario where the blackening function changes abruptly at time $t=0$
\begin{equation}
f(t,u) = 1- \Theta(t) \, \bigg(\frac{u}{u_H} \bigg)^{d_\theta+z} \, .
\end{equation}
In this case, thermodynamics is well-defined before and after time $t=0$ and the respective bulk geometries are in fact static. If we repeat the our derivation, we will see that there is no hindrance and now temperature will be given by equation \eqref{eq:temp_explicit} times $\Theta(t)$. Hence the energy density for this time-dependent bulk solutions will be
\begin{equation}
\epsilon(t) =   \frac{d_\theta}{16 \, \pi G_N} \, \frac{\Theta(t)}{u_H^{d_\theta+z}} \,  . 
\end{equation}
Thus for adiabatic time-dependent bulk geometries or for the ones with abrupt changes, the expression for the energy density is the same. This suggests that for general time-dependence in the blackening function, the energy density may be given by the same expression. In particular, we expect the energy density to be
\begin{equation}
\label{eq:HSV_energy_density}
 \epsilon(t) = \frac{d_\theta}{16 \, \pi G_N} \, \frac{g(t)}{u_H^{d_\theta+z}} \, ,
\end{equation}
where $g(t)$ is now a general time-dependent function that characterizes the time-dependent bulk geometry.

\hrulefill

\bibliographystyle{JHEP}

\bibliography{draft-biblio}

\end{document}